\documentclass[ALICE,manyauthors]{cernphprep}
\usepackage[comma,square,numbers,sort&compress]{natbib}
\usepackage{hyperref}

\usepackage{lineno}
\usepackage{xspace}
\usepackage{multirow}
\usepackage{graphicx}
\usepackage[loose]{units}
\usepackage{upgreek}
\usepackage{csquotes}
\usepackage{enumerate}
\usepackage{amsmath,amssymb,amsbsy,mathrsfs} % for advanced math typesetting
\usepackage[T1]{fontenc}
\usepackage{orcidlink}

\begin{document}
%%%%%%%%%%%%%%%%%%%%%%%%%%%%%%%%%%%%%%%%%%%%%%%%%%
% These are some new commands that may be useful 
% for paper writing in general. If other newcommands
% are needed for your specific paper, please feel 
% free to add them here. 
%
% The currently available commands are organized in: 
% 1) Systems
% 2) Quantities
% 3) Energies and units
% 4) Detectors
% 5) particle species 
%%%%%%%%%%%%%%%%%%%%%%%%%%%%%%%%%%%%%%%%%%%%%%%%%%

% 1) SYSTEMS 
\newcommand{\pp}             {pp\xspace}
\newcommand{\ppbar}        {\mbox{$\mathrm {p\overline{p}}$}\xspace}
\newcommand{\XeXe}         {\mbox{Xe--Xe}\xspace}
\newcommand{\PbPb}         {\mbox{Pb--Pb}\xspace}
\newcommand{\pA}             {\mbox{pA}\xspace}
\newcommand{\pPb}           {\mbox{p--Pb}\xspace}
\newcommand{\AuAu}         {\mbox{Au--Au}\xspace}
\newcommand{\dAu}           {\mbox{d--Au}\xspace}

% 2) QUANTITIES 
\newcommand{\s}                 {\ensuremath{\sqrt{s}}\xspace}
\newcommand{\snn}             {\ensuremath{\sqrt{s_{\mathrm{NN}}}}\xspace}
\newcommand{\pt}                {\ensuremath{p_{\rm T}}\xspace}
\newcommand{\xt}                 {\ensuremath{x_{\rm T}}\xspace}
\newcommand{\meanpt}       {$\langle p_{\mathrm{T}}\rangle$\xspace}
\newcommand{\ycms}           {\ensuremath{y_{\rm CMS}}\xspace}
\newcommand{\ylab}             {\ensuremath{y_{\rm lab}}\xspace}
\newcommand{\etarange}[1] {\mbox{$\left | \eta \right |~<~#1$}}
\newcommand{\yrange}[1]    {\mbox{$\left | y \right |~<~#1$}}
\newcommand{\dndy}           {\ensuremath{\mathrm{d}N_\mathrm{ch}/\mathrm{d}y}\xspace}
\newcommand{\dndeta}        {\ensuremath{\mathrm{d}N_\mathrm{ch}/\mathrm{d}\eta}\xspace}
\newcommand{\avdndeta}    {\ensuremath{\langle\dndeta\rangle}\xspace}
\newcommand{\dNdy}          {\ensuremath{\mathrm{d}N_\mathrm{ch}/\mathrm{d}y}\xspace}
\newcommand{\Npart}          {\ensuremath{N_\mathrm{part}}\xspace}
\newcommand{\Ncoll}           {\ensuremath{\langle N_\mathrm{coll}\rangle}\xspace}
\newcommand{\dEdx}           {\ensuremath{\textrm{d}E/\textrm{d}x}\xspace}
\newcommand{\RpPb}          {\ensuremath{R_{\rm pPb}}\xspace}

% 3) ENERGIES, UNITS
\newcommand{\snineH}     {$\sqrt{s}~=~0.9$~Te\kern-.1emV\xspace}
\newcommand{\sseven}     {$\sqrt{s}~=~7$~Te\kern-.1emV\xspace}
\newcommand{\stwoH}      {$\sqrt{s}~=~0.2$~Te\kern-.1emV\xspace}
\newcommand{\stwo}         {$\sqrt{s}~=~2.76$~Te\kern-.1emV\xspace}
\newcommand{\sfive}         {$\sqrt{s}~=~5.02$~Te\kern-.1emV\xspace}
\newcommand{\sthirteen}   {$\sqrt{s}~=~13$~Te\kern-.1emV\xspace}
\newcommand{\snntwo}     {$\sqrt{s_{\mathrm{NN}}}~=~2.76$~Te\kern-.1emV\xspace}
\newcommand{\snnfive}     {$\sqrt{s_{\mathrm{NN}}}~=~5.02$~Te\kern-.1emV\xspace}
\newcommand{\LT}             {L{\'e}vy-Tsallis\xspace}
\newcommand{\GeVc}        {Ge\kern-.1emV/$c$\xspace}
\newcommand{\MeVc}        {Me\kern-.1emV/$c$\xspace}
\newcommand{\TeV}           {Te\kern-.1emV\xspace}
\newcommand{\GeV}          {Ge\kern-.1emV\xspace}
\newcommand{\MeV}          {Me\kern-.1emV\xspace}
\newcommand{\GeVmass}  {Ge\kern-.2emV/$c^2$\xspace}
\newcommand{\MeVmass} {Me\kern-.2emV/$c^2$\xspace}
\newcommand{\lumi}           {\ensuremath{\mathcal{L}}\xspace}

% 4) DETECTORS 
\newcommand{\ITS}           {\rm{ITS}\xspace}
\newcommand{\TOF}          {\rm{TOF}\xspace}
\newcommand{\ZDC}          {\rm{ZDC}\xspace}
\newcommand{\ZDCs}        {\rm{ZDCs}\xspace}
\newcommand{\ZNA}          {\rm{ZNA}\xspace}
\newcommand{\ZNC}          {\rm{ZNC}\xspace}
\newcommand{\SPD}          {\rm{SPD}\xspace}
\newcommand{\SDD}          {\rm{SDD}\xspace}
\newcommand{\SSD}          {\rm{SSD}\xspace}
\newcommand{\TPC}          {\rm{TPC}\xspace}
\newcommand{\TRD}          {\rm{TRD}\xspace}
\newcommand{\VZERO}     {\rm{V0}\xspace}
\newcommand{\VZEROA}   {\rm{V0A}\xspace}
\newcommand{\VZEROC}   {\rm{V0C}\xspace}
\newcommand{\Vdecay} 	   {\ensuremath{V^{0}}\xspace}
\newcommand{\EMCal}       {\rm{EMCal}\xspace}
\newcommand{\DCal}          {\rm{DCal}\xspace}

% 4) PARTICLE SPECIES 
\newcommand{\ee}            {\ensuremath{e^{+}e^{-}}} 
\newcommand{\piz}           {\ensuremath{\pi^{0}}\xspace}
\newcommand{\zz}           {Z\ensuremath{^{0}}\xspace}
\newcommand{\ww}           {W\ensuremath{^{\pm}}\xspace}
\newcommand{\pip}           {\ensuremath{\pi^{+}}\xspace}
\newcommand{\pim}          {\ensuremath{\pi^{-}}\xspace}
\newcommand{\kap}          {\ensuremath{\rm{K}^{+}}\xspace}
\newcommand{\kam}         {\ensuremath{\rm{K}^{-}}\xspace}
\newcommand{\pbar}         {\ensuremath{\rm\overline{p}}\xspace}
\newcommand{\kzero}       {\ensuremath{{\rm K}^{0}_{\rm{S}}}\xspace}
\newcommand{\lmb}          {\ensuremath{\Lambda}\xspace}
\newcommand{\almb}        {\ensuremath{\overline{\Lambda}}\xspace}
\newcommand{\Om}          {\ensuremath{\Omega^-}\xspace}
\newcommand{\Mo}           {\ensuremath{\overline{\Omega}^+}\xspace}
\newcommand{\X}              {\ensuremath{\Xi^-}\xspace}
\newcommand{\Ix}             {\ensuremath{\overline{\Xi}^+}\xspace}
\newcommand{\Xis}           {\ensuremath{\Xi^{\pm}}\xspace}
\newcommand{\Oms}        {\ensuremath{\Omega^{\pm}}\xspace}
\newcommand{\degree}    {\ensuremath{^{\rm o}}\xspace}
\newcommand{\gpro}        {\ensuremath{\gamma^{\rm ~prompt}}\xspace}
\newcommand{\gdec}       {\ensuremath{\gamma^{\rm ~decay}}\xspace}
\newcommand{\gfra}         {\ensuremath{\gamma^{\rm ~frag.}}\xspace}

\newcommand{\raa}  {\ensuremath{R_{\rm AA}}\xspace}
\newcommand{\rcp}  {\ensuremath{R_{\rm CSP}}\xspace}
\newcommand{\rcs}  {\ensuremath{R_{\rm CSC}}\xspace}
\newcommand{\ptg}  {\ensuremath{p_{\rm T}^{\gamma}}\xspace}
\newcommand{\xtg}   {\ensuremath{x_{\rm T}^{\gamma}}\xspace}
\newcommand{\etag}  {\ensuremath{\eta^{\gamma}}\xspace}
\newcommand{\xe}       {\ensuremath{x_{\rm E}}\xspace}
\newcommand{\sigmalong}{\ensuremath{\sigma_{\rm long}^{2}}\xspace}
\newcommand{\sigmalongPb}{\ensuremath{\sigma_{\rm long,~5\times5}^{2}}\xspace}
\newcommand{\tr}{L1-\ensuremath{\gamma}\xspace}

\newcommand{\evt} {\ensuremath{\N_{\rm evt}}\xspace}
\newcommand{\lint}{\ensuremath{\mathcal{L}_{\rm int}}\xspace}

\newcommand{\alphaf}{\ensuremath{\alpha_{\rm MC}}\xspace}
\newcommand{\ptIsoChUE} {\ensuremath{p_{\rm T}^{\rm iso,~UE}}\xspace}
\newcommand{\ptIsoUE}      {\ensuremath{p_{\rm T}^{\rm iso,~ch,~UE}}\xspace}
\newcommand{\ptIsoCh}      {\ensuremath{p_{\rm T}^{\rm iso,~ch}}\xspace}
\newcommand{\ptIso}          {\ensuremath{p_{\rm T}^{\rm iso}}\xspace}
\newcommand{\ptT}   {\ensuremath{p_{\rm T}^{\rm trig}}\xspace}
\newcommand{\ptA}   {\ensuremath{p_{\rm T}^{\rm assoc}}\xspace}

\newcommand {\mom}   {\mbox{\rm  GeV$\kern-0.15em /\kern-0.12em c$}}
\newcommand {\gmom} {\mbox{\rm  GeV$\kern-0.15em /\kern-0.12em c$}}
\newcommand {\mass} {\mbox{\rm  GeV$\kern-0.15em /\kern-0.12em c^2$}}
\newcommand{\slfrac}[2]{\left.#1\right/#2}

%%%%%%%%%%%%%%%  Title page %%%%%%%%%%%%%%%%%%%%%%%%
\begin{titlepage}
% The dates below correspond to CERN approval
% please don't touch: EB chairs will take care
\PHyear{2024}       % required, will be obtained from CERN
\PHnumber{244}      % required, will be obtained from CERN
\PHdate{17 September}  % required, will be obtained from CERN
%%%%%%%%%%%%%%%%%%%%%%%%%%%%%%%%%%%%%%%%%%%%%%%%%%%%

%%% Put your own title + short title here:
\title{Measurement of the inclusive isolated-photon production cross section in \pp and \PbPb collisions at $\mathbf{\sqrt{\textit{s}_{\textup{NN}}} = 5.02}$~TeV}
\ShortTitle{Isolated-$\gamma$ production in \pp \& \PbPb col.  at \snnfive }   % appears on left page headers

%%% Do not change the next lines
\Collaboration{ALICE Collaboration\thanks{See Appendix~\ref{app:collab} for the list of collaboration members}}
\ShortAuthor{ALICE Collaboration} % appears on the right page headers, do not change

\begin{abstract}

The ALICE Collaboration at the CERN LHC has measured the inclusive production cross section of isolated photons at midrapidity as a function of the photon transverse momentum (\ptg), in \PbPb collisions in different centrality intervals, and in \pp collisions, at centre-of-momentum energy per nucleon pair of \snnfive.  
The photon transverse momentum range is between 10--14 and 40--140~\GeVc, depending on the collision system and on the \PbPb centrality class.
The result extends to lower \ptg than previously published results by the ATLAS and CMS experiments at the same collision energy. 
The covered pseudorapidity range is $|\etag| <0.67$.
The isolation selection is based on a charged particle isolation momentum threshold $\ptIsoCh = 1.5$~\GeVc within a cone of radii $R=0.2$ and $0.4$.
The nuclear modification factor is calculated and found to be consistent with unity in all centrality classes, and also consistent with the HG-PYTHIA model, which describes the event selection and geometry biases that affect the centrality determination in peripheral \PbPb collisions.
The measurement is compared to next-to-leading order perturbative QCD calculations and to the measurements of isolated photons and \zz bosons from the CMS experiment, 
which are all found to be in agreement.

\end{abstract}
\end{titlepage}

\setcounter{page}{2} %please do not remove this line

%%%%%%%%%%%%%%%%%%%%%%%%%%%%%%%%
% begin the main text
%%%%%%%%%%%%%%%%%%%%%%%%%%%%%%%%

\section{Introduction}

Heavy-ion collisions (AA) at ultrarelativistic energies produce a quark--gluon plasma (QGP)~\cite{ALICE:2022wpn,Jacak:2012dx,LHC1review,Braun-Munzinger:2015hba,TheBigPicture,PhenixQGP,StarQGP,PhobosQGP,BrahmsQGP}, a state of deconfined quarks and gluons. The properties of the QGP can be investigated by measuring the different observables related to final-state particles, such as transverse momentum (\pt) or angular distributions, as a function of parameters like the plasma volume, density, temperature, or lifetime. 
The range of such parameters can be changed by varying the collision energy, the size of the colliding nuclei, or the collision centrality.
The observables measured in heavy-ion collisions are compared to the same observables measured in proton--proton (\pp) or proton--nucleus reference collision systems to obtain estimations of the QGP properties via the comparison to theoretical models. 

AA collisions occur with different values of the impact parameter between the trajectories of the nuclei, ranging from central collisions with small impact parameter, to peripheral collisions. 
Experimentally, centrality classes are defined in terms of percentiles of the hadronic cross section~\cite{ALICE:PbPb5TeVMBCrossSec}: 
0--10\%, for example, is the class of the most central collisions in the analyses presented in this article.

The high-energy quarks and gluons produced by partonic hard scatterings, which occur at the early stages of the collision, lose energy via collisional and radiational processes in the presence of a QGP.
As a consequence, the high-\pt\ jet and hadron production, scaled by the average number of nucleon--nucleon binary collisions \Ncoll, is modified with respect to their production in \pp collisions: this effect is known as ``jet quenching''~\cite{Bjorken:1982tu,d'Enterria:2009am}. 
This modification can be quantified by the nuclear modification factor 
\begin{equation}
\label{eq:raa}
\raa =  \frac{1}{ \Ncoll } \frac{\sigma_{\rm NN}^{\rm INEL}}{N_{\rm AA}^{\rm evt}} \frac{{\rm d}^{2}N_{\rm AA}^{\rm particle,~jet} /  ({\rm d}p_{\rm T}~{\rm d}\eta) }{ {\rm d}^{2}\sigma_{\rm \pp}^{\rm particle,~jet}  /  ({\rm d}p_{\rm T}~{\rm d}\eta)},
\end{equation}
where $N_{\rm AA}^{\rm evt}$ is the number of AA minimum bias (MB) collisions, $N_{\rm AA}^{\rm particle,~jet}$ is the number of particles or jets measured in AA collisions, $\sigma^{\rm particle,~jet}_{\rm pp}$ is the jet or particle production cross section in \pp collisions, $\eta$ is the pseudorapidity, and $\sigma_{\rm NN}^{\rm INEL}$ is the nucleon--nucleon inelastic cross section. The value of \Ncoll is obtained from a Glauber model calculation~\cite{ALICE:2013hur, PhysRevC.97.054910}.  

The electroweak bosons -- photons, \zz, and \ww -- do not interact strongly with the QGP. Therefore, while the \raa\ of jets and high-\pt\ hadrons is expected to be smaller than unity because of the energy loss of the parent parton in the plasma, that of the electroweak bosons produced before the QGP formation should be equal to unity when only the \Ncoll scaling of their production in hard scatterings is considered.
However, deviations from unity can arise from cold nuclear matter effects. These include modification of the parton distribution functions in the nuclei (nPDF) compared to the proton PDF, such as shadowing at small Bjorken-$x$ values, as well as isospin effects~\cite{Arleo_2007,Aad2016PbPb,Arleo:2011gc,ZHANG_2009}. Small-$x$ PDF modifications can be probed by low-\pt\ jets, hadrons, and photons.
As expected, the LHC and RHIC experiments have reported a strong suppression of the production of jets and hadrons for $\pt \gtrsim 5$~\GeVc in central \PbPb and Au--Au collisions, which has been attributed to jet quenching~\cite{Adams_2003,PhysRevLett.101.232301,PhysRevC.87.034911,ALICE:2018vuu,Khachatryan_2017,ALICE:2019hno,ALICE:2015mjv,ALICE:2019qyj,ATLAS:2019108,PhysRevC.96.015202}. 
In contrast, it has been shown that in AA collisions there is no modification of either the \ww and \zz boson production at the LHC~\cite{Aad:2019sfe,Chatrchyan:2012nt,ALICE:2022cxs,Aad:2012ew,Chatrchyan:2014csa,Aad:2019lan,PhysRevLett.127.102002} or of high-\pt direct photons, i.e. photons which are directly produced in elementary processes, and as such are not products from hadronic decays~\cite{Afanasiev:2012dg,Chatrchyan:2012vq,Aad2016PbPb,Sirunyan:2020}.

Direct photons include thermal photons (QGP thermal radiation), which are a significant contribution only for $\pt \lesssim 4$~\GeVc, and prompt photons originating from hard scatterings. 
At the leading order (LO) in perturbative Quantum Chromodynamics (pQCD), prompt photons are produced via $2 \rightarrow 2$ processes: (i)
quark--gluon Compton scattering $\rm{q g} \rightarrow \rm{q} \gamma$, and (ii) quark--antiquark annihilation $\rm{q \overline{q}} \rightarrow g \gamma$ and, with a much smaller contribution, $\rm{q \overline{q}} \rightarrow \gamma \gamma$. 
In addition, prompt photons are produced by higher order processes like parton fragmentation or bremsstrahlung.
The collinear part of such processes has been shown to contribute effectively also at LO~\cite{Aurenche:1993}. 
A clean separation of the different prompt photon sources is neither experimentally achievable nor possible theoretically.
However, requiring the photons to be ``isolated'' allows suppression of the contributions from fragmentation and bremsstrahlung~\cite{Ichou:2010wc}, which are commonly accompanied by other parton fragments.
The isolation criterion typically consists of requiring that the sum of the transverse momenta of the produced particles (\ptIso) in a cone with angular radius $R$ around the photon direction is smaller than a given threshold value. 
The advantage of this selection is that it can be implemented both in the experimental measurements and in the theoretical calculations. A strong additional motivation for applying an isolation selection is to reduce the background due to photons originating from hadron decays,
as hadrons at reasonably high \pt\ would, in general, be produced in jet fragmentation and accompanied by other fragments nearby.

Since isolated prompt photons do not interact strongly and are produced before the QGP formation, they can be used as a calibrated reference for the rate of hard processes.
Given that \Ncoll is not measured directly, but rather linked to the centrality by means of the Glauber model, measurements of prompt photons via isolation as a function of centrality, compared with \pp\ measurements at the same centre-of-momentum energy per nucleon pair (\snn) and high \ptg with negligible cold nuclear matter effects, allow the test of the \Ncoll scaling.
It has indeed been shown that the Glauber model does not fully capture the experimental biases on the centrality selection, which are significant for peripheral collisions~\cite{LOIZIDES2017408}. 
These biases are due to initial-state geometry effects and to correlations between the hard processes producing jets and the soft particle yield, which is used for estimating the centrality.
These biases can be understood and modelled, e.g. via simulations with the HG-PYTHIA event generator~\cite{LOIZIDES2017408}, such that their effect
on the hadron \raa\ measurements can be disentangled from the energy loss~\cite{Acharya:2018njl}.
High-precision measurements of electroweak bosons can allow to further pinpoint the bias and provide an
experimental baseline for the \raa calculations of hadrons and jets.
The \zz-boson measurement in \PbPb collisions at \snnfive by the CMS experiment~\cite{PhysRevLett.127.102002} quantified the bias and observed that the cross section for \zz bosons in the 70--90\% centrality class is approximately 25\% lower than the unbiased cross section in the 0--90\% centrality class, in agreement with the HG-PYTHIA calculations~\cite{LOIZIDES2017408,Jonas:2021xju}.

The measurement of the isolated-photon production rate can also be used to test pQCD theory calculations, in particular, the need to include higher orders than leading order and next-to-leading order (NLO).
A detailed discussion of the dependence on the isolation-cone radius and of the different isolation-momentum definitions in \pp collisions at the LHC can be found in Ref.~\cite{Chen_2020}. 
It is found that decreasing the cone radius for a fixed isolation-momentum threshold increases the cross section at higher orders than LO since part of the QCD radiation out of the cone is not vetoed.
Due to the better description of the QCD radiation of the additional higher-order external partons, such an increase can still be of the order of 5\% from NLO to NNLO (next-to-next-to-leading order) for the isolation criteria used in this study.
The ATLAS Collaboration performed this measurement in \pp collisions at centre-of-momentum energy \sthirteen for $\ptg > 250$~\GeVc, and found a good agreement with the NNLO calculations~\cite{Aad:2023}.
Such studies of the dependence of the measured isolated photon cross section on the isolation-cone radius value can further constrain the treatment of the QCD radiation isolation in the models. In particular, 
the effect can be more significant at the lower \ptg reached in the measurement presented in this paper since the fraction of fragmentation photons is larger. 
Hence, this measurement tests the model predictions
in an unexplored momentum regime.

Isolated photons can also be used to constrain the (n)PDF in the proton and in the nucleus, in particular via their measurement at $\ptg < 20$~\GeVc, where shadowing effects are more significant~\cite{Arleo:2011gc}.
The dominant contribution to the prompt photon production at the LHC is the quark--gluon Compton diagram, which is directly sensitive to the gluon density.
The high $\sqrt s$ of collisions at the LHC allows access to very small values of the longitudinal momentum fraction $x$ of the initial-state partons, which are essentially gluons. The gluon PDF has a much larger uncertainty than the quark
PDFs~\cite{Arleo:2011gc,DENTERRIA2012311,Arleo:PhotonLHCYellow,Ichou:2010wc,Perez:2012um}.
Therefore, isolated-photon measurements allow probing the low-$x$ gluon content of one of the incoming protons or nuclei and thus constrain the PDF and nPDF~\cite{ALICE:2019rtd}. 

\newpage

Measurements of differential \pt\ spectra of isolated photons and direct photons have been performed at SPS~\cite{Appel:1986ix}, Tevatron~\cite{Abazov:2005wc,Aaltonen:2009ty}, and RHIC~\cite{PhysRevLett.98.012002,PhysRevLett.94.232301,adare2015centrality,Afanasiev:2012dg,PhysRevC.109.044912}, and also at 
fixed target experiments~\cite{Ferbel:1984ef}. The measurements by the ATLAS and CMS Collaborations at the LHC in \pp and \PbPb collisions at various energies can be found in Refs.~\cite{Khachatryan:2010fm,CMS2011,Chatrchyan:2012vq,Sirunyan2019,Sirunyan:2020,Aad:2010sp,Aad:2011tw,Aad:2013zba,Aad:2016xcr,Aad:2017,Aad:2019,Aad:2023}. 
The ALICE Collaboration has measured the isolated-photon yield in pp collisions at $\s =$~7~TeV~\cite{ALICE:2019rtd} and 13~TeV~\cite{ALICE:2024kgy}, as well as the direct-photon yield: via the excess of the inclusive-photon yield versus decay-photon yields in pp collisions at $\s =$~2.76 and 8~TeV~\cite{ALICE:2018mjj}, and in Pb--Pb collisions at $\sqrt{s_{\rm NN}} =$~2.76~TeV~\cite{ALICE:2015xmh}; and via dielectron measurements in \PbPb collisions at  $\sqrt{s_{\rm NN}} =$~5.02~TeV~\cite{ALICE:2023jef}. 

This paper presents the isolated-photon cross section in \PbPb and \pp collisions at \snnfive measured by ALICE, using a data sample collected in the years 2015 and 2018 for \PbPb collisions, and a data sample collected in the year 2017 for \pp collisions. 
The results from \PbPb collisions are provided in the centrality classes 0--10\%, 10--30\%, 30--50\%, 50--70\%, and 70--90\%. 
These analyses have been performed with isolated photons measured at midrapidity ($|\etag|<0.67$) with a transverse momentum range $10-14<\ptg<40-140$~GeV/$c$ (depending on the collision system and centrality class), which corresponds to $(3.8-5.4) \times 10^{-3}<\xtg<(15.4-26.9) \times 10^{-3}$, with \xtg $=2\ptg/\sqrt{s} \approx \text{Bjorken-}x$ at midrapidity.
The measurement follows closely the analysis strategy presented in the previous ALICE measurements in \pp collisions at \sseven~\cite{ALICE:2019rtd} and \sthirteen~\cite{ALICE:2024kgy}.
The isolated-photon nuclear modification factor is also calculated, together with the ratio of cross sections in \pp collisions with different \s.
The latter was already measured in \pp collisions by ALICE~\cite{ALICE:2024kgy} and ATLAS~\cite{ATLAS:2019drj} at different collision energies.
For the first time with ALICE, and for the first time at the LHC for $\ptg < 250$~\GeVc, the ratio of the cross sections obtained with different isolation-cone angular radii values $R=0.4$ (used for the previous measurements) and $R=0.2$ is presented.
This paper is divided into the following sections: Section~\ref{sec:detector} presents the detector setup and the data sample used for the analysis; Section~\ref{sec:analysis} describes the analysis procedure; the systematic uncertainties are presented in Sect.~\ref{sec:sys_unc}; the final results compared to model calculations and conclusions are presented in Sect.~\ref{sec:results} and~\ref{sec:conclusion}, respectively.

\section{Detector description and data selection\label{sec:detector}}

The ALICE experiment and its performance during the LHC Run 2 (2015--2018) are described in Refs.~\cite{Aamodt:2008zz,Abelev:2014ffa}.
Photon reconstruction was performed using the Electromagnetic Calorimeter (EMCal)~\cite{ALICE:2022qhn} while charged particles used in the photon isolation were reconstructed with the combination of the Inner Tracking System (ITS)~\cite{Aamodt:2010aa} and the Time Projection Chamber (TPC)~\cite{Alme:2010ke}, which are the main components the ALICE central tracking detectors.  

The ITS is composed of six cylindrical layers of silicon detectors with full azimuthal acceptance and surrounds the interaction point. 
The different layers provide a pseudorapidity coverage of $|\eta| <$ 2 (inner) to  $|\eta| <$ 0.9 (outer).
The two innermost layers have fine granularity and small radial distances (3.9 and \unit[7.6]{cm}) from the beam line providing high spatial precision for tracking close to the primary vertex. 
The high-precision points and the low material budget of the ITS guarantee excellent resolution on the charged-particle track parameters in the vicinity of the primary vertex and on the reconstructed position of the primary vertex of the collision.

The TPC is a large ($\approx$ \unit[90]{m$^3$}) cylindrical drift detector filled with gas. It covers $|\eta| <$ 0.9 over the full azimuth angle, with a maximum of 159 reconstructed space points along the track path. The TPC and ITS tracking points are matched when possible, forming reconstructed charged particle tracks. Since the ITS and TPC are placed in a longitudinal magnetic field, track momentum can be calculated from the measured track curvature radius. 

The EMCal is a lead-scintillator sampling electromagnetic calorimeter used to measure photons and electrons via the electromagnetic showers they create in the calorimeter. 
The scintillation light is collected by optical fibres coupled to Avalanche Photo Diodes that amplify the signal.
The energy resolution is $\sigma_E/E = (1.4 \pm 0.1)\%\oplus (9.5 \pm 0.2)\%/\sqrt{E}\oplus (2.9 \pm  0.9) \%/E$, with energy $E$ in units of~GeV.
The EMCal is installed at a radial distance of \unit[4.28]{m} from the ALICE interaction point.
The basic unit of the EMCal is called ``cell''. There are 17664 cells installed in total. 
Cells have a transverse size of $6 \times 6~\mbox{cm}^2$, which corresponds to $\Delta \varphi~\mathrm{(rad)}\times \Delta \eta \simeq 0.0143\times0.0143$, 
approximately twice the Moli\`ere radius.
The calorimeter consists of twenty supermodules (SM) with different number of cells: twelve of them at $80^\circ<\varphi<187^\circ$ form the EMCal top section;
and the other eight SMs at $260^\circ<\varphi<327^\circ$ 
form the EMCal bottom section, also called DCal (di-jet calorimeter).
The pseudorapidity coverage is $|\eta|<0.7$, although DCal does not cover $|\eta|<0.22$ for most of its $\varphi$ coverage. Details on the SM configuration can be found in Refs.~\cite{ALICE:2024kgy,ALICE:2022qhn}. 

The V0 detector consists of two arrays of 32 plastic scintillators located at $2.8 < \eta < 5.1$ (V0A) and  $-3.7 < \eta < -1.7$ (V0C)~\cite{ALICE:2013axi}. 
Each of the V0 arrays consists of 32 channels and is segmented in four rings in the radial direction, and each ring is divided into eight sectors in the azimuthal direction. The V0 detector signals, which are proportional to the charged-particle multiplicities, are used to divide the \PbPb event sample into centrality classes. 
A Glauber Monte Carlo model is fitted to the V0 amplitude distribution to compute the fraction of the hadronic cross section corresponding to any given range of V0 amplitudes. 

The data were taken with a minimum bias interaction trigger and EMCal Level-1 photon-dedicated triggers (\tr). 
The MB trigger is defined as a coincidence between the V0A and the V0C trigger signals.
In the 2015 \PbPb sample, the MB triggered data were taken so that the centrality distribution was uniform, but for the 2018 data sample, the 0--10\% and 30--50\% centrality classes were enhanced with dedicated V0 triggers. 
Events above 90\% centrality are excluded, since there are substantial contributions from electromagnetic processes, and their low multiplicity results in an inefficient trigger.
The \tr\ triggers are based on energy depositions in $4\times4$ calorimeter cells larger than 4~GeV in \pp collisions in the year 2017, and larger than 10~GeV in \PbPb for the year 2015. For the 2018 \PbPb collisions, the threshold has been set at 10~GeV for the 50\% more central collisions (\tr-high), and at 5~GeV otherwise (\tr-low).
A detailed description of the \tr\ triggers can be found in Refs.~\cite{Bourrion_2013,ALICE:2022qhn}. 

An offline event selection based on the V0 timing information is applied to remove beam-induced background events. In addition, in \PbPb collisions further beam-background reduction is obtained using the information from two zero-degree calorimeters (ZDCs) positioned at 112.5 m on either side of the nominal interaction point. In particular, a selection is applied on the correlation between the sum and the difference of times measured in each of the ZDCs~\cite{Abelev:2014ffa}.
Furthermore, in \pp collisions only events with one reconstructed primary vertex are accepted in the analysis
to exclude pileup events within the same bunch crossing. Out-of-bunch pileup is removed with cuts on the V0 timing~\cite{Abelev:2014ffa}. 
In \PbPb collisions, the same event pileup is negligible and such rejection is not applied. 
Finally, only events with a primary vertex along the beam direction within $\pm 10$~cm from the centre of the apparatus are considered in this analysis, to grant a uniform pseudorapidity acceptance.

The measurements in \PbPb collisions  presented here are performed in five centrality classes:  
0--10\%, 10--30\%, 30--50\%, 50--70\%, and 70--90\%. The corresponding \Ncoll values are:  1572~$\pm$~17,  783~$\pm$~7, 265~$\pm$~3, 65.9~$\pm$~1.2, and 10.9~$\pm$~0.2, respectively, obtained from~\cite{ALICE:PbPb5TeVMBCrossSec}. 
The integrated luminosity per each centrality class, collision system, and trigger combination are discussed in Sect.~\ref{sec:trigger}.

Note that the TPC was not included in the data sample of \pp collisions at \sfive triggered by the EMCal taken in the year 2017. 
A lightweight readout approach with only the EMCal and ITS detectors was used, which allowed an enhanced sampled luminosity by reading out at a higher rate.
This data sample has been used in previous isolated-photon measurements~\cite{ALICE:2020atx}.

\newpage
\section{Isolated-photon reconstruction and corrections}
\label{sec:analysis}
 
 The analysis procedure followed to measure isolated photons  consists of the following steps:
 (a) reconstruction of clusters of cells in the calorimeter and of tracks with the ITS and the TPC;
 (b) photon identification via charged particle
vetoing (CPV) using track--cluster matching and via the cell energy spread (shower shape); 
 and (c) selection of isolated-photon candidates. 
 A more detailed description and discussion of the steps presented here can be found in Refs.~\cite{ALICE:2022qhn,ALICE-PUBLIC-2024-003,ALICE:2019rtd,ALICE:2024kgy}.

In order to obtain correction factors to the raw isolated photon spectra,
the detector response is modelled by Monte Carlo (MC) simulations reproducing the detector conditions of the data-taking periods.
The corrections discussed in the next subsections are obtained using PYTHIA~8 (version 8.210~\cite{Sjostrand:2014zea} using the Monash 2013 tune~\cite{Skands_2014}) as a particle generator, creating \pp collisions in intervals of transverse momentum of the hard scattering with two jets (jet--jet, background events) or with a prompt photon and a jet ($\gamma$--jet, mainly Compton and annihilation processes, signal events) in the final state. The transport of the generated particles in the detector material is done using GEANT3~\cite{Geant3}.   
For the $\gamma$--jet event generation, the event is accepted when the prompt photon enters the EMCal acceptance. 
For the jet--jet event generation, the event is accepted when at least one jet produces a high-\pt\ photon, requested to originate from a hadron decay, in the EMCal acceptance. 
This enables to enhance the number of such photons, which are the main 
background in this analysis. Two samples with different trigger thresholds ($\ptg >3.5$ or 7~\GeVc) have been used in the jet--jet event generation. 

For the calculation of the correction factors for \PbPb collisions, each simulated \pp collision is embedded into a real \PbPb minimum bias triggered event selected within the different centrality classes considered in this analysis, so that the effect of the underlying event (UE) low-energy particles is properly taken into account.
For the calorimeter, the embedding is performed at the cell level by summing the cell energy of the data and the simulation. 
For the charged particles measured with the tracking systems, the embedding is done at the track level, adding to the list of available tracks from the simulation those coming from the data.

In the analysis procedure, the outputs of the $\gamma$--jet and jet--jet simulations are combined to calculate the prompt-photon purity (see Sect.~\ref{sec:purity}). 
To take into account 
the suppression of high-\pt hadron production due to jet quenching
in heavy-ion collisions, the contribution of the particles of hadronic origin in the jet--jet simulation is scaled by the nuclear modification factor of charged particles, obtained by combining the ALICE~\cite{ALICE:2018vuu} and CMS~\cite{Khachatryan_2017} results, so as to cover the full \pt range of this measurement.
Finally, we note that fragmentation photons contained in jet--jet samples have not been included in the aforementioned corrections to improve computational efficiency. Their impact on the calculated purities and efficiencies is found to be marginal and well within the reported uncertainties.

\subsection{Cluster reconstruction and selection \label{sec:clust}}

Particles deposit their energy in several calorimeter cells, forming a cluster. The different cluster reconstruction algorithms used in the EMCal are described in detail in Ref.~\cite{ALICE:2022qhn} together with the calibration procedure and corrections.
Clusters are obtained by grouping all cells with common sides whose energy is above an aggregation threshold, starting from a seed cell. Clusterisation thresholds are given in Table~\ref{tab:cuts}.
Because of the large particle multiplicity of the UE, contributions from several particles are likely to be merged into the same cluster in central heavy-ion collisions. To avoid this, an additional condition is applied with respect to previous ALICE measurements of isolated photons to restrict the growth of the cluster: cells are added to the cluster only if the energy of the cell to be added is lower than the already added neighbouring cell in the direction of the seed cell~\cite{ALICE:2022qhn}.
Although this condition was targeted to \PbPb collisions, it has also been applied to the \pp measurements presented here for consistency.

The cluster quality selection criteria applied in this measurement are listed in Table~\ref{tab:cuts}. A more detailed description can be found in Ref.~\cite{ALICE-PUBLIC-2024-003}. 
As the charged particle veto needs TPC tracks, this selection criterion was only applied in \PbPb collisions.
In addition, the two calorimeter sections located at the highest $\varphi$ are excluded and
lead to the calorimeter acceptances also listed in Table~\ref{tab:cuts}. Clusters that pass these selection criteria are called ``inclusive clusters''. 

%~table~%%%%%%%%%
\begin{table}[ht]
\begin{center}
\caption{\label{tab:cuts} Cluster reconstruction and selection criteria. Description and discussion can be found in Ref.~\cite{ALICE-PUBLIC-2024-003}.}
\setlength{\tabcolsep}{7mm}{
\begin{tabular}{ c c  }
\hline
Cluster seed threshold  & $E_{\rm seed}>500$~\MeV \\
Cluster aggregation threshold   & $E_{\rm agg}>100$~\MeV  \\
\hline
Number of cells         & $N_{\rm cell}>1$        \\
$N$ cells from highest $E$ cell to SM border    & $N_{\rm border}>1$        \\
Cluster time - bunch crossing time            & $|\Delta t_{\rm cluster}|<20$~ns  \\       
Abnormal signal removal & $F_{+} = 1-\frac{\sum_{\rm cell} E_{{\rm adjacent~to~highest}~E}}{E_{{\rm highest}~E~{\rm cell}}} < 0.95$          \\    
\hline
Charged particle veto (\PbPb only):  & \\
~~~~ when                    & $E_{\rm cluster}/p^{\rm track}<1.7$ \\
~~~~ track--cluster $\eta$ residual & $\Delta \eta^\text{residual} > 0.010 +(\pt^\text{track} +4.07)^{-2.5}$\\
~~~~ track--cluster $\varphi$ residual & ~~~~~~$\Delta \varphi^\text{residual} > 0.015 +(\pt^\text{track} +3.65)^{-2~~}$~rad \\
\hline
Acceptance: & \\
Top section   &  $~~81.2^{\circ}<\varphi<185.8^{\circ}$~~~~~~~~~~~~~~$|\eta|<0.67$ \\   
Bottom section    & $261.2^{\circ}<\varphi<318.8^{\circ}$~~$0.25<|\eta|<0.67$ \\
\hline
\end{tabular}}
\end{center}
\end{table}
%%%%%%%%%%

\subsection{Photon identification via cluster shower shape measurement}
\label{sec:photonident}
Inclusive clusters can have a wider elongated shape if one or several additional particles deposit their energy nearby in the detector. 
The most frequent case in \pp collisions is a two-particle merged cluster when the distance between 
them is larger than two cells and their electromagnetic showers overlap partially. 

In particular, the neutral-meson decays to two photons generate elongated clusters when the opening angle between the decay photons is larger than the angular size of an EMCal cell (otherwise, both showers completely overlap), but smaller than the electromagnetic shower size.
These limits translate into the ranges $8<\pt<20$~\GeVc and $40<\pt< 60$~\GeVc for \piz and $\eta$ mesons, respectively~\cite{ALICE:2022qhn}.

Merged and single photon clusters can be discriminated by the variable \sigmalong, called ``shower shape'', which is the square of the larger eigenvalue of the cluster cell spatial distribution weighted by the cell energy in the $\eta-\varphi$ plane~\cite{ALICE:2022qhn}, and can be calculated as
%%%%
\begin{equation}
\label{eq:ss1}
\sigma_{\rm long}^{2} = (\sigma_{\varphi\varphi}^{2}+\sigma_{\eta\eta}^{2})/2+\sqrt{(\sigma_{\varphi\varphi}^{2}-\sigma_{\eta\eta}^{2})^{2} / 4+\sigma_{\eta\varphi}^{4}},\\
\end{equation}
%%%%
where ${\sigma^2_{xz}} = \big \langle xz \big \rangle - \big \langle x \big \rangle \big \langle z \big \rangle$ and $\big \langle x \big \rangle = (1/w_{\rm tot}) \sum w_i x_i$ ($x_i$ are in cell units, and therefore ${\sigma^2_{xz}}$ are dimensionless)
are weighted over all cells associated with the cluster in the $\varphi$ or $\eta$ direction. The weights $w_i$ depend logarithmically on the ratio of the energy $E_i$ of the $i$-th cell to the cluster energy $E_{\rm cluster}$ as $w_i=\mathrm{max}(0,4.5+\ln(E_{i}/E_{\rm cluster}))$, and 
$w_{\rm tot} = \sum w_i$~\cite{Awes:1992}.

In the previous isolated-photon measurements in \pp and \pPb collisions made by ALICE~\cite{ALICE:2019rtd,ALICE:2024kgy, ALICE:2020atx}, 
the limitation on the aggregation of the cells to the cluster (Sect.~\ref{sec:clust})
was not applied, which allowed the use of the shower-shape 
parameter to reject efficiently the clusters from \piz\ and $\eta$ mesons decaying into two photons for meson energies up to 20 and 60~GeV, respectively. 
The cell aggregation restriction applied in the measurements presented in this article significantly decreases this rejection power, since the two showers from meson decays are reconstructed as two different clusters. 
To increase the rejection of the decay photons contribution to a similar level as in previous measurements while leaving untouched the other cluster reconstruction performances in \PbPb collisions, the selection of cells used for the \sigmalong\ calculation is enlarged with respect to the set of cells used to calculate the cluster energy and position -- these latter parameters would otherwise be affected by the underlying event. The cells used for the shower shape calculation are those with an energy deposit above the aggregation threshold, which share a common side, 
and are located in a window of $5\times5$ cells centred at the highest energy cell of the cluster~\cite{ALICE-PUBLIC-2024-003}. The shower shape obtained this way is denoted \sigmalongPb. 

\begin{figure}[ht]
\begin{center}
\includegraphics[width=1\textwidth]{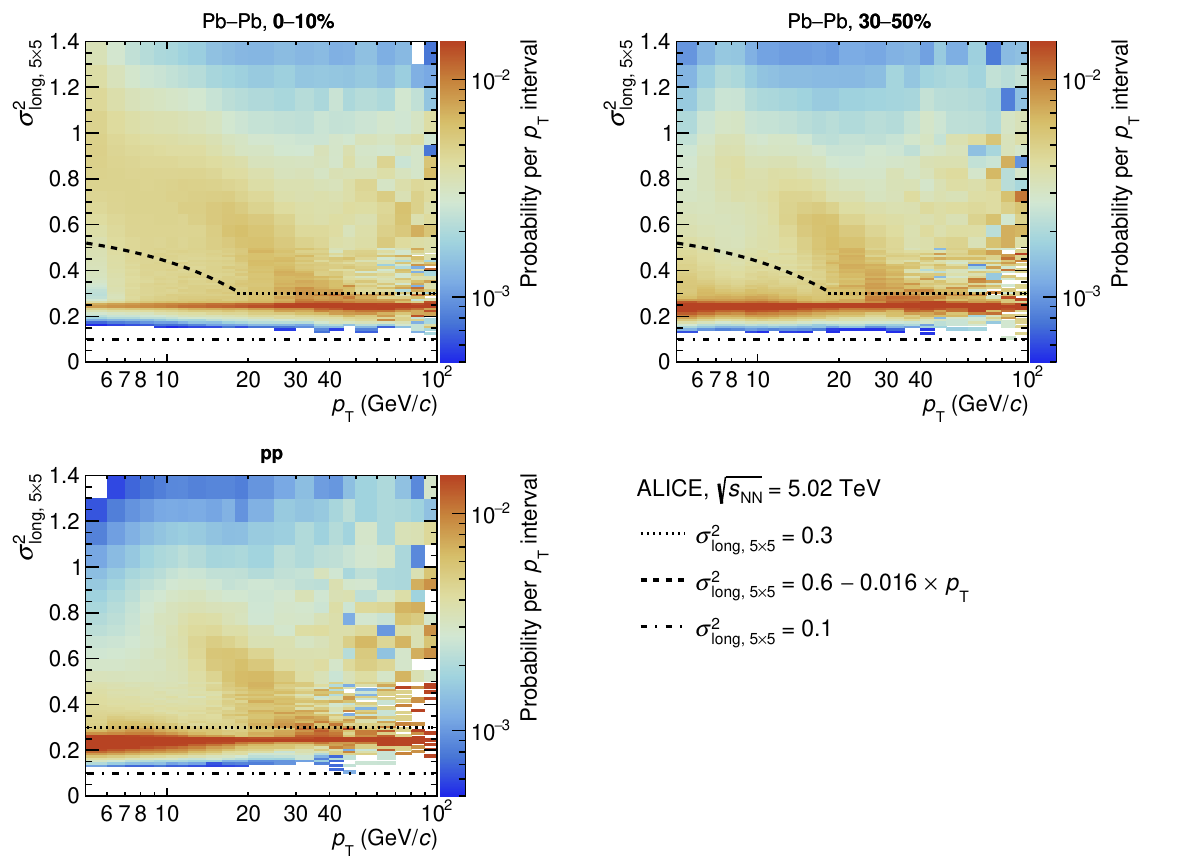}
\end{center}
\caption{\label{fig:M02} (colour online) Inclusive-cluster \sigmalongPb\ distribution as a function of  \pt\ in data for \pp (bottom left frame) 
and \PbPb collisions for two different centrality classes 0--10\% central (top left frame) and  30--50\% semi-central (top right frame). 
The dotted line corresponds to the tight value of the upper selection limit for single photon candidate clusters (narrow clusters)
and the dashed line corresponds to a looser photon upper selection 
used in \PbPb collisions below 18~\GeVc. The dotted-dashed line corresponds to the narrow cluster's lower limit.}
\end{figure}

The inclusive-cluster \sigmalongPb\ distributions as a function of \pt\ are shown in Fig.~\ref{fig:M02} for data in \pp and \PbPb collisions in the 0--10\% and 30--50\% centrality classes (other centrality classes can be found in Ref.~\cite{ALICE-PUBLIC-2024-003}). 
Most of the single photons are reconstructed as clusters with \sigmalongPb~$ \approx 0.25$. The presence of the collision underlying event has a tendency to enlarge the \sigmalongPb\ value at low \pt.
At higher \sigmalongPb, a clear \pt-dependent band is observed between 8 and 20~\GeVc: This band is populated by two \piz-decay photons contributing to a single cluster. 
Due to the kinematic boost and resulting opening angle decrease, the value of \sigmalongPb\ for this type of cluster decreases with increasing energy, which leads to a progressive overlap with the single photon band for $20 < \pt < 40$~\GeVc. 
Another fainter band, due to the merged $\eta$ meson decays, appears above 40~\GeVc.  

%%%%%%%%
%%%%%%%%

In this analysis, ``photon candidates'' refer to clusters with a narrow shape, i.e.~a small value of \sigmalongPb. In \pp collisions, they can be distinguished from the merged meson decays by applying an upper limit \sigmalongPb~$<$~0.3. In \PbPb collisions, this limit is also used, but only for $\pt > 18$~\GeVc: below, a looser \pt\ dependent upper limit  $\sigma_{\rm max}^{2} (\pt)=0.6-0.016\times \pt$ is applied, so that single photon clusters with a significant UE contribution can still be selected, without increasing the number of accepted merged decay-photon clusters. 
A lower limit at $\sigmalongPb~=~0.1$ is used in addition to cleaning the cluster sample from anomalous high-energy depositions~\cite{ALICE-PUBLIC-2024-003, ALICE:2022qhn}. 

\begin{figure}[hb]
\begin{center}
\includegraphics[width=1\textwidth]{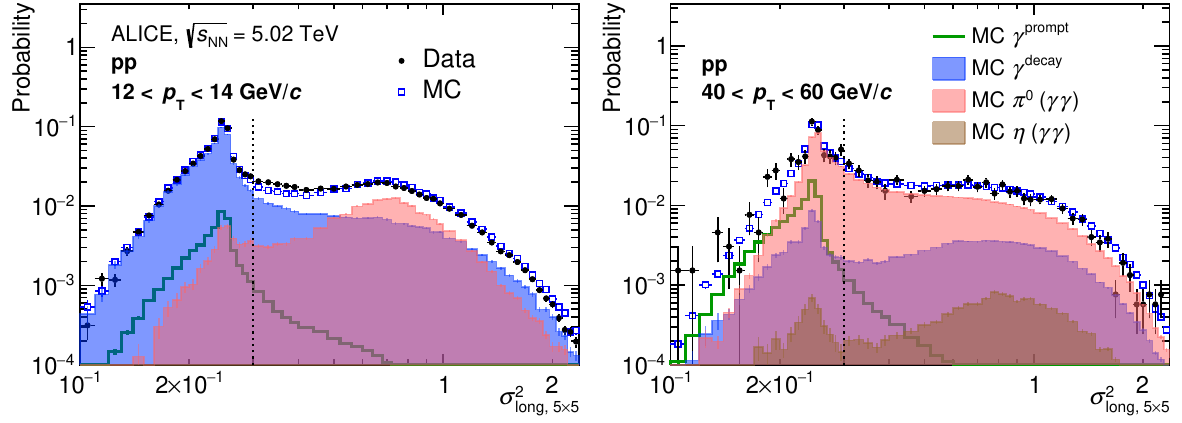} \\
\includegraphics[width=1\textwidth]{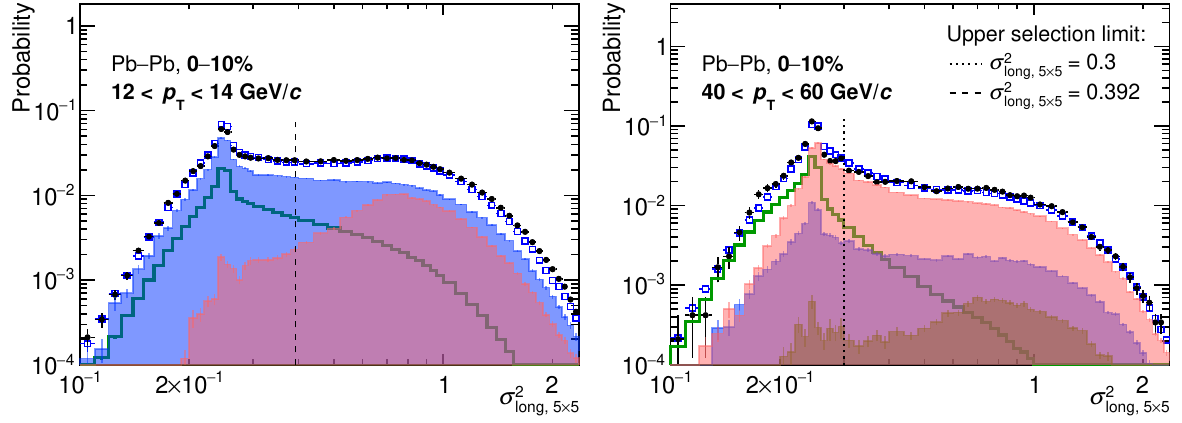}
\end{center}
\caption{\label{fig:M02DataMCProj} (colour online) Inclusive-cluster \sigmalongPb\ distribution in data (black bullets) and PYTHIA~8 simulation (jet--jet+$\gamma$--jet processes, blue squares).  
The four panels display these distributions for two selected cluster \pt\ ranges, $12<\pt<14$~\GeVc on the left and $40<\pt<60$~\GeVc on the right, and two collision systems:  \pp (top)  and \PbPb 0--10\% central (bottom). 
The simulation is decomposed in its different particle origins: prompt $\gamma$ ($\gamma^{\rm prompt}$, green line), not merged decay $\gamma$ ($\gamma^{\rm decay}$, blue area), merged decay photon clusters ($\gamma\gamma$) from \piz (red area) or $\eta$ (brown area).
The threshold value $\sigmalongPb~=~0.3$ or 0.392 (corresponding to $\sigma_{\rm max}^{2}$(13~\GeVc)) is shown on all plots as a dotted or dashed vertical line, respectively. }
\end{figure}

Figure~\ref{fig:M02DataMCProj} shows a projection of the inclusive-cluster \sigmalongPb distribution shown for data in Fig.~\ref{fig:M02}, and from simulation ($\gamma$--jet plus jet--jet PYTHIA~8  cross-section distributions) for a low- and a high-\pt interval and for central \PbPb collisions and \pp collisions: a reasonable description is achieved in simulation after including a modelling, at the cell energy level, of the electronics cross talk. The model consists of the addition of a small fraction of energy (at the per cent level) from a given cell into the surrounding cells, depending on the reference cell energy and location in the calorimeter. 
This modelling is the same one used in previous ALICE measurements~\cite{ALICE:2022qhn,ALICE:2019rtd,ALICE:2024kgy}, but an updated parameterisation of the model has been used for this analysis to better describe the calorimeter performance in \PbPb collisions~\cite{ALICE-PUBLIC-2024-003}. 
This same parameterisation is used in the \pp collisions measurement for consistency, improving the performance as well. The results do not change significantly compared to the previous parameterisation for \pp collisions.\\

%%%%%%%%
%%%%%%%%

\newpage

Figure~\ref{fig:M02DataMCProj} also shows the contributions from the simulations for different particles creating the clusters.
At low \pt, the dominant contributions to the narrow shower shape region are from single \piz-decay photons, while the merged photon clusters from \piz-decay photons contribute more at high \pt.
Prompt photons show a peaked distribution at 0.25, which has in central \PbPb collisions a significant tail at high \sigmalongPb\ due to the UE contribution.
Decay-photon clusters without contribution from a second decay photon in the cluster show a similar distribution to prompt photons, but the tail is more significant due to nearby particles originating from the same jet and overlapping with the cluster. 
Merged clusters from \piz\ (resp. $\eta$) meson decays have two maxima in the \sigmalongPb\ distribution for $12<\pt<14$~\GeVc (resp. $40<\pt<60$~\GeVc): the maximum in the range $\sigmalongPb = 0.6-0.9$ is due to the merging of rather symmetric energy photon decays; the maximum at 
 $\sigmalongPb =0.25$ is due to clusters for which most of the energy comes from one of the decay photons, while the contribution of the second one does not affect the shower shape parameter.
For  $\pt>20$~\GeVc, the merged clusters from \piz\ meson decays have only one maximum at $\sigmalongPb =  0.25$ with a significant tail at high \sigmalongPb.

\subsection{Isolated-photon selection}
\label{sec:isolation}

Direct prompt photons are mostly  isolated, i.e.\ have no hadronic activity in their vicinity except 
for the underlying event of the collision, in contrast to other photon sources 
like photons from parton fragmentation or from decays of hadrons, which have a high probability 
to be accompanied by other fragments~\cite{Ichou:2010wc}. 
An isolation criterion is applied to the photon candidate to suppress the contribution by fragmentation and decay photon production.
An equivalent isolation criterion is commonly included in theoretical calculations to account for the suppression of the fragmentation contribution to the total prompt photon cross section~\cite{Chen_2020,Ichou:2010wc}.
The isolation criterion is based on the so-called ``isolation momentum'' $p_{\rm T}^{\rm iso}$, i.e. the transverse momentum sum of all particles measured inside a cone of radius $R$ around the photon candidate, located at coordinates $\eta^{\gamma}$ and $\varphi^{\gamma}$ in the angular space. 
A particle of coordinates $\eta$ and $\varphi$ is inside the cone when 
%%%%%%%%%%%%%%%%%%%%
\begin{equation}
\label{eq:rsize}
\sqrt{ (\eta - \eta^{\gamma})^2 + (\varphi-\varphi^{\gamma})^2} \quad < \quad R.
\end{equation}
%%%%%%%%%%%%%%%%%%%%

The cone radius value $R=0.4$ is commonly used for \pp and \pPb collisions as it contains the dominant fraction of the jet energy~\cite{PhysRevD.71.112002}. 
However, in \PbPb collisions the number of UE particles entering the cone is considerable, 
so a smaller cone radius can be considered to give better control over the UE contribution. 
In this article, both $R=0.4$ and 0.2 are used. 

\begin{figure}[ht]
\centering
%\begin{center}
\includegraphics[width=1\textwidth]{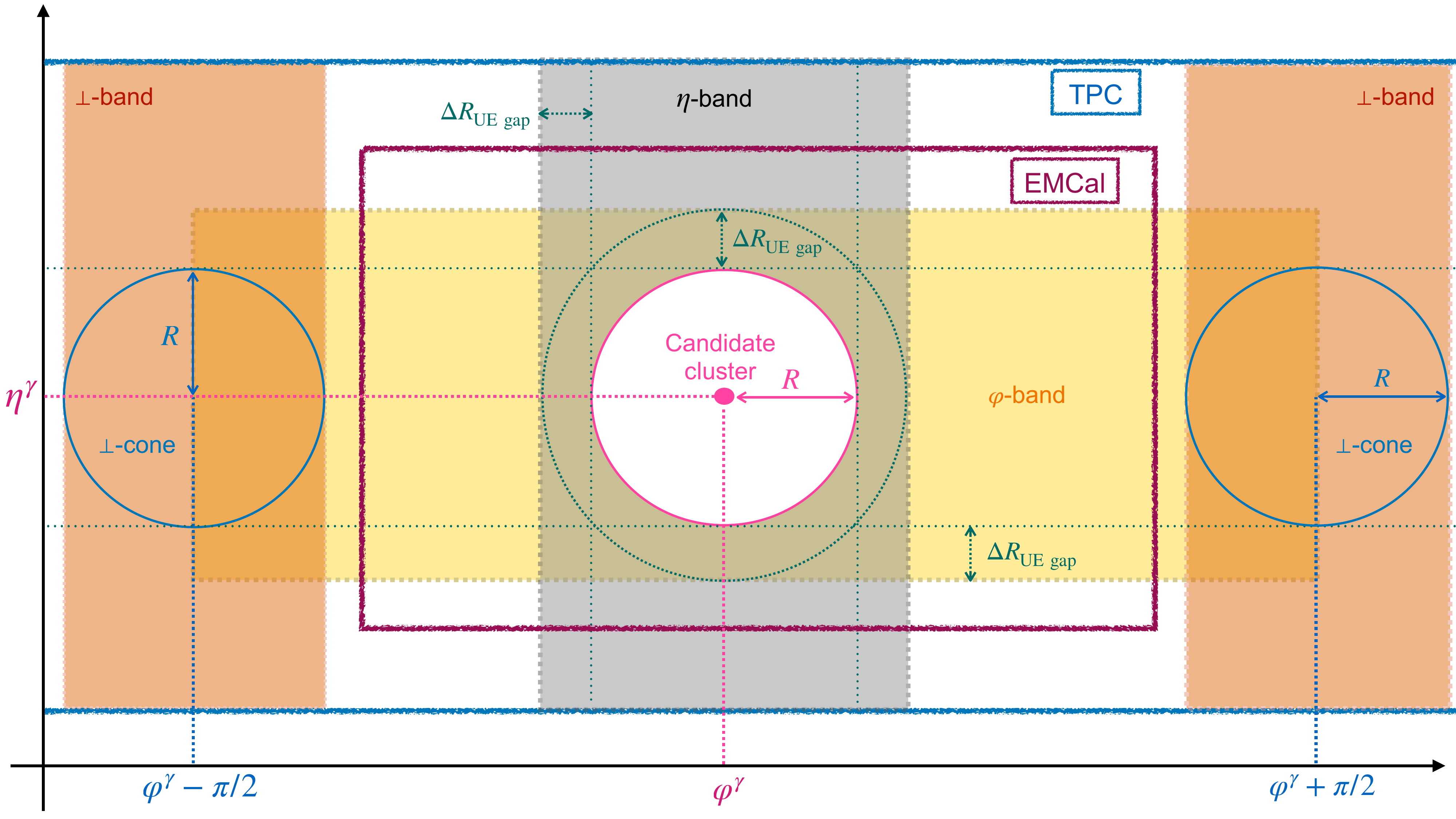} 
%\end{center}
\caption{(colour online) 
Schematic view of the UE estimation areas considered in the analysis and for the estimation of the associated systematic uncertainty. The radius gap $\Delta R_{\rm UE~gap}$ (see text) is also illustrated.
}
\label{fig:UEareasSketch}
\end{figure}

Accepted tracks in the cone are required to satisfy $|\eta^{\rm track}|< 0.9$ and $\pt^{\rm track}>0.15$~\GeVc, the track definition is given in Refs.~\cite{ALICE:2024kgy,ALICE-PUBLIC-2024-003}. Note that in the \pp collision data sample triggered by the EMCal, the TPC was not included in the readout, and therefore ITS-only tracks are used for the isolation, like in Ref.~\cite{ALICE:2020atx}. The same $\eta$ acceptance as in \PbPb collisions is used, along with a transverse momentum selection $0.15<\pt^{\rm ITS~track}<15$~\GeVc to reduce the fake-track rate at high \pt.
The isolation momentum is calculated as the sum of the transverse momenta of all the charged tracks (ch) that fall into the cone, 
from which an estimate of the transverse momentum due to the UE inside the cone is subtracted

%%%%
\begin{equation}
\label{eq:etiso}
p_{\rm T}^{\rm iso,~ch}=\sum p_{\rm T}^{\rm track}-\pi \times R^{2} \times \rho_{\rm UE},
\end{equation}
%%%%
where $\rho_{\rm UE}$ is the estimation of the UE track \pt\ density. 
The density $\rho_{\rm UE}$ is estimated event by event by summing the track \pt in a rectangular area called ``$\eta$-band'' centred around the azimuth $\varphi^{\gamma}$ of the candidate cluster. 
The width $\Delta \varphi$ of the rectangular area along the azimuth depends on the analysis parameters, 
while the width $\Delta \eta$ along the pseudorapidity covers the full track acceptance: $|\eta|<0.9$.
The area covered by this band is shown schematically in Fig.~\ref{fig:UEareasSketch}.
This band is chosen because this area should be affected by the same elliptic flow~\cite{ALICE:2016ccg} as the isolation cone area in \PbPb collisions since the $\varphi$ region is the same in the isolation cone and the band. 
Other bands, depicted schematically also in Fig.~\ref{fig:UEareasSketch}, 
which cover other $\varphi$ values, have been tested and used in the estimation of the systematic uncertainty (see Sect.~\ref{sec:sys_unc}), with similar final results~\cite{ALICE-PUBLIC-2024-003}.

The isolation cone is excluded from this $\eta$-band, but when the photon cluster is the result of jet fragmentation, jet hadronic remnants can still be found outside the selected isolation-cone radius $R$. They can thus contribute to the track-\pt\ measured in the band used for the UE estimation, biasing it to a higher value. To get rid of this possible bias, an additional parameter is introduced: a gap $\Delta R_{\rm UE~gap}$ such that the region excluded from the band is a cone of radius $R + \Delta R_{\rm UE~gap}$ as shown in Fig.~\ref{fig:UEareasSketch}.
The width of the $\eta$-band along the azimuth is chosen to be $\Delta \varphi = 2 \times (R + \Delta R_{\rm UE~gap})$, and the UE density is then 
$\rho_{\rm UE}=(\Sigma p_{ {\rm track~in~} \eta {\rm~band} } )/(\Delta \varphi \times \Delta \eta - \pi (R + \Delta R_{\rm UE~gap})^{2})$. 
In this measurement, $\Delta R_{\rm UE~gap}=0.1$ is used as default, but the values $\Delta R_{\rm UE~gap}=0$, as well as 0.3 for $R=0.2$, are also considered for a systematic uncertainty evaluation. The default gap value for $R=0.2$ is chosen the same as for $R=0.4$ so that the inspected area for the $\rho_{\rm UE}$ estimation is larger than for the $R=0.4$ case: it reduces the UE fluctuations while still excluding the jet core.

%%%%%%%%%
\begin{figure}[t]
\begin{center}
\includegraphics[width=1.\textwidth]{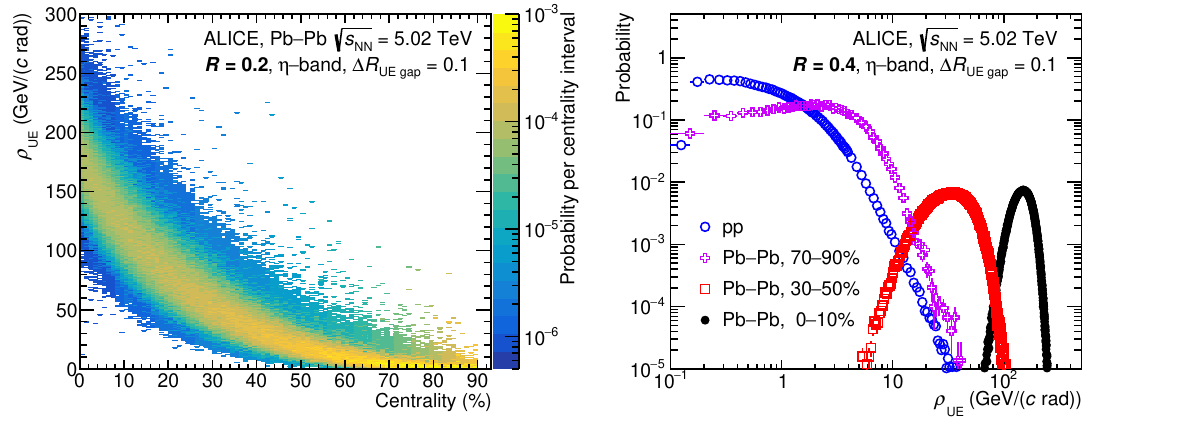} 
\end{center}
\caption{\label{fig:RhoUE} (colour online) $\rho^{\rm UE}$ distributions normalised by the number of events in each centrality class in data 
calculated in the $\eta$-band with $\Delta R_{\rm UE~gap}=0.1$ out of an isolation cone centred at inclusive clusters with $\pt~>~10$~\GeVc. 
Left: for \PbPb collisions and the cone radius $R=0.2$ as a function of centrality. 
Right: for \pp and \PbPb collisions for different centrality classes for the cone radius $R=0.4$.  }
\end{figure}

%%%%%%%%%
As expected, the UE track density, shown in Fig.~\ref{fig:RhoUE} for events with high-\pt\ inclusive clusters at the centre of the isolation cone, strongly depends on the centrality. For a given centrality percentile, its distribution has a large width due to the UE event-by-event fluctuations.
The density is beyond 100~GeV/($c$ rad) in central (0--10\%) \PbPb collisions, still reaches several tens~of~GeV/($c$ rad) in semi-central (30--50\%) collisions, but its value is only a few GeV/($c$ rad) for the most peripheral \PbPb collisions, and less than 1~GeV/($c$ rad) in \pp collisions. 
%%%%%%%%%
%%%%%%%%%

When the cluster candidate for isolation has a pseudorapidity $0.5<|\eta|<0.67$, a small fraction of the isolation cone of radius $R=0.4$ is out of the tracking acceptance $|\eta^{\rm track}|<0.9$. 
To maximise the photon acceptance, such candidate clusters are kept in the analysis, but the measured isolation momentum is scaled up to account for the cone area that is out of the tracking acceptance~\cite{ALICE:2020atx,ALICE:2024kgy}.

\begin{figure}[ht]
\begin{center}
\includegraphics[width=1.0\textwidth]{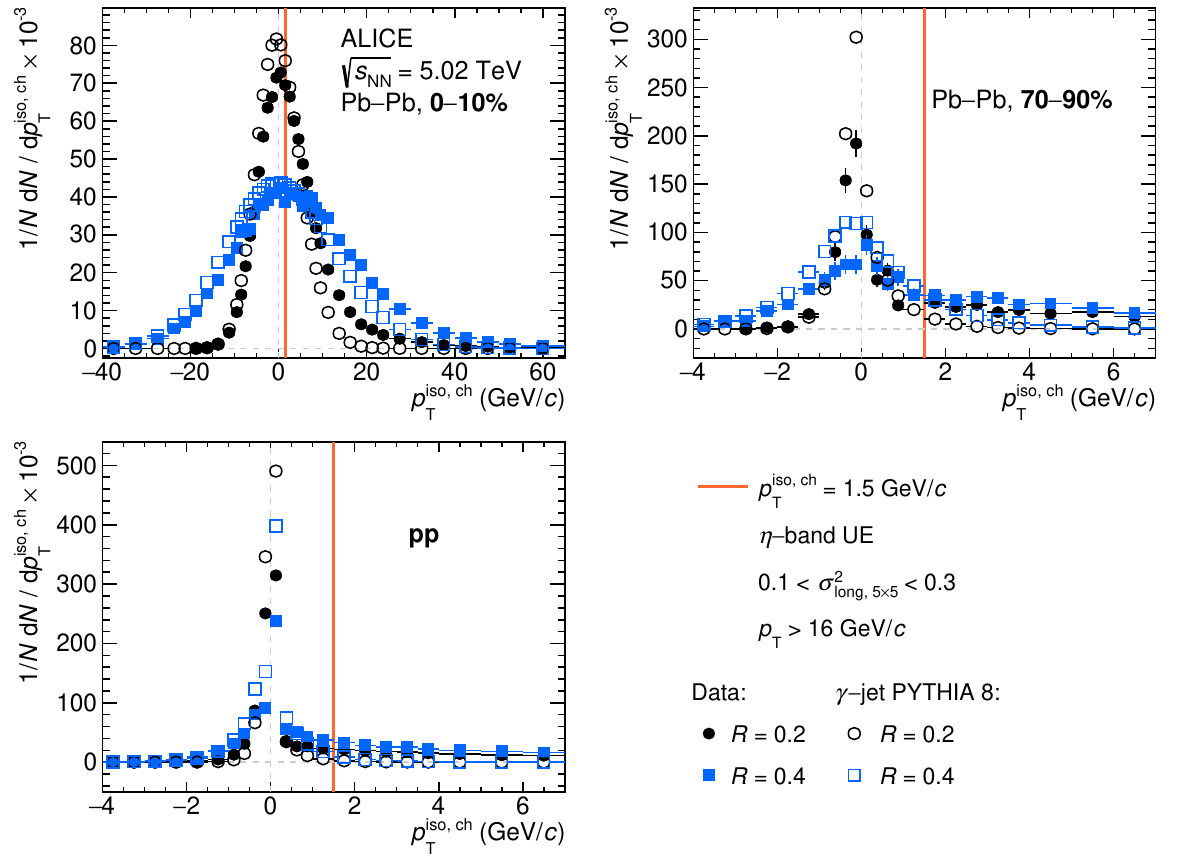} 
\end{center}
\caption{\label{fig:IsoPtCh} (colour online) \ptIsoCh\ distribution 
for narrow clusters with $0.1<\sigmalongPb<0.3$ for $\pt~>~16$~GeV/$c$,
in \pp (bottom left frame) and \PbPb collisions in two centrality classes, 0--10\% central (top left frame) and 70--90\% peripheral (top right frame), 
with $R=0.2$ (black bullets) and $R=0.4$ (blue squares),
in data (full markers) and simulated PYTHIA~8 $\gamma$--jet  (open markers), embedded into data in the considered centrality class for the \PbPb collision case.  }
\end{figure}

Figure~\ref{fig:IsoPtCh} shows the \ptIsoCh\ distribution for both $R$ values, for clusters with a shower shape between $0.1<\sigmalongPb<0.3$ and $\pt > 16$~\GeVc, in data as well as in PYTHIA~8 simulations of prompt photons ($\gamma$--jet process), either native \pp collisions, or embedded into real \PbPb collision data in two extreme centrality classes (other centrality classes are reported in Ref.~\cite{ALICE-PUBLIC-2024-003}). Even though the UE energy to be subtracted is large, the distributions are centred around zero, even for the most central events. 
In the prompt-photon simulation, the distribution is symmetric since there is no jet contribution. On the contrary, the data contain a jet contribution when the cluster does not originate from a prompt photon. 
This contribution induces a widening of the distribution tail at positive values of \ptIsoCh.
The width of the distribution is larger for $R=0.4$ than for $R=0.2$, due to the larger UE fluctuations in the isolation cone. For the same reason, the width decreases when moving to more peripheral collisions.

%%%%%%%%%
%%%%%%%%%
%\newpage

The candidate photon is declared isolated if $p_{\rm T}^{\rm iso,~ch}<$~1.5\,\GeVc, following previous ALICE measurements~\cite{ALICE:2020atx,ALICE:2024kgy}. 
For the most central Pb--Pb collisions, the chosen threshold value for the isolation momentum is smaller than the width of the \ptIsoCh distribution, which has an r.m.s.~of about 5~(12)~\GeVc for $R=0.2$~(0.4) in the 0--10\% centrality class~\cite{ALICE-PUBLIC-2024-003}. This may suggest increasing the threshold value for central collisions in order to preserve more signal. However, 
the use of the same value for all the considered centrality classes is preferred to ease the comparison with other collision systems, collision energies, or models.

\subsection{Purity of the isolated-photon candidate sample\label{sec:purity}}

The isolated-photon candidate sample still contains a sizeable contribution from background clusters, mainly from neutral-meson decay photons.  
To estimate the background contamination, the same procedure as in Refs.~\cite{ALICE:2019rtd, ALICE:2024kgy, ALICE-PUBLIC-2024-003} is followed, also known as the ABCD method.

Different classes of measured clusters are used:
(1) classes based on the shower shape \sigmalongPb, i.e.\ \textit{narrow} (photon-like) or \textit{wide} (most often elongated, i.e.\ non-circular), and 
(2) classes defined by the isolation momentum \ptIsoCh, i.e.\ \textit{isolated} (iso) and \textit{anti-isolated} ($\overline{\rm iso}$).  
The different classes are denoted by sub- and superscripts, e.g.~narrow isolated clusters are denoted $X_{\rm n}^{\rm iso}$, and wide anti-isolated clusters as $X_{\rm w}^{\overline{\rm iso}}$. The yield of isolated-photon candidates in this nomenclature is $N_{\rm n}^{\rm iso}$. 
It consists of signal ($S$) and background ($B$) contributions: 
$N_{\rm n}^{\rm iso} = S_{\rm n}^{\rm iso} + B_{\rm n}^{\rm iso}$.
The contamination of the candidate sample is then $C = B_{\rm n}^{\rm iso}/N_{\rm n}^{\rm iso}$, and the purity $P$ is $P \equiv 1 - C$. 

The \sigmalongPb-parameter values for narrow and wide clusters correspond to the signal and background clusters introduced in Sect.~\ref{sec:photonident}: the wide clusters (mostly background) correspond to clusters with $ 0.4 <~$\sigmalongPb~$ <2$ in \pp collisions, and in \PbPb collisions when $\pt > 18$~\GeVc.
When $\pt < 18$~\GeVc in \PbPb collisions, clusters are considered wide when $0.1+\sigma_{\rm max}^{2}(\pt)  <~$\sigmalongPb~$ <2$. 
The narrow clusters (containing most of the signal) are defined in Sect.~\ref{sec:photonident}. 
The anti-isolation criterion is set to $4<~\ptIsoCh~<25$\,\GeVc for all collision systems: the lower limit is placed far from the signal isolation momentum threshold at $\ptIsoCh<1.5$~\GeVc, to have a gap available for systematic studies.
The same isolation and anti-isolation regions are used for both cone radii used in this measurement.

Considering the partial assumption that the ratios of isolated over anti-isolated background are the same in the narrow cluster range and in the wide cluster range and that the signal contribution to the background classes is negligible, the purity is calculated in a semi-data-driven way as
%%%%
\begin{equation}
\label{eq:ABCDpurityMC}
P = 1-\bigg(\frac {N_{\rm n}^{\overline{\rm iso}}/N_{\rm n}^{\rm iso}} {N_{\rm w}^{\overline{\rm iso}}/N_{\rm w}^{\rm iso}}\bigg)_{\rm data} \times \bigg(\frac {B_{\rm n}^{\rm iso}/N_{\rm 
n}^{\overline{\rm iso}}} {N_{\rm w}^{\rm iso}/N_{\rm w}^{\overline{\rm iso}}}\bigg)_{\rm MC}. 
\end{equation}
%%%%
The MC factor corrects the correlation between isolation momentum and shower shape. It is calculated via the addition of jet--jet 
 (background) and $\gamma$--jet (signal) counts scaled to their respective cross sections.
This difference between the degree of the correlation between isolation momentum and shower shape distribution in data and simulation is a potential source of bias 
and is discussed in Ref.~\cite{ALICE-PUBLIC-2024-003}. 
A similar approach as in previous ALICE isolated-photon measurements is followed to estimate this difference~\cite{ALICE:2019rtd,ALICE:2024kgy}. 

\begin{figure}[hb]
\begin{center}
\includegraphics[width=1.0\textwidth]{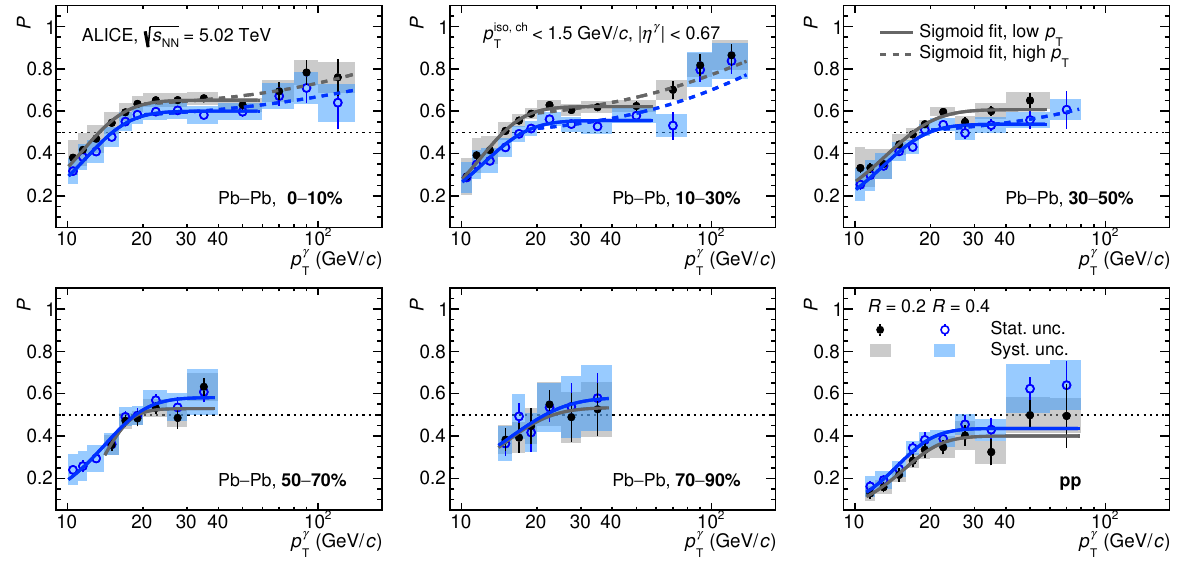} 
\end{center}
\caption{\label{fig:purityIsoPhoton} (colour online) Purity of the isolated-photon sample as a function of \ptg\ calculated using Eq.~\eqref{eq:ABCDpurityMC} and with the statistical and systematic uncertainty discussed in Sect.~\ref{sec:sys_unc} for $R=0.2$ and $0.4$. The curves (plain for low \ptg, dashed for high \ptg) 
are the two sigmoid functions as defined in Eq.~\eqref{eq:sigmoid}, obtained by fitting the points as explained in the text.
}
\end{figure}

Figure~\ref{fig:purityIsoPhoton} shows the purity calculated using Eq.~\eqref{eq:ABCDpurityMC}. 
The boxes indicate the systematic uncertainty, whose estimation is explained in Sect.~\ref{sec:sys_unc}. 
The purity found at low \ptg is small, due to a large contamination from \piz:
 at $\ptg = 10-12$\,\GeVc\ the contamination reaches 70--80\% for \PbPb collisions and approximately 90\% for pp collisions.
For higher \ptg, the contamination decreases and stabilises around $\ptg \simeq 18$\,\GeVc at 40--50\% in \PbPb collisions and  60\% in pp collisions.
It then decreases again above 40~GeV/$c$, reaching about 20\% above 80~GeV/$c$ for central \PbPb collisions. 
The purity for pp collisions calculated for $R~=~0.4$ is consistent with the previous ALICE isolated photon--hadron correlation measurements~\cite{ALICE:2020atx} 
since the differences in the analysis procedure, such as shower shape and cluster definition, larger acceptance, and different energy density calculation methods, do not lead to changes within the uncertainties.
In \PbPb collisions, the purity decreases when moving from central to peripheral collisions, due to the fact that the \pt of the main contamination background -- photons from neutral-meson decays --  is shifted down due to the jet quenching in the more central collisions. Still, the purity for most peripheral (70--90\%) collisions  remains larger than that measured in pp collisions (about 50\% in the range $20<\ptg <40$~\GeVc instead of 40\%), in part because of the lack of TPC information in \pp collisions that impedes doing cluster--track association.

In 0--10\% \PbPb collisions, the purity is larger by a factor of 1.1 for $R=0.2$ with respect to $R=0.4$. The values get closer for less central collisions, and become almost identical for peripheral \PbPb collisions. In \pp collisions, the situation is reversed: the purity is larger for $R=0.4$ than for $R=0.2$ by a factor of about 1.2.
This ordering in pp collisions is due to the fact that the larger the cone radius, the more jet fragments can enter when one triggers on decay photons, and thus the larger is \ptIsoCh as seen in Fig.~\ref{fig:IsoPtCh}. 
The change in ordering in the more central \PbPb collisions is due to the larger UE fluctuations in \ptIsoCh\ for $R=0.4$, also seen in Fig.~\ref{fig:IsoPtCh}, 
that allow more background clusters produced by jet particles to be considered isolated.\\

%%%%%%%%%
%%%%%%%%%

The \ptg\ dependence of the purity is caused by an interplay of physics and detector effects. 
Most of the contamination is due to \piz-decay photons.
On the one hand, the \pt\ spectra of prompt photons are harder than those of neutral pions, mainly because the latter undergo fragmentation, as is also found in pQCD calculations~\cite{Abeysekara:2010ze, Arleo:PhotonLHCYellow}. 
For this reason, the $N_{\gamma^{\rm prompt}} / N_{\gamma^{\rm decay}(\pi^{0})}$ 
yield ratio rises with \ptg, and therefore, the photon purity increases with \ptg.
Also, the probability of tagging a photon as isolated varies with \ptg. 
At higher decay-photon \ptg, isolation is less probable for a fixed isolation momentum. 
On the other hand, the rejection of clusters from \piz\ and $\eta$ decays 
at high \pt becomes less effective due to the decreasing decay-photon opening angle when increasing the meson \pt. 
Below 18\,\GeVc, the contamination is dominated by single (i.e.\ unmerged) decay photons from {\piz} mesons, as shown by Fig.~\ref{fig:M02DataMCProj}-left, the remaining contributors being mainly photons from $\eta$ meson decays. 
Above 18\,\GeVc, a large fraction of the $\pi^{0}\rightarrow\gamma\gamma$ decays produces two photons with narrow opening angle and gives rise to merged clusters in the EMCal with a narrow shower shape that satisfies the condition for the single photon signal, as can be appreciated in Fig.~\ref{fig:M02} and Fig.~\ref{fig:M02DataMCProj}-right. 
The clusters produced by merged photons from $\eta$-meson decays contribute to the narrow shower shape region for $\pt>60$~\GeVc but they remain subdominant compared to merged \piz-decay clusters. Instead, in the range $40<\pt<60$~\GeVc, most of the merged $\eta$-decay clusters have wide shower shapes, which is in part the reason for the increase of purity in this \ptg\ region since the contribution of single photon clusters from $\eta$ decays to the narrow clusters decreases.
The combined effect of these mechanisms leads to the rise of the purity at low \ptg, followed by a plateau for $18<\ptg <  40$\,\GeVc, then by a rise above $\ptg = 40$~\GeVc. Above 80 GeV/$c$, another plateau is expected, as observed in the ALICE measurement in \pp collisions at $\s=13$~TeV~\cite{ALICE:2024kgy}.

To reduce the  point-to-point statistical fluctuations in the purity used to correct the isolated-photon raw yield, the distribution is fitted by one or two sigmoid functions to reproduce the trend of the purity with~\ptg
%%%%
\begin{equation} \label{eq:sigmoid}
f_{i, ~\rm fit-sigm}(\ptg) = \frac{a_i}{1+\exp(-b_i \times (\ptg - c_i))},
\end{equation}
%%%%
where $i$ indicates the different fitting ranges, which depend on the collision system. 
The first fit is done from $\ptg =10$--14 to 40--60~\GeVc. In most of the \PbPb centrality classes between 0 and 50\%, enough points are available beyond 60~\GeVc to reliably describe the tendency by a second fit function from $\ptg~=20$~\GeVc to $\ptg~=80-140$~\GeVc. 
Although these fits start at 20~\GeVc, they are used for the purity correction only above 60~\GeVc. 
In \pp collisions, although purity values are obtained up to 80~\GeVc, the uncertainties are too large to obtain a reliable fit at high \ptg. 
The fit is therefore done up to \ptg = 40~\GeVc, and extrapolated up to 80~\GeVc. 
In the extrapolation interval, a slow rise with \ptg is observed in simulation as well as in \pp collisions at \sthirteen, 
but the estimated size of the rise is covered by the assigned uncertainties.
The fit results are shown in Fig.~\ref{fig:purityIsoPhoton} and the fit parameters are provided in Ref.~\cite{ALICE-PUBLIC-2024-003}.
The systematic uncertainties on the purity are used during the fitting as discussed in Sect.~\ref{sec:sys_unc}.

\subsection{Isolated-photon efficiency \label{sec:eff}}

The photon reconstruction, identification and isolation efficiencies have been computed using PYTHIA~8 simulations of $\gamma$--jet processes
in which, for each event, a prompt photon from a $2\rightarrow 2$ Compton or annihilation process is emitted in the EMCal acceptance.  Only those falling in the fiducial acceptance defined in Table~\ref{tab:cuts} are considered in the efficiency calculation. 

\begin{figure}[hb]
\begin{center}
\includegraphics[width=1.0\textwidth]{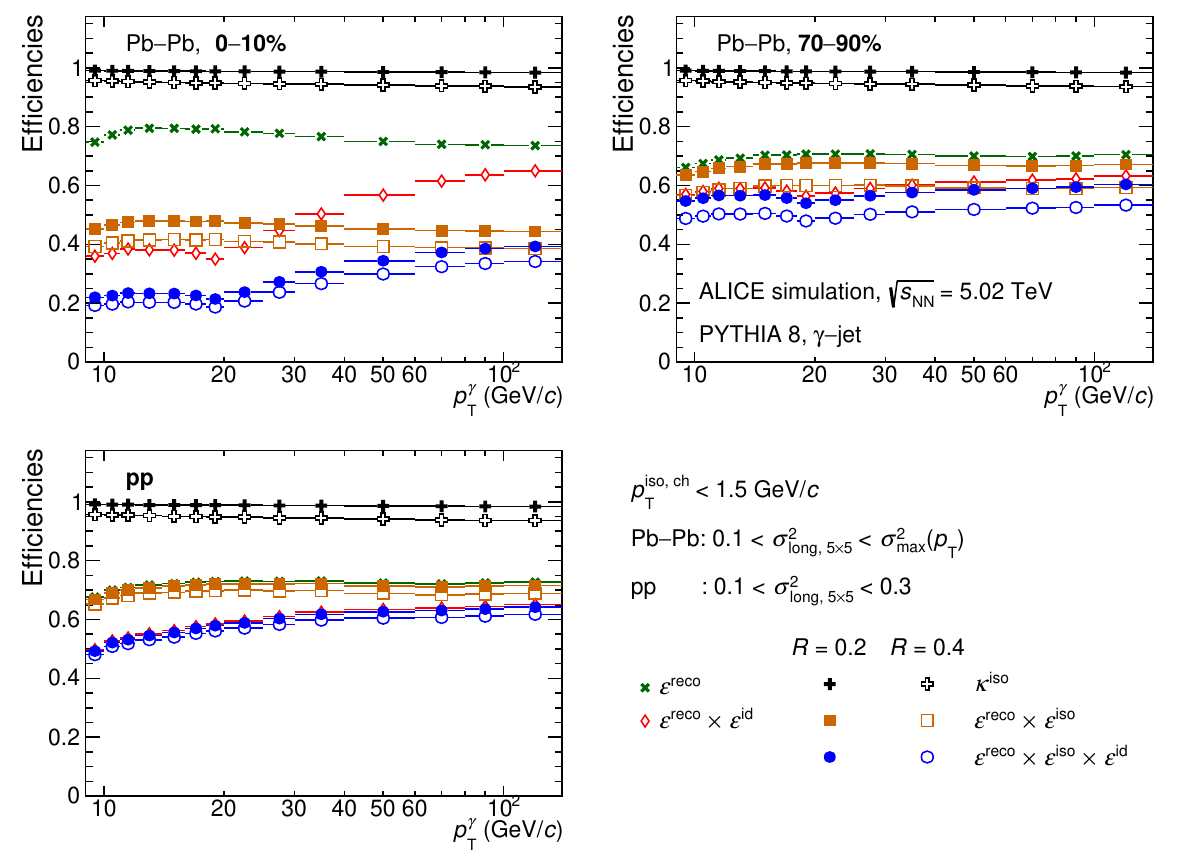} 
\end{center}
\caption{\label{fig:EffIsoPhotonComp} (colour online) 
Contributions from reconstruction, identification, and isolation to the total efficiency calculated using Eq.~\eqref{eq:efficiency}, 
as a function of the reconstructed photon \ptg for \pp (bottom left frame) collisions and \PbPb collisions for two centrality classes: 0--10\% (upper left frame) and 70--90\% (upper right frame). 
Green markers: reconstruction efficiency $\varepsilon^{\mathrm{rec}}$.
Red or brown markers: efficiency additionally due to the photon identification by shower shape selection $\varepsilon^{\mathrm{rec}} \times \varepsilon^{\mathrm{id}}$ or the isolation criterion $\varepsilon^{\mathrm{rec}} \times \varepsilon^{\mathrm{iso}}$.
Blue markers: efficiency due to the isolation criterion and shower shape selection $\varepsilon^{\mathrm{rec}} \times \varepsilon^{\mathrm{id}} \times \varepsilon^{\mathrm{iso}}$.
Black markers: fraction $\kappa^{\mathrm{iso}}$ of generated photons which are isolated.
The efficiency is obtained from PYTHIA~8 simulations of \pp collisions $\gamma$--jet processes, embedded into data in the considered centrality class for the \PbPb collision case. }
\end{figure}

Different efficiencies can be considered depending on the selection criteria: reconstruction $\varepsilon^{\mathrm{rec}}$ (inclusive cluster selection), photon identification $\varepsilon^{\mathrm{id}}$ (shower shape selection), and isolation $\varepsilon^{\mathrm{iso}}$. They are calculated as the ratio of \ptg\ spectra, where the denominator is the number of generated photons ${\rm d}N ^{\rm gen}_{\gamma} / {\rm d}p_{\rm T}^{\rm gen}$, 
and the factors in the numerator are the reconstructed spectra after different selection criteria, ${\rm d}N ^{\rm rec}_{\rm cut} / {\rm d}p_{\rm T}^{\rm rec}$. Figure~\ref{fig:EffIsoPhotonComp} presents the different contributions as a function of \ptg\ for \pp collisions and \PbPb collisions in the centrality classes 0--10\% and 70--90\%:
\begin{enumerate}[(i)]
\item The reconstruction efficiency of photons is $\varepsilon^{\mathrm{rec}} \approx 70-80 \%$; the efficiency loss is mainly due to excluded regions in the calorimeter and 
exclusion of clusters close to the border of the EMCal supermodules. This efficiency is higher for more central collisions due to the shift of the spectrum to higher \ptg induced by the additional UE energy.
\item The photon identification by shower shape selection induces a strong decrease of the efficiency $\varepsilon^{\mathrm{rec}} \times \varepsilon^{\mathrm{id}}$  in 0--10\% central collisions, by about 40\% below $\ptg = 40$~\GeVc, because the UE enlarges the photon cluster shape. 
In peripheral \PbPb and in pp collisions, the efficiency is only reduced by 10--20\%.
\item Applying the isolation criterion on top of the previous selections further decreases the overall efficiency,
as the isolation cone radius is large. The efficiency is then $\varepsilon^{\mathrm{rec}} \times \varepsilon^{\mathrm{id}} \times \varepsilon^{\mathrm{iso}} \approx 20-40 \%$ for the most central \PbPb collisions, and 50--60\% in the most peripheral \PbPb and in \pp collisions. 
\end{enumerate}
In addition, the generator-level isolation fraction 
$\kappa^{\rm iso} = \frac{{\rm d}N^{\rm gen}_{\gamma,~{\rm iso}}/ {\rm d}p_{\rm T}^{\rm gen} } {  {\rm d}N^{\rm gen}_{\gamma}/ {\rm d}p_{T}^{\rm gen}}$, 
where $N^{\rm gen}_{\gamma}$ is the number of generated prompt photons and $N^{\rm gen}_{\gamma,~{\rm iso}}$ is the number of generated photons which pass the same isolation selection criteria as at the detector level, has to be considered. 
It varies from low to high \ptg at 95.5--93.5\% for $R=0.4$, and 99--98.5\% for $R=0.2$, identically for all the collision systems considered. 

%%%%%%%%%
%%%%%%%%%

\newpage

\begin{figure}[ht]
\begin{center}
\includegraphics[width=1.02\textwidth]{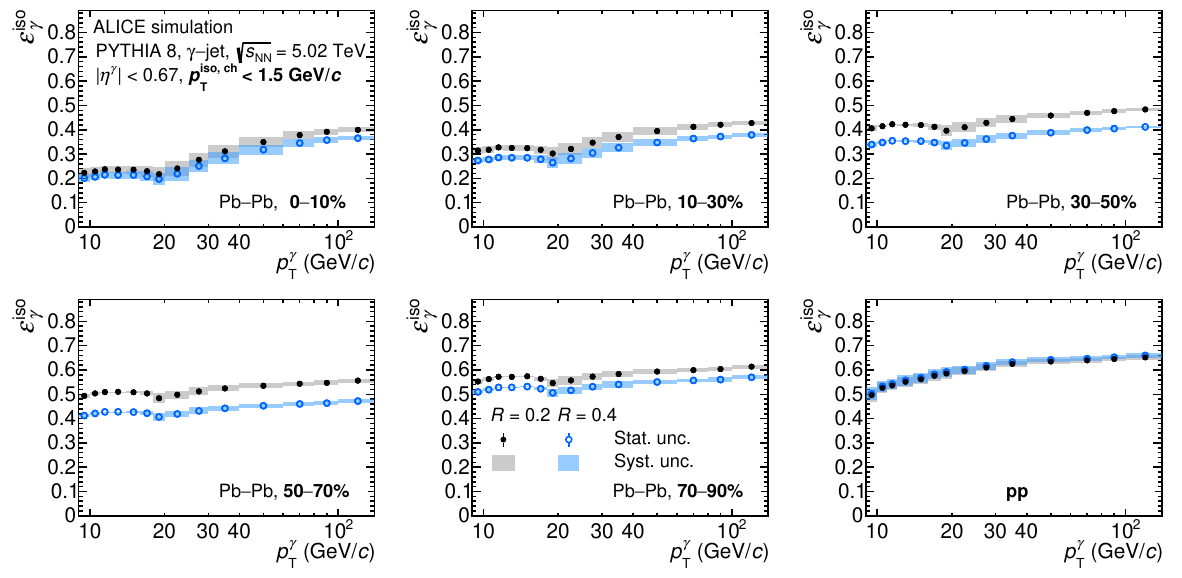} 
\end{center}
\caption{\label{fig:EffIsoPhoton} (colour online) Total isolated-photon efficiency as a function of \ptg\ calculated using Eq.~\eqref{eq:efficiency} 
with the systematic uncertainty discussed in Sect.~\ref{sec:sys_unc} for $R=0.2$ and $0.4$, 
 for \pp collisions and \PbPb collisions for five centrality classes. 
The efficiency is obtained from PYTHIA~8 simulations of \pp collisions $\gamma$--jet processes, embedded into data in the considered centrality class for the \PbPb collision case.}
\end{figure}

The total efficiency corresponds to the ratio of the reconstruction, identification, and isolation efficiency to the isolated generated photon fraction and is calculated as follows
%%%%%%%%%%%%%%%%%%%%%%%%%%%%%%%%%%%
\begin{equation}
\label{eq:efficiency}
     \varepsilon_{\gamma}^{\rm iso} = \frac{{\rm d}N ^{\rm rec}_{\rm n,\,iso}} { {\rm d}p_{\rm T}^{\rm rec}} {\Bigg /} \frac{{\rm d}N ^{\rm gen}_{\gamma,\,\rm iso}} { {\rm d}p_{\rm T}^{\rm gen}}
     \equiv \frac{\varepsilon^{\mathrm{rec}} \times \varepsilon^{\mathrm{id}} \times \varepsilon^{\mathrm{iso}}}{\kappa^{\mathrm{iso}}},
\end{equation}
%%%%%%%%%%%%%%%%%%%%%%%%%%%%%%%%%%%
where $N^{\rm rec}_{\rm n,~iso} $ is the number of 
clusters which are reconstructed and identified as isolated photons and which are produced by a prompt photon. 
Figure~\ref{fig:EffIsoPhoton} shows the $\varepsilon_{\gamma}^{\rm iso}$  
with the corresponding systematic uncertainties discussed in Sect.~\ref{sec:sys_unc}. The kink observed at $\ptg = 20$~\GeVc is due to the kink which separates the two shower shape selection criteria used (Fig.~\ref{fig:M02}).
In all \PbPb collision centralities, the efficiency for $R = 0.4$ is lower by a factor of about 0.85--0.9 than that for $R=0.2$. This is a consequence of using the same \ptIsoCh isolation threshold value for both cone radii, which makes isolation less efficient when larger cones are used.
In \pp collisions, the efficiency for $R=0.2$ is much closer to the one for $R=0.4$ due to the small contribution from the UE in such collisions. \\

%%%%%%%%%
%%%%%%%%%

\subsection{Trigger efficiency, rejection factor and luminosity}
\label{sec:trigger}
The isolated-photon yield correction needs to take into account the performance of the calorimeter trigger, 
in particular when calculating the event normalisation and luminosity.
The EMCal \tr-low and -high trigger efficiency $\varepsilon_{\rm trig}$ is the probability that the trigger selects events 
in which a high-energy cluster is reconstructed in the EMCal acceptance above a given trigger energy threshold. The threshold values are listed in Sect.~\ref{sec:detector}.  
This trigger efficiency does not reach 100\% above the trigger threshold because 
of the reduced geometric coverage of the trigger compared to the EMCal acceptance: some trigger cell tiles ($2\times 2$ cells) and even full TRU cards (Trigger Region Unit, $24\times16$ cells along $\varphi \times \eta$) were inactive or masked during the data taking. 
Furthermore, a \pt dependence of the trigger is observed since higher energy clusters cover more cells (owing to nearby jet particles in the event and meson decay merging). These are less affected by small masked regions.

\begin{figure}[hb]
\begin{center}
 \includegraphics[width=0.497\textwidth]{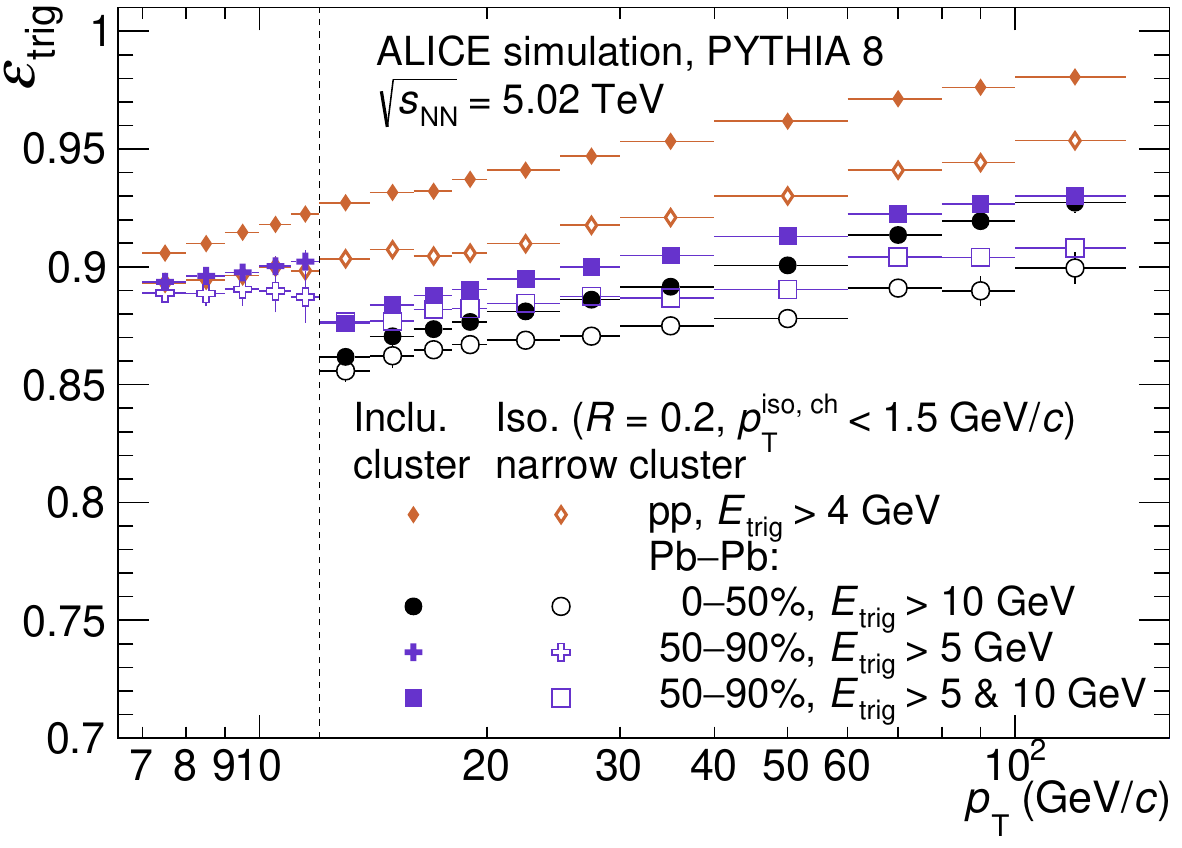}
 \includegraphics[width=0.497\textwidth]{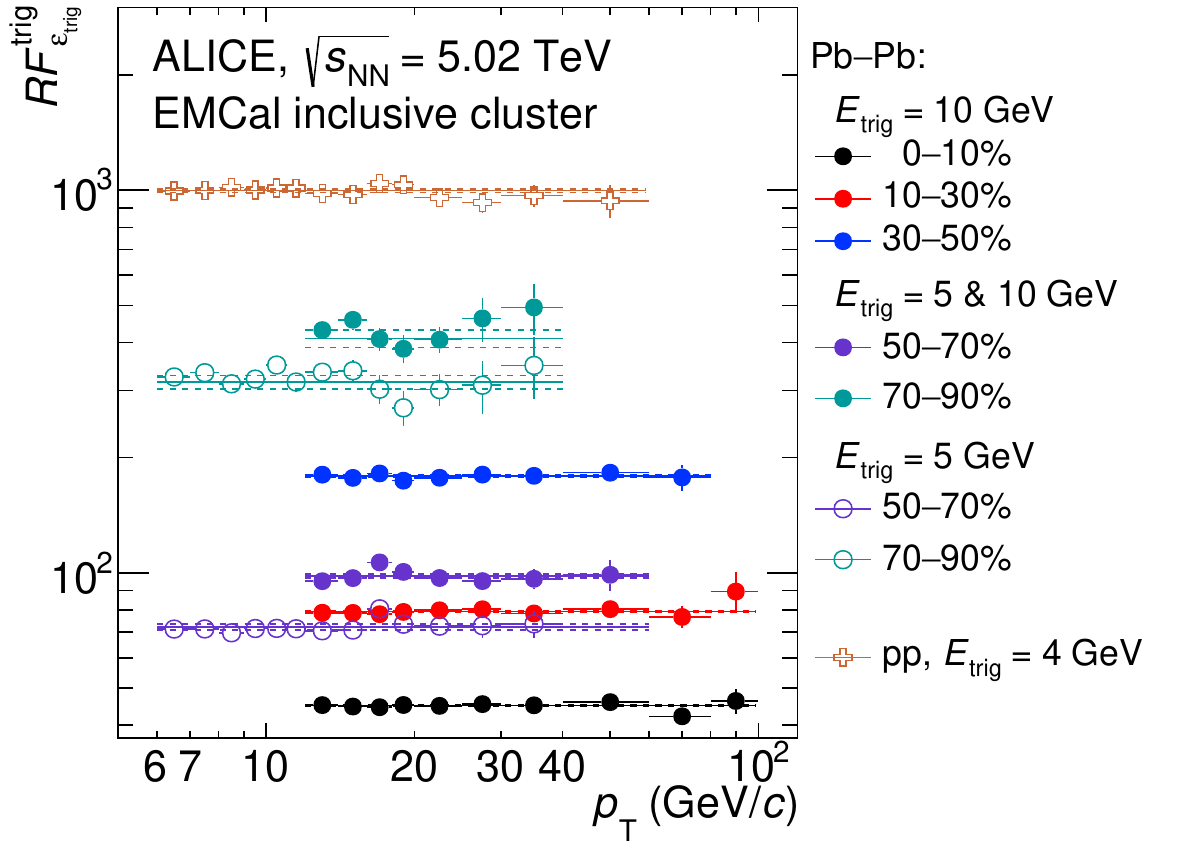}
\end{center}
\caption{\label{fig:RF_effiTrig}(colour online) 
Left: \tr trigger efficiency for inclusive clusters and isolated-narrow clusters with $R=0.2$ (similar for $R=0.4$) 
obtained with PYTHIA~8 simulations combining $\gamma$--jet and jet--jet processes, 
embedded in data in the considered centrality class for \PbPb collisions and considering the corresponding trigger thresholds ($E_{\rm trig}$) in each system. 
Right: \tr trigger rejection factor calculated by applying the trigger efficiency for \pp collisions and for each of the \PbPb centrality classes considered. 
Solid lines over points result from a constant fit, values given in Table~\ref{tab:Lumi}, 
dashed lines indicate the fit uncertainty obtained with the procedure explained in Sect.~\ref{sec:sys_unc}. 
For clusters above 12~\GeVc and peripheral \PbPb collisions, the rejection factor and trigger efficiency are calculated for the sum of the two triggered samples with thresholds at 5 and 10 GeV.
}
\label{fig:EffTrig_RF}
\end{figure}

The trigger efficiency is calculated from simulation,  combining the jet--jet and $\gamma$--jet PYTHIA~8 simulations, by applying the same trigger logic as in the data, and it is shown in Fig.~\ref{fig:RF_effiTrig}-left.
In \pp collisions, the trigger efficiency for inclusive clusters $\varepsilon_{\rm trig}^{\rm clus}$  
varies from nearly 90\% at $\pt = 7$~\GeVc to close to 97\% at 80~\GeVc. 
In \PbPb collisions, a dependence on the trigger threshold is observed, but not on the centrality. For the lower threshold (\unit[5]{GeV}), below $\pt =12$~\GeVc, the efficiency is indeed found to be close to the efficiency in \pp collisions, which had a similar trigger threshold.
For the higher threshold in \PbPb collisions (\unit[10]{GeV}), the efficiency for inclusive clusters rises from about 85\% at $\pt =12$~\GeVc to about 93\% at $\pt =140$~\GeVc. 
The trigger efficiency for isolated and narrow clusters $\varepsilon_{\rm trig}^{\rm iso}$  is lower than $\varepsilon_{\rm trig}^{\rm clus}$ by 1--3\% for all trigger thresholds, 
since narrow clusters are less likely to trigger near masked regions due to their smaller size.
For peripheral collisions, both \tr-low and -high triggers are combined for $\pt > 12$~\GeVc. 
Figure~\ref{fig:EffTrig_RF}-left also shows the trigger efficiencies for the combined sample: the points are overall 2\% higher than for the high threshold alone.

The EMCal trigger rejection factor $RF^{\rm trig}_{\varepsilon_{\rm trig}}$ quantifies the enhancement fraction of calorimeter triggers with respect to MB triggers. 
It is calculated via the ratio of the inclusive-cluster \pt spectra measured in data corrected by the inclusive-cluster trigger efficiency obtained from simulation and discussed above
%%%%
\begin{equation}
RF^{\rm trig}_{\varepsilon_{\rm trig}} = \frac{1}{\varepsilon_{\rm trig}^{\rm clus}} \frac{1/N_{\rm evt}^{\rm L1\text{-}\gamma} \times {\rm d}N^{\rm L1\text{-}\gamma}/{\rm d}p_{T} }{ 1/N_{\rm evt}^{\rm MB} \times {\rm d}N^{\rm MB}/{\rm d}p_{T}},
\label{eq:rfeff}
\end{equation}
%%%%
where $N_{\rm evt}^{\rm trig}$ is the number of events and $N^{\rm trig}$ is the number of inclusive clusters, each for a given trigger. 

Figure~\ref{fig:RF_effiTrig}-right shows the trigger rejection factors calculated with Eq.~\eqref{eq:rfeff} for the different trigger configurations in the analysed samples,
and Table~\ref{tab:Lumi} lists the results of the fit in the plateau region, with an uncertainty explained in the next Section. 
Note that for the calculation in \pp collisions, the MB sample contained  $8.41 \times 10^{8}$ events and
it was collected not at the same time but some days before since it included the TPC. 
Although this sample is used for calculating the rejection factor, it is not included in the isolated photon analysis 
since these events are negligible compared to the EMCal \tr triggered sample. \\

%~fig~%%%%%%%%%
%%%%%%%%%%

\newpage

The rejection factor depends on the trigger threshold and on the centrality: 
it is more likely to find a high-energy and large-size cluster in central compared to peripheral \PbPb collisions due to the larger number of nucleon--nucleon binary collisions.
The rejection factor ranges from about 45 for the 0--10\% \PbPb collisions to about 300 (\tr-low) and 1000 (\tr-high, not shown in Fig.~\ref{fig:EffTrig_RF}-right) for 70--90\% \PbPb collisions, and close to 1000 for \pp collisions. Combining the two trigger thresholds in peripheral \PbPb collisions, a factor around 100 (400) is obtained for the 50--70\% (70--90\%) centrality class.

The integrated luminosity collected with each trigger 
($\lint^{\rm trig}$) 
has been determined using the expression
%%%%
\begin{equation}
\lint^{\rm trig} =  \frac{N_{\rm evt}^{\rm trig}~RF^{\rm trig}_{\varepsilon_{\rm trig}}} { \sigma_{\rm NN}^{\text{col. system}}} \times \Ncoll = \mathscr{L}_{\rm NN}^{\rm trig} \times \Ncoll
\label{eq:linttrig}
\end{equation}
%%%%
where $\sigma_{\rm NN}^{\text{col. system}}$ is the measured nucleon--nucleon cross section, that corresponds to $\sigma_{\rm MB}^{\rm \pp}=50.9 \pm 1.1$~mb for \pp collisions~\cite{ALICE:pp5TeVMBCrossSec} and to $\sigma^{\rm INEL}_{\rm NN} = 67.6 \pm 0.6$~mb for \PbPb collisions~\cite{ALICE:PbPb5TeVMBCrossSec}, and $\mathscr{L}_{\rm NN}^{\rm trig}$ is the cross-section normalisation factor used in Eq.~\ref{eq:cs} of Sect.~\ref{sec:results}.
The final production cross section is measured as a function of \ptg, thus, the different triggers are combined depending on the trigger threshold, except in \pp collisions, where only the \tr triggered data are used.
In \PbPb collisions, the \tr-high trigger threshold is at $E=10$~GeV but a satisfactory efficiency is reached only above slightly larger energies.
The spectrum was therefore measured in the following way:
\begin{itemize}
\item below $\ptg=12$~GeV/$c$, using only the MB  trigger for the centrality classes within 0--50\%, and
a combination of MB  plus \tr-low trigger for the peripheral centrality classes; 
\item above 12~\GeVc, using the combination of the MB and \tr-high trigger for the centrality classes within 0--50\%, and a combination of the three triggers for the peripheral centrality classes.
\end{itemize}
The corresponding values of the integrated luminosity per trigger combination are presented in Table~\ref{tab:Lumi}. \\ 

%%%%%%%%%%%
\begin{table}[ht]
\begin{center}
\caption{\label{tab:Lumi}
Trigger $RF^{\rm trig}_{\varepsilon_{\rm trig}}$ (Eq.~\eqref{eq:rfeff}) fits to a constant in Fig.~\ref{fig:RF_effiTrig}-right, $\mathscr{L}_{\rm NN}^{\rm trig}$, and $\lint^{\rm trig}$ (Eq.~\eqref{eq:linttrig}),
for \pp and \PbPb collisions per centrality class and per trigger inclusive cluster \pt range.
The $\mathscr{L}_{\rm NN}^{\rm trig}$ uncertainty contains both the $\sigma_{\rm NN}^{\text{col. system}}$ and rejection factor uncertainties. The integrated luminosity uncertainty includes in addition the \Ncoll uncertainty.}

\begin{tabular} {l*{5}{c}r}  Trigger & System &\pt\ (\GeVc) & $RF^{\rm trig}_{\varepsilon_{\rm trig}}$  & $\mathscr{L}_{\rm NN}^{\rm trig}$ (nb$^{-1}$)    &$\lint^{\rm trig}$ (nb$^{-1}$) \\
\hline
\hline
\tr                               & pp            &  $\pt>11$  & 997 $\pm$ 10 &  265 $\pm$   7 & 265 $\pm$   7 \\
\hline
& Pb--Pb:  &~ &~ \\
\hline
MB      &~~0--10\% & $\pt<12$     &     &  1.189  $\pm$  0.011 & 1869   $\pm$  26  \\
MB      &10--30\% &  $\pt<12$     &     &  0.522  $\pm$  0.005  &   409  $\pm$  5     \\
MB      &30--50\% &  $\pt<12$     &     & 1.163   $\pm$  0.010  &  308   $\pm$  5     \\
\hline
MB+\tr-high  &~~0--10\% &  $\pt>12$     &   45.0 $\pm$  0.2  &  2.50    $\pm$  0.02   &  3936 $\pm$  55   \\
MB+\tr-high  &10--30\% &  $\pt>12$     &  79.2 $\pm$  0.4   & 4.90     $\pm$  0.05    &  3834 $\pm$  51   \\ 
MB+\tr-high  &30--50\% &  $\pt>12$     &   179.3 $\pm$  1.5   &  5.01    $\pm$  0.05      & 1325  $\pm$  21   \\
\hline
MB+\tr-low  & 50--70\% & $\pt<12$  & 72.2 $\pm$  1.2   & 3.5       $\pm$  0.5         & 230  $\pm$  5    \\
MB+\tr-low  & 70--90\% &  $\pt<12$ & 315 $\pm$  13 & 3.62    $\pm$  0.11       & 39.5 $\pm$  1.3   \\
\hline
MB+\tr-high+low  & 50--70\% &  $\pt>12$     &    98.2 $\pm$  1.2   & 4.88     $\pm$  0.07       & 322  $\pm$  7   \\ 
MB+\tr-high+low  & 70--90\% &  $\pt>12$     &   410 $\pm$  20   & 5.1      $\pm$  0.2          & 55    $\pm$  2      \\
\hline

\end{tabular}
\end{center}
\end{table}
%%%%%%%%%%%

\newpage
\section{Systematic uncertainties}
\label{sec:sys_unc}

\begin{figure}[ht]
\centering
\includegraphics[width=1\textwidth]{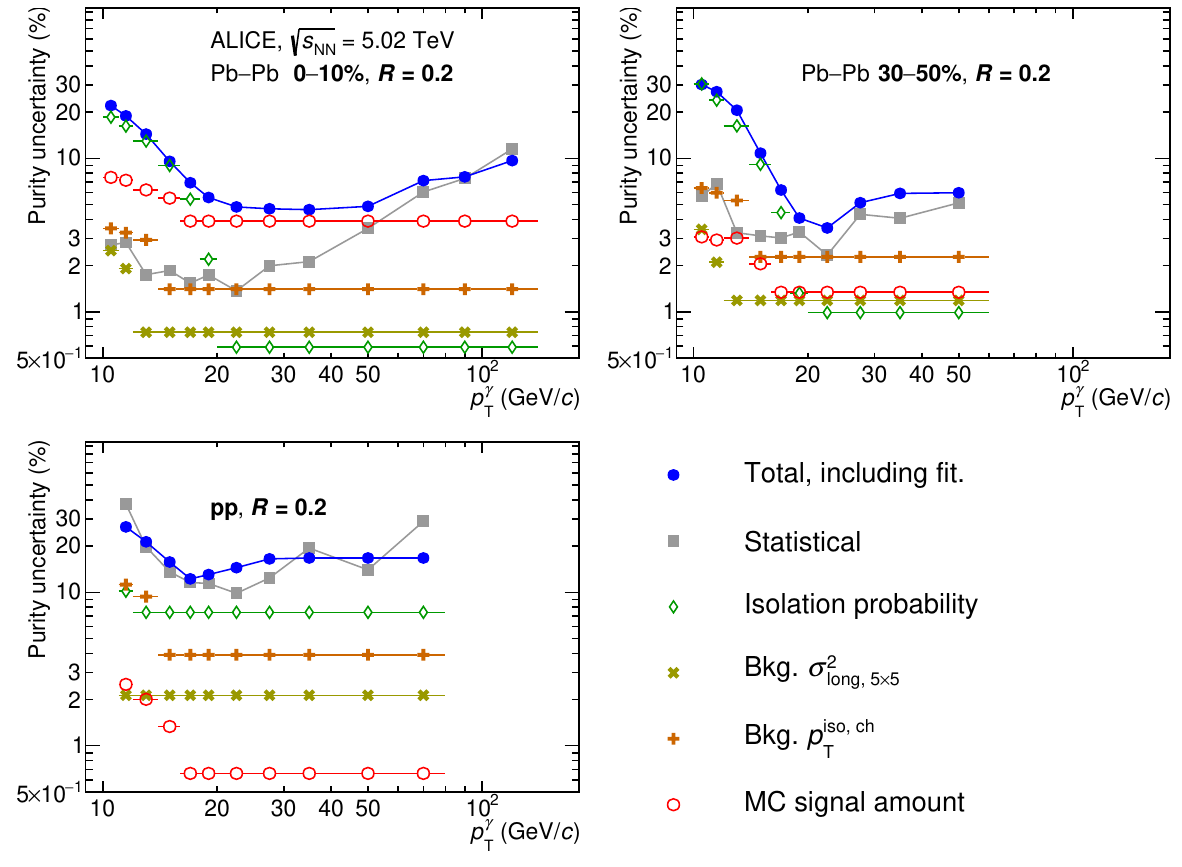} 
\caption{(colour online) 
Contributions to the systematic uncertainty of the isolated-photon purity and their quadratic sum as a function of \ptg for $R=0.2$,  in \pp collisions (bottom left frame) and two \PbPb collision centrality classes: 0--10\% (top left frame) and 30--50\% (top right frame). 
The statistical uncertainty is also shown and may appear larger than the total uncertainty as a result of the smoothing done by the fit over the purity (see~text).
}
\label{fig:pur_sys_R02}
\end{figure}

\begin{figure}[ht]
\centering
\includegraphics[width=1.\textwidth]{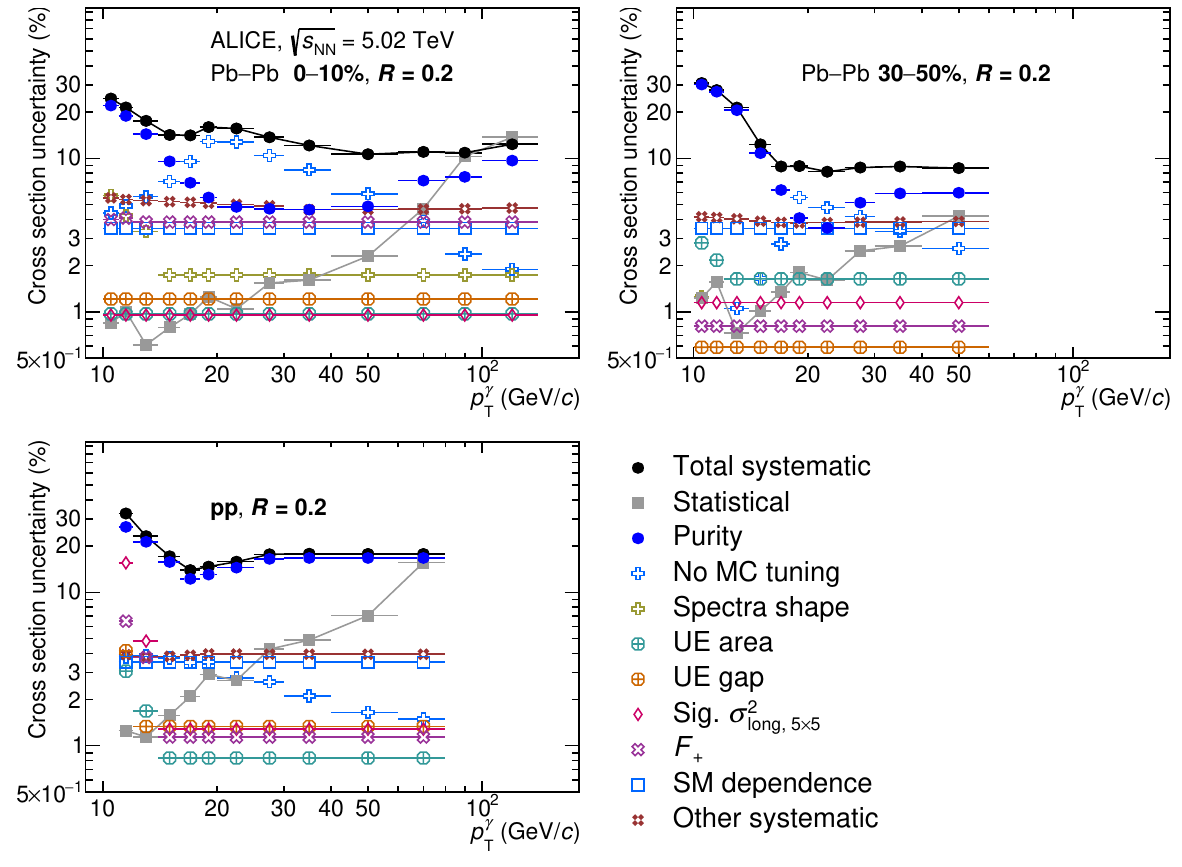} 
\caption{(colour online) 
Contributions to the systematic uncertainty of the isolated-photon cross section
and their quadratic sum as a function of \ptg for $R=0.2$ in \pp collisions (bottom left frame) and two \PbPb collision centrality classes: 0--10\% (top left frame) and 30--50\% (top right frame). 
Statistical uncertainty is also shown. }
\label{fig:spe_sys_R02}
\end{figure}

Figure~\ref{fig:pur_sys_R02} displays
the estimated relative systematic uncertainties for all the considered sources for the purity calculation
for $R=0.2$ in \pp collisions and \PbPb collisions in two centrality classes.
Equivalently,  Fig.~\ref{fig:spe_sys_R02} collects all the estimated relative systematic uncertainty sources considered for the cross section measurement.
The uncertainty contributions from all the sources are added in quadrature, and the individual contributions and their sum are shown in the figures.  
All sources are considered uncorrelated.
The contributions to the cross section include the total uncertainty for the purity. 
Summary tables and figures for all the centrality classes and both cone radii can be found in Ref.~\cite{ALICE-PUBLIC-2024-003}. 

The uncertainty contributions assigned to the purity correction using the ABCD method described in Sect.~\ref{sec:purity} are estimated from variations of the anti-isolation momentum (labelled bkg. \ptIsoCh in Fig.~\ref{fig:pur_sys_R02}) and the shower shape for wide-cluster (bkg. \sigmalongPb) ranges, 
their correlation effect on the MC correction (isolation probability), 
the amount of signal in the simulation with respect the background (MC signal amount), 
and from the errors of the fit to the purity including the statistical uncertainty. 

The uncertainty due to the choice of the background wide-cluster \sigmalongPb\ range is investigated by comparing the results obtained for various \sigmalongPb\ selections.
The lower limit is moved between 0.35 and 0.6, and the upper limit is chosen below or equal to 2 such that the interval width is at least 0.5. 
The largest value for the upper limit is chosen to coincide with the end of the shower shape distribution, and the values for the lower limit are chosen in such a way that a wide range of contamination by single prompt photons was covered (see Fig.~\ref{fig:M02DataMCProj}), while maintaining a small gap with the upper limit of the signal \sigmalongPb\ range.
The anti-isolation \ptIsoCh\ background range is also varied: the lower \ptIsoCh\ limit is chosen between 2 and 6~\GeVc  
and the upper limit is chosen below or equal to 70~\GeVc such that the range size is at least 10~\GeVc. 
Again, the largest value for the upper limit was chosen to coincide with the end of the distribution. The smallest value for the lower limit was set slightly above the upper limit of the signal \ptIsoCh\ range, and its largest value was set to the width that the \ptIsoCh\ distribution for $R=0.2$ has in the worst case, which is for the most central collisions. See the Gaussian fit width values for each centrality class and $R$ value in Ref.~\cite{ALICE-PUBLIC-2024-003}.
For both the wide-cluster and anti-isolation range variation, the average of the differences due to these variations is used to estimate each uncertainty.
 
The systematic uncertainty related to the correlation effects between  \ptIsoCh and \sigmalongPb
mentioned in Sect.~\ref{sec:purity}, labelled as ``isolation probability'', 
is obtained by the difference between the 
variations of the MC factors in Eq.~\eqref{eq:ABCDpurityMC} according to the procedure explained in Refs.~\cite{ALICE-PUBLIC-2024-003,ALICE:2024kgy}.
These variations test how sensitive the MC correction is to the assumption of having a constant or a linear dependence 
of the ratio of the shower shape distributions for isolated and non-isolated candidates, in the wide cluster region.

The signal-to-background ratio in the simulation influences the aforementioned leakage of signal into the background regions used to estimate the purity. 
This uncertainty is labelled as ``MC signal amount'' in the figures and is quantified by varying by $\pm 20$\% in the simulation the amount of signal events
($\gamma$--jet) with respect to the background events (jet--jet). 
This variation is chosen considering about a 5\% uncertainty in the measured nuclear modification factor of charged particles~\cite{ALICE:2018vuu,Khachatryan_2017} used to scale the jet--jet simulations, and the expected approximate 15\% contribution from fragmentation photons after isolation from NLO pQCD calculations~\cite{Ichou:2010wc,DENTERRIA2012311}.

The purity total uncertainty is calculated by
adding all the systematic-uncertainty sources together with the statistical uncertainty in quadrature
to obtain an uncertainty $\sigma_{\rm P}$. The purity points are then shifted up and down by $1\sigma_{\rm P}$ and fitted again by the sigmoid functions. In each \ptg interval, the total purity uncertainty is calculated as the average of the difference between the middle purity fit value and each of both shifted fit values. This procedure allows to also naturally take into account possible biases due to the fitting.
 
 \newpage
 
 %%%%%%%%%
%%%%%%%%%
 
The statistical uncertainty in the purity determination dominates over the whole \ptg range in \pp and peripheral \PbPb collisions and at high \ptg in central and semi-central \PbPb collisions.
Among the systematic uncertainty sources, the  ``isolation probability'' by far dominates the others in \pp and peripheral \PbPb collisions, and it is also the dominant uncertainty at low \ptg in the 0--10\% and  10--30\% centrality classes in \PbPb collisions. In the latter collisions though, the ``MC signal amount'' uncertainty source dominates at intermediate to high \ptg, especially for $R=0.4$.
Overall, the uncertainties for both radii are comparable, although slightly smaller for $R=0.2$ 
compared to $R=0.4$ at intermediate to high \ptg. 

For the systematic uncertainty on the cross section, different sources of uncertainty are evaluated on top of the uncertainty due to the purity. 
The uncertainties due to the choice of the neutral cluster selection criteria 
are evaluated via variations with respect to the default selections reported in Table~\ref{tab:cuts}:  
the track--cluster matching (CPV), distance to masked channels $d_{\rm mask}$, cluster time $\Delta t_{\rm cluster}$, and the abnormal signal removal parameter $F_{+}$. 
For each variation of those parameters and other parameters discussed later, 
the efficiency and purity are reevaluated and applied to the spectrum.
In all those cluster quality selection variations, nearly no dependence on $R$ is observed.

The uncertainty due to the charged particle veto is estimated by varying the parameters of the track \pt-dependent selection criteria to looser ones: 
$\Delta \eta^{\rm residual}>0.025$ and $\Delta \varphi^{\rm residual}>0.03$ radians. 
The variation corresponds to close to two calorimeter cell sizes instead of the nominal value that tends to cover one cell size or less at high \pt.
The resulting uncertainty on the cross section for central \PbPb events is at 2\% with a small decrease with $\ptg$, and decreases to 0.5\% for peripheral \PbPb events.

%%%%%%%%%
%%%%%%%%%

Unlike in previous ALICE isolated-photon measurements~\cite{ALICE:2019rtd,ALICE:2024kgy}, there is no requirement on the distance to a masked bad or dead channel from the highest energy cell in the cluster since it has a large impact on cluster acceptance, but it is considered as a systematic uncertainty.
The yields considering $d_{\rm mask}>2$ cells or no such requirement give a constant uncertainty of 2\% for all colliding systems. 

The cluster time selection window is varied between $\Delta t_{\rm cluster} = 10$ and 40~ns to study the effect of pileup and cells with anomalous depositions that pass the $F_{+}$ selection. 
The minimum value of 10~ns remains above the time resolution of the detector, and the choice of a 40~ns upper value allows the inclusion of one pileup bunch, as they have a 25~ns time shift with respect to the considered collision.
Details about the timing and bunches can be found in Ref.~\cite{ALICE:2022qhn}.
The uncertainty is found to be about 2--3\% in central \PbPb collisions and decreases below 1\% for peripheral \PbPb and \pp collisions. 

The $F_{+}$ selection value is varied from 95\% to 93\%. In Ref.~\cite{ALICE:2022qhn}, the $F_{+}$ distribution has a minimum at 0.95 between signal and background clusters for central \PbPb collisions. But since the minimum is rather wide, the variation is moved down to 0.93 to avoid even more the background region without removing signal significantly. An uncertainty of about 4\% is observed for central \PbPb collisions, that decreases to 1\% in peripheral collisions, with almost no \ptg dependence. In \pp collisions, approximately 1\% uncertainty is estimated above $\ptg=12$~\GeVc, while at lower \ptg\ it increases to 4--6\%. 

Two sources of systematic uncertainties are considered for the efficiency in Fig.~\ref{fig:EffIsoPhoton}. 
First, the description of the shower shape in simulations is considered via an uncertainty
estimated from the difference between standard simulations and those 
including modelling of the cross talk observed in the EMCal readout cards and is labelled as ``No MC tuning''. 
Second, depending on the shape of the  PYTHIA~8 generated prompt photon \ptg distribution in the simulation, the efficiency can change due to \ptg bin-to-bin migrations and is labelled as ``Spectral shape''. 
This uncertainty is calculated by applying a \ptg-dependent weight to the generated signal 
so as to reproduce the spectra from the JETPHOX NLO calculation presented in the next Section, 
which includes prompt and fragmentation photons.

The choice of the \sigmalongPb\ range for narrow photon-like showers (signal) is
important for the efficiency and purity of the measurement. 
The uncertainty is estimated by varying the upper limit of the signal range by $-0.03$ and $+0.05$, and is found to lie at 1--3\% with no \ptg dependence. 
Only in \pp collisions below $\ptg = 12$~\GeVc a large uncertainty is found: of 7\% or 15\% depending on the $R$ value. 

The estimation of the UE density is checked in different areas shown schematically in Fig.~\ref{fig:UEareasSketch}: 
a $\varphi$-band that covers the same $\Delta \eta$ than the isolation cone but covers $\Delta \varphi =\pi$, 
limited to avoid the jet emitted in the opposite direction to a high-energy particle;  
perpendicular bands that cover the same area as the $\varphi$-band but centred at $\varphi = \pm \pi/2$ from the photon;
cones perpendicular to the isolated-photon candidate direction; and the  FASTJET jet area/median package~\cite{Cacciari_2010}. 
The \ptIsoCh distributions obtained with the different estimators can be found in Ref.~\cite{ALICE-PUBLIC-2024-003}. 
The results obtained with the different methods are consistent with each other
and the average of the difference between the default and the alternative areas is used as uncertainty, 
excluding the perpendicular cones and bands for \PbPb collisions since their particle density is different due to the anisotropic transverse flow~\cite{ALICE:2016ccg}.

Another uncertainty assigned to the UE density determination is due to the choice of the gap between the $\eta$-band and the cone. 
For $R=0.4$ and $R=0.2$, the UE density is estimated with and without the gap of $\Delta R_{\rm UE~gap}=0.1$ used as default. 
For $R=0.2$, an additional gap of $\Delta R_{\rm UE~gap}=0.3$ (with the same $\eta$-band area as for $R=0.4$ and $\Delta R_{\rm UE~gap} = 0.1$) is used, 
the average of the variations with respect the default case is used as uncertainty. 

The uncertainty on the energy scale of the EMCal is estimated to be 0.5\%~\cite{ALICE:2022qhn}.
The effect of this uncertainty on the measured cross section amounts to 2.1\%. 
A material budget uncertainty accounting for the material of the different detectors traversed by photons before they reach the EMCal has been previously determined in Ref.~\cite{ALICE:2018mjj} and amounts to 2.1\%. 

Due to the different hardware and electronics performances of the calorimeter supermodules, the result can potentially change depending on the SM where the cluster is measured. The dispersion of the inclusive cluster yields is calculated via double ratios of data over simulation yields in single SM over full SM  and is found to be 3.5\%,  
labelled as the ``SM dependence'' uncertainty.

The uncertainty on the trigger normalisation has two sources: the use of the trigger efficiency to estimate the trigger rejection factor and correct the yields, 
and the fitting used to calculate the trigger rejection factor. 
For the first source, the comparison of the yields 
calculated with or without the trigger efficiency is considered, and half of the difference is taken as the uncertainty. 
The trigger rejection factor is calculated by fitting with a constant above the trigger threshold when it is fully efficient:  
above $\ptg = 12$~\GeVc for \PbPb collisions, with the higher \tr threshold, and above  $\ptg = 6$~\GeVc \pp and \PbPb collisions with the lower \tr threshold. 
The fitting range is varied, the calculated standard deviation of all the variations gives less than  0.2--0.6\% uncertainty for central and semi-central \PbPb collisions 
(lower the lower the centrality). For peripheral collisions, it increases to above 1\% in centrality 50--70\% to 3--4\% in the 70--90\% centrality class: 
the uncertainty is higher in peripheral events due to the lower number of MB-triggered events. 
In \pp collisions, the uncertainty is found to be 1.6\%.
This uncertainty is considered as a normalisation uncertainty and not added to the \ptg-differential yield systematic uncertainty.
The other normalisation uncertainties are those associated with $\sigma_{\rm MB}$ and \Ncoll. 
These uncertainties are relatively small, of the order of 1.5\% for central and semi-central \PbPb collisions and of 2\% for 50--70\% \PbPb and \pp collisions, and between 3\% and 5\% for 70--90\% collisions.  
The total normalisation uncertainties can be found in Table~\ref{tab:Lumi}.

Figure~\ref{fig:spe_sys_R02} includes
points labelled as ``other systematic'' that correspond to the sum in quadrature of the uncertainty sources with small or no dependence on \ptg and values lower than 2.5\%: material budget, cluster time, trigger efficiency, energy scale,  CPV, and distance to masked channels.

The total systematic uncertainty on the cross section is obtained 
by adding the contributions of the different sources described above in quadrature,
as well as the purity uncertainty. 
The resulting uncertainties range between 10\% and 30\%. 
In \pp and \PbPb peripheral collisions, as well as in central and semi-central \PbPb collisions at low \ptg, the dominant uncertainty is the one on the purity.
At intermediate \ptg and central \PbPb collisions, the dominant uncertainty is the ``No MC tuning'' uncertainty, and at $\ptg > 80$~\GeVc the statistical uncertainty.

The systematic uncertainties on the \raa (Eq.~\eqref{eq:raa}) and the ratio of cross sections with different $R$ are calculated from the effect of the previously described variations on those ratios. For both,
the statistical uncertainty dominates above $\ptg~=~40$~\GeVc for \PbPb central and semi-central collisions and above $\ptg = 20$~\GeVc for the other collision systems.
The statistical uncertainty dominates in all the reported \ptg\ ranges for 70--90\% \PbPb collisions in the ratio of cross sections with different radii. 
For the \raa, the systematic-uncertainty sources that are fully correlated between \PbPb and \pp collisions -- the energy scale, distance to masked channels, material budget, and SM-dependence -- cancel out in the ratio. 
The other sources partially cancel, except CPV since there is no such selection in \pp collisions.  
The ``isolation probability'' source dominates on all centrality classes at low \ptg, and at intermediate \ptg for semi-central collisions and $R=0.4$.
In central collisions at intermediate \ptg, the ``No MC tuning'' uncertainty dominates for $R=0.2$, in a similar proportion as the isolation probability for $R=0.4$. 

For the ratio of spectra with  $R=0.4$ over $R=0.2$, the cross-section normalisation uncertainties cancel.
The same systematic uncertainty sources which cancel completely for the \raa cancel also in these ratios. In addition, also the ``cluster time'' uncertainty source, being correlated between the results with different radii, cancels out in the ratio.  
For the rest of the systematic-uncertainty sources, there is a stronger partial cancellation than for the \raa. 
The overall main contributions to the total systematic uncertainty are the ``UE area'' and the anti-isolation ``\ptIsoCh~background range'' for all the collision systems and in addition the ``MC signal amount'' source in central \PbPb collisions. In the lower \ptg intervals,
the ``isolation probability'' in central \PbPb and \pp collisions dominates or contributes significantly. The total systematic uncertainty stays at the level of 3--5\% for all \ptg, except in \pp collisions where it rises below 20~\GeVc, reaching up to 15\% in the 11--12~\GeVc \ptg interval.

\section{Results\label{sec:results}}
This section presents the main results of the isolated-photon measurement at midrapidity ($|\eta^{\gamma}| < 0.67$), with a charged particle isolation momentum threshold of $\ptIsoCh = 1.5$~\GeVc, for two cone radii ($R = 0.2$ and 0.4) around the photon candidates, which were selected by requiring a narrow shape in the calorimeter: $0.1 < \sigmalongPb < 0.3$ for pp collisions and for \PbPb collisions above 18~\GeVc, and $0.1 < \sigmalongPb < 1.6 - 0.06 \times \ptg$ for \PbPb collisions below 18~\GeVc.
The isolated-photon differential production cross section can be obtained from the following equation for a given triggered data sample
%%%%%%%%%%%%%%%%%%%%%%
\begin{equation}
\label{eq:cs}
\frac{{\rm d}^{2}\sigma^{\gamma~{\rm iso}}}{{\rm d}\ptg~{\rm d}\eta} =  \frac{1}{\mathscr{L}_{\rm NN}^{\rm trig}} \times \frac{{\rm d}^{2}N^{\rm iso}_{\rm n}}{{\rm d}\ptg~{\rm d}\eta} \times \frac{P}{\varepsilon_{\rm trig}^{\rm iso} \times \varepsilon_{\gamma}^{\rm iso} \times \text{Acc}}
\end{equation}
%%%%%%%%%%%%%%%%%%%%%
where all the terms were described in the previous Sections 
and Acc~$=\Delta \eta \times \Delta \varphi / 2\pi$ is the acceptance area obtained from the values in Table~\ref{tab:cuts}. 
The luminosities per collision system and centrality class are listed in Table~\ref{tab:Lumi} with the corresponding normalisation uncertainties discussed in Sect.~\ref{sec:sys_unc}. The triggered data samples are combined depending on the \ptg range and centrality class as discussed in Sect.~\ref{sec:trigger}.

\newpage

\begin{figure}[h]
\begin{center}
\includegraphics[width=0.497\textwidth]{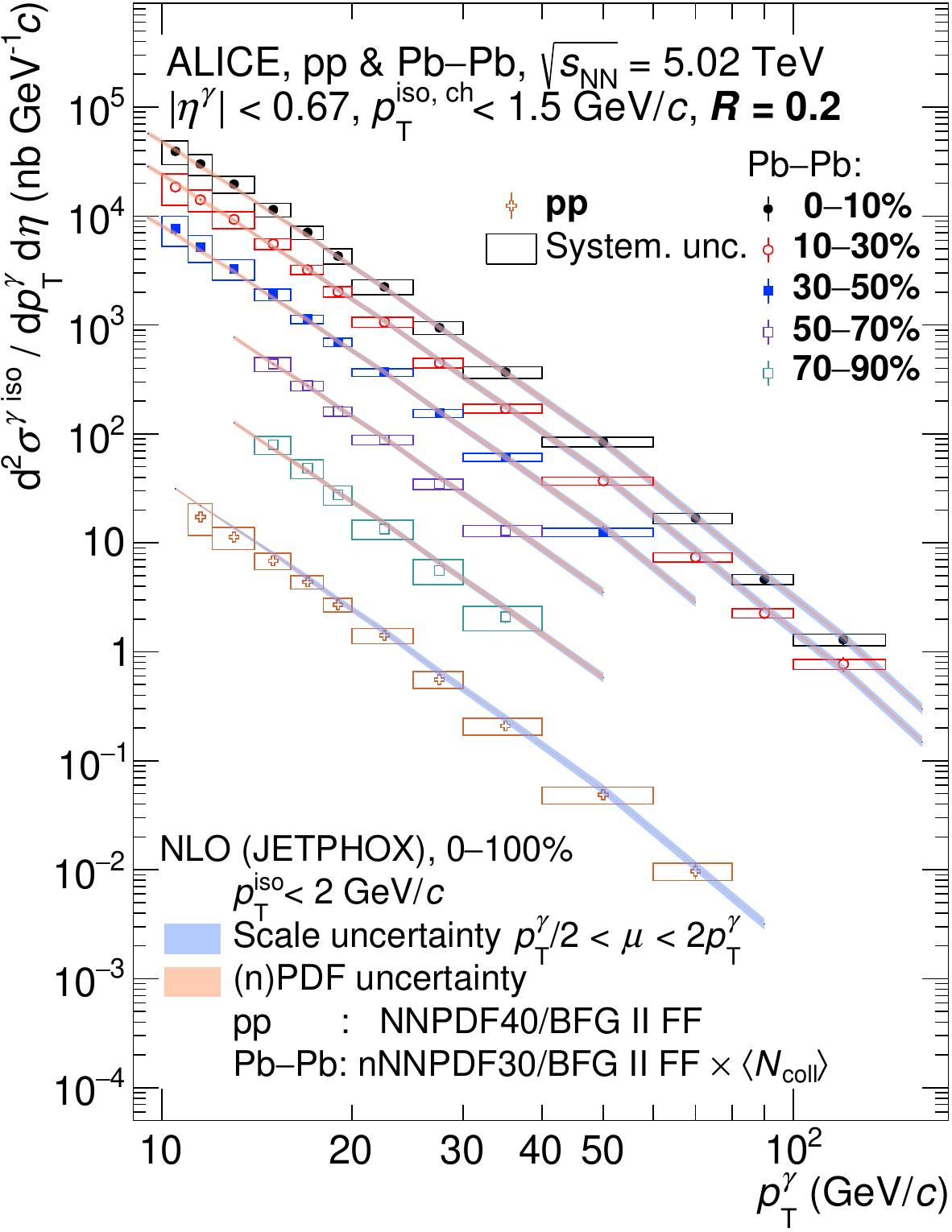}
\includegraphics[width=0.497\textwidth]{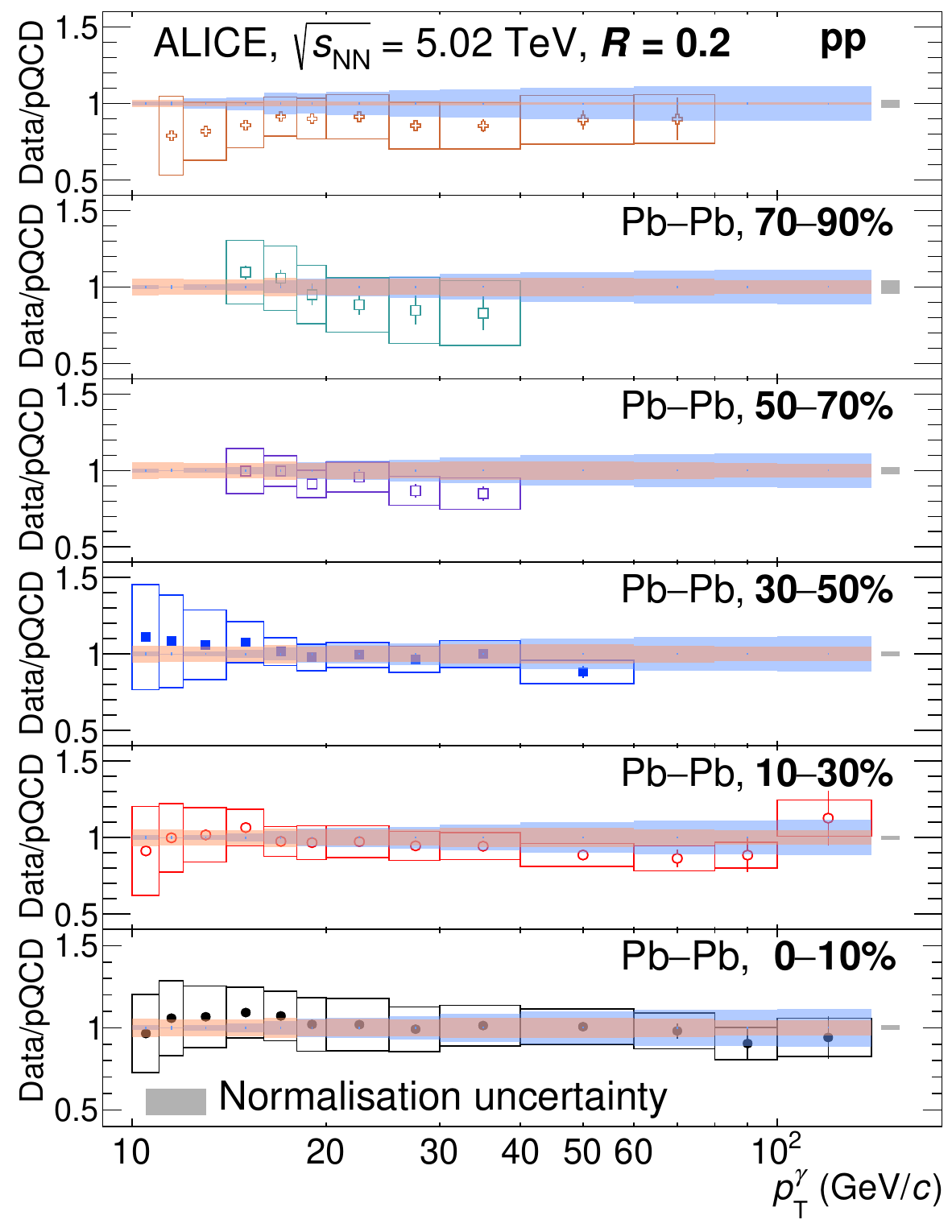}
\end{center}
\caption{\label{fig:SpecR02} (colour online) 
Left: Isolated-photon differential cross section measured in \pp and \PbPb collisions at \snnfive for five \PbPb centrality classes for $R=0.2$. Error bars and boxes are the statistical and systematic uncertainties, respectively. The bands correspond to NLO pQCD calculations with JETPHOX, for \PbPb collisions calculated for the 0--100\% centrality class and scaled by \Ncoll. Right: Ratio of data over JETPHOX NLO pQCD calculations. 
The bands centred at unity correspond to the JETPHOX pQCD calculations, their width represents the scale (blue) uncertainty and PDF (orange) uncertainty. 
The normalisation uncertainties are not included in the left panel but they are shown in the right panel as a grey box on the right of each of the frames around unity.
 }
\label{fig:isoPhotonCrossSectionR02}
\end{figure}

\begin{figure}[ht]
 \begin{center}
 \includegraphics[width=0.497\textwidth]{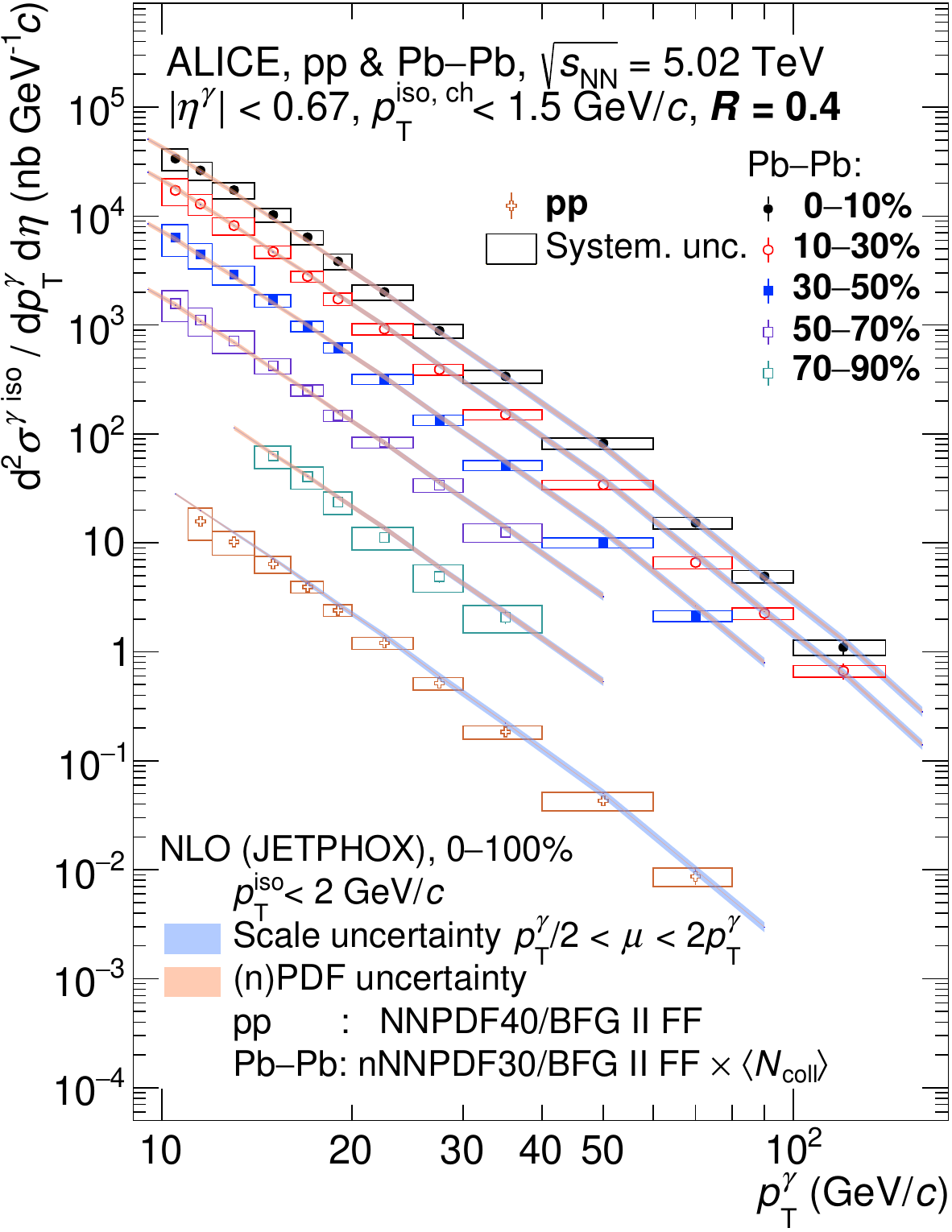}
 \includegraphics[width=0.497\textwidth]{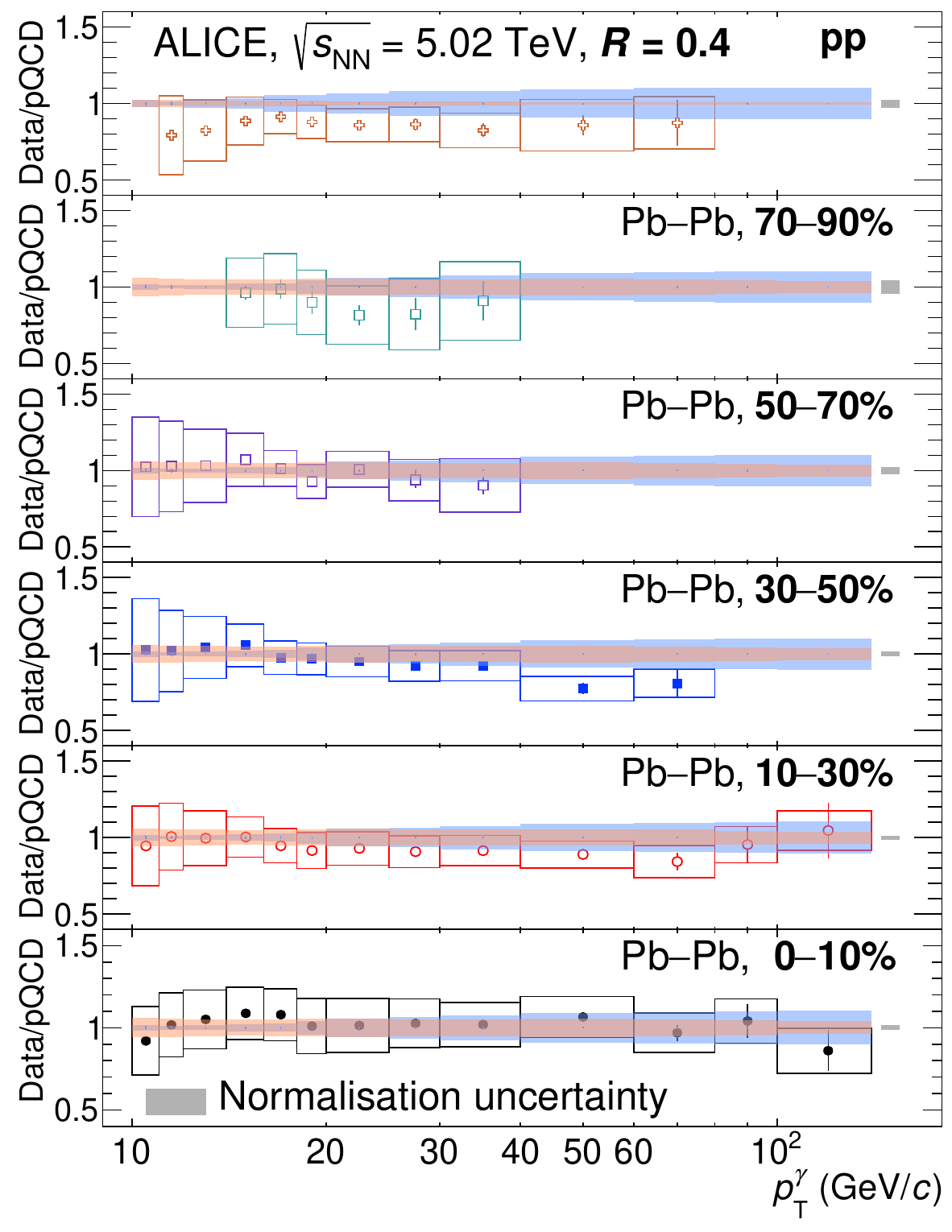}
\end{center}
\caption{ \label{fig:SpecR04} (colour online) 
Left: Isolated-photon differential cross section measured in \pp and \PbPb collisions at \snnfive for five \PbPb centrality classes for $R=0.4$. Error bars and boxes are the statistical and systematic uncertainties, respectively. The bands correspond to NLO pQCD calculations with JETPHOX, for \PbPb collisions calculated for the 0--100\% centrality class and scaled by \Ncoll. 
Right: Ratio of data over JETPHOX NLO pQCD calculations. 
The bands centred at unity correspond to the JETPHOX pQCD calculations, their width represents the scale (blue) uncertainty and PDF (orange) uncertainty.  
The normalisation uncertainties are not included in the left panel but they are shown in the right panel as a grey box on the right of each of the frames around unity.
 }
\label{fig:isoPhotonCrossSectionR04}
\end{figure}

Figures~\ref{fig:SpecR02}-left and~\ref{fig:SpecR04}-left show for $R=0.2$ and $R=0.4$, respectively, the measured isolated-photon cross section as a function of \ptg for each of the colliding systems. 
The measurement is compared to next-to-leading order pQCD calculations using the JETPHOX 1.3.1 Monte Carlo program~\cite{Fontannaz, Aurenche}. 
The fragmentation function (FF) used is BFG II~\cite{PFFth}. 
The PDF and nPDF parameterisations for protons and Pb nuclei are  NNPDF4.0~\cite{Ball_2022} and  nNNPDF3.0~\cite{khalek2022nnnpdf30}, respectively.\\

%~fig~%%%%%%%%%
%%%%%%%%%%

%~fig~%%%%%%%%%
%%%%%%%%%%

The JETPHOX+nPDF theoretical calculations were performed using the nPDF for 0--100\% centrality and without including hot-medium modifications.
To compare to \PbPb data, the JETPHOX+nPDF theoretical calculations are scaled by the number of binary collisions \Ncoll, 
calculated using the Glauber model~\cite{ALICE:PbPb5TeVMBCrossSec,PhysRevC.97.054910},  
listed in Sect.~\ref{sec:detector}.  
The central values of the predictions were obtained by choosing factorisation, 
normalisation, and fragmentation scales equal to the photon transverse momentum ($\mu_{f}=\mu_{R}=\mu_{F}=\ptg$). 
Scale uncertainties were determined varying all scales simultaneously to 0.5 and 2 times their nominal values.
Uncertainties related to the (n)PDFs are given at 90\% confidence level and were obtained by performing the calculations with each of the 101 eigenvector sets of NNPDF4.0  and 201 eigenvector sets of nNNPDF3.0.
The isolation criterion in pQCD calculations corresponds to a restriction of the phase space available to final-state radiation in a cone of $R < 0.2$ or $0.4$~\cite{Fontannaz}.
The isolation threshold used is $p_{\rm T}^{\rm iso} < $~2 \GeVc, where both charged and neutral particles momenta are used in $p_{\rm T}^{\rm iso}$. This criterion is equivalent to the $\ptIsoCh < 1.5$~\GeVc criterion used in data with only charged particles, it was determined using the neutral energy fraction in the isolation cone observed in PYTHIA~8 simulations.
Theoretical predictions are computed in the same \ptg\ intervals as the data. 

Figures~\ref{fig:SpecR02}-right and~\ref{fig:SpecR04}-right 
display the data-over-theory ratio as a function of \ptg  for $R=0.2$ and $R=0.4$, respectively.
These ratios show that the measured isolated-photon cross section  
and the one obtained with the 0--100\% JETPHOX+nPDF calculation scaled by \Ncoll 
are in agreement for the full transverse momentum range measured in \pp and in each of the \PbPb centrality classes, for both cone radii. 
The additional normalisation uncertainty, coming from the measured minimum bias cross section and from the EMCal trigger rejection factors, is not added to the systematic uncertainties on the data points, but rather shown as a separate grey box on each panel with the theory-to-data ratio. The $\lint^{\rm trig}~=~\mathscr{L}_{\rm NN}^{\rm trig} ~\times~\Ncoll$ (Eq.~\eqref{eq:linttrig}) uncertainty, provided in Table~\ref{tab:Lumi}, enters into this normalisation uncertainty box. 

%~fig~%%%%%%%%%
\begin{figure}[ht]
   \begin{center}
   \includegraphics[width=1\textwidth]{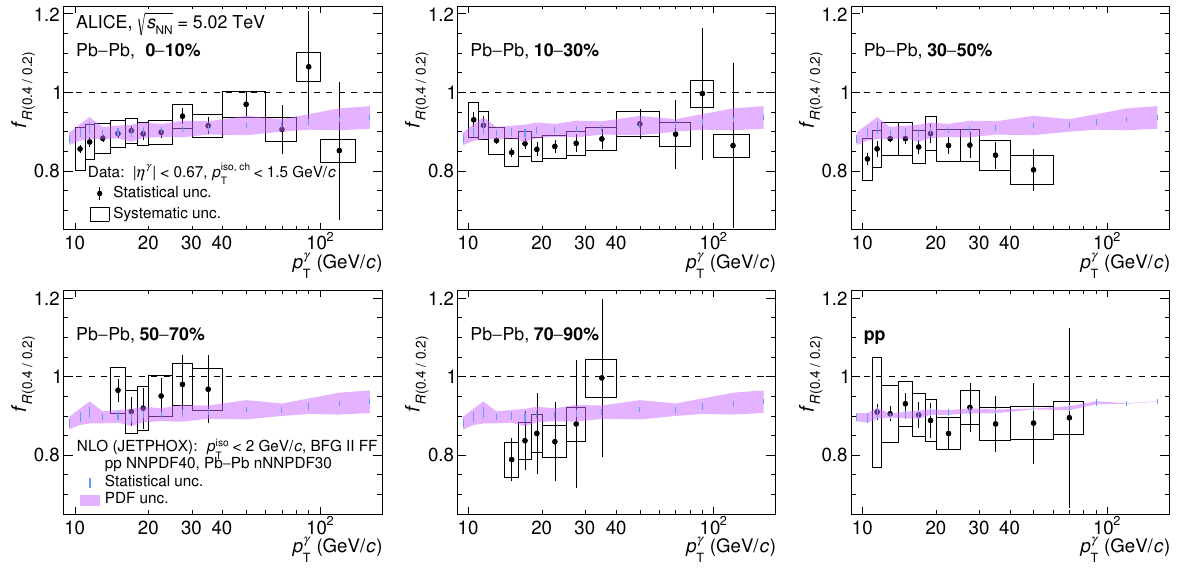}
   \end{center}
   \caption{\label{fig:Rratio}(colour online) Ratio of isolated-photon cross section measured with $R=0.4$ over $R=0.2$ for \PbPb and \pp collisions at \snnfive. 
   Each panel for each \PbPb collisions centrality class, bottom right panel for \pp collisions. 
   Error bars and boxes are the statistical and systematic uncertainties, respectively. 
   The violet band corresponds to pQCD calculations with JETPHOX, the width represents the  PDF uncertainty, and the blue vertical bars indicate 
   the statistical uncertainty from the Monte Carlo approach.
   }
\end{figure}
%%%%%%%%%%

The ratio of the \ptg-differential cross-sections measured with $R=0.2$ and $R=0.4$,
was reported above 250~\GeVc by the ATLAS Collaboration in \pp collisions at \sthirteen~\cite{Aad:2023} where an agreement between data and theory was observed.
Figure~\ref{fig:Rratio} shows ratio $f_{R(\frac{0.4}{0.2})}$ of the \ptg-differential cross-sections measured with $R=0.4$ and $R=0.2$ for the first time in \pp collisions between 11 and 80~\GeVc, and for the first time in \PbPb collisions between 10--14 and 40--140~\GeVc (depending the centrality class).
This ratio will further constrain the non-perturbative part of the FF, since it is sensitive to the fraction of fragmentation photons passing the isolation criterion~\cite{Chen_2020,Aad:2023}. 
The ratio in data ranges between 0.8 and 1, with no clear trend depending on \ptg, and the ratio in NLO pQCD calculations is around 0.9, with a small increase for increasing \ptg. 
The NLO pQCD calculations scale uncertainty cancel out in the ratio. 
Due to partial uncertainty cancellations, the PDF uncertainty on the ratio is significantly smaller than that on the spectra, and ranges from 1.5\% to 0.5\% from low to high \ptg in \pp collisions, and from 3\% to 1.7\% in \PbPb collisions. 
The measured ratios of cross sections with different $R$ are described by the JETPHOX NLO pQCD calculations in all collision systems. 
The dependence on isolation-cone size is well captured by NLO pQCD calculations incorporating an isolation criterion.
Also, the ratios measured in \pp and \PbPb collisions for different centrality classes agree with each other:
no modification of the ratio is observed in central \PbPb collisions compared to peripheral \PbPb and \pp collisions. 

\begin{figure}[ht]
\begin{center}
\includegraphics[width=0.497\textwidth]{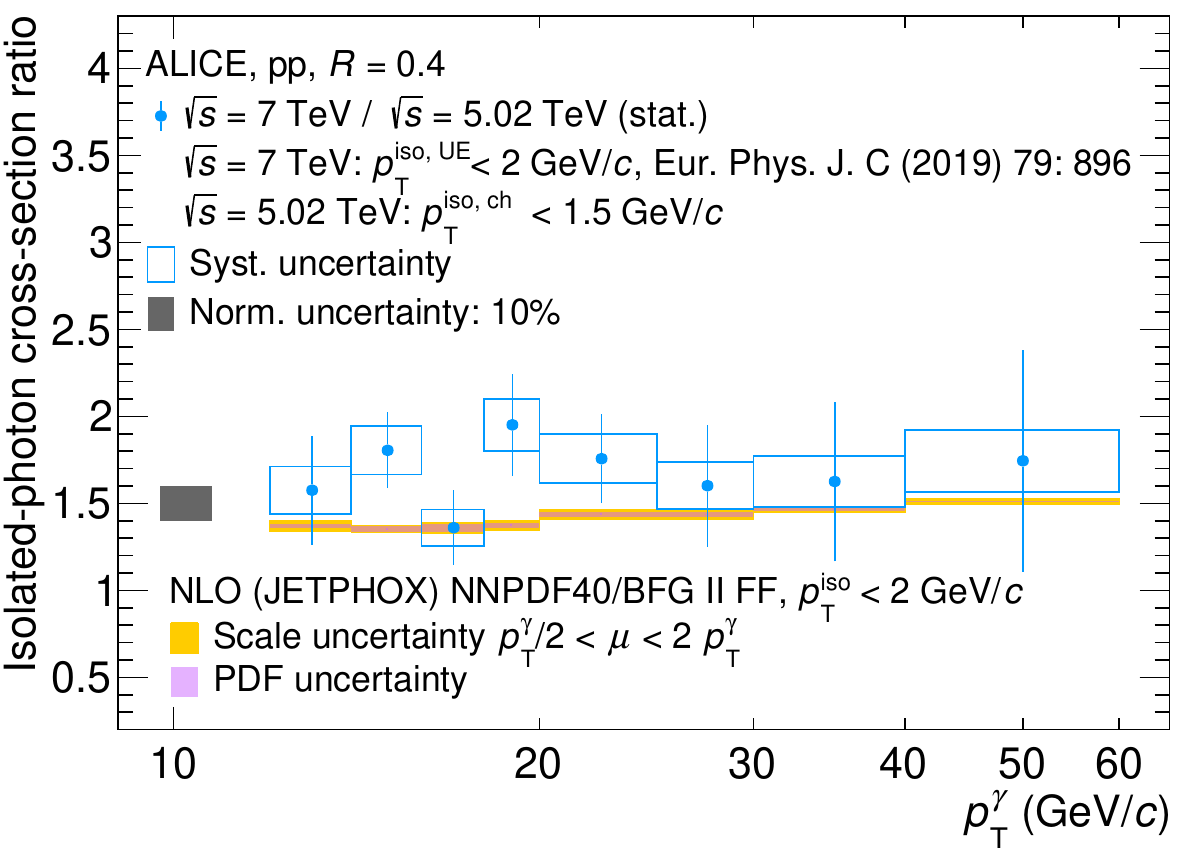}
\includegraphics[width=0.497\textwidth]{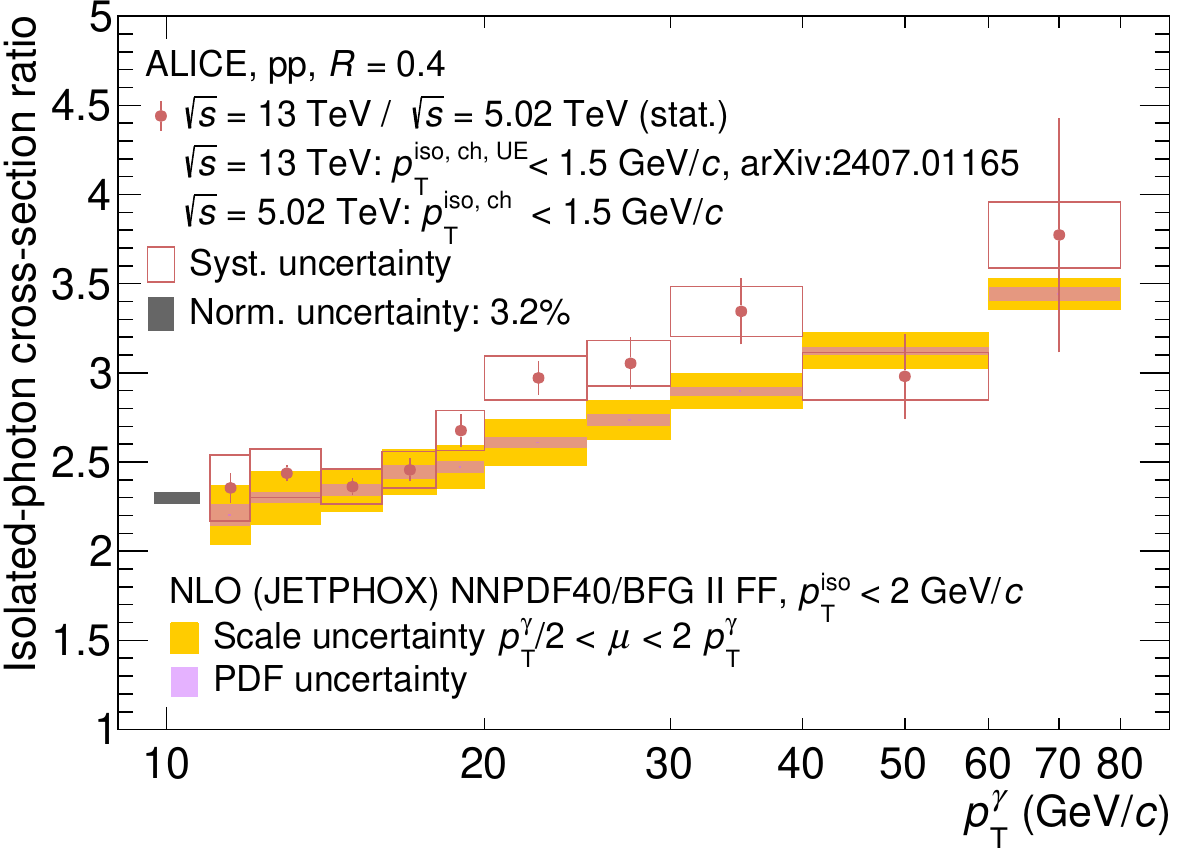}
\end{center}
\caption{\label{fig:sqrts7_13_Over5}(colour online) 
Isolated-photon cross section ratio of pp collisions at  \sseven from~\cite{ALICE:2019rtd} (left)  and \sthirteen from~\cite{ALICE:2024kgy} (right) over \sfive in data and NLO calculation from JETPHOX for $R=0.4$. Error bars and empty boxes are the data statistical and systematic uncertainties, respectively. Filled boxes represent the theory scale (orange) and PDF (pink) uncertainties. }
\label{fig:ppSqrtRatios}
\end{figure}

Figure~\ref{fig:sqrts7_13_Over5} shows the ratios of the isolated-photon cross section measured by ALICE in pp collisions at \sthirteen (from Ref.~\cite{ALICE:2024kgy}, right panel) and at \sseven (from Ref.~\cite{ALICE:2019rtd}, left panel) over the one at \sfive. The measured ratios are compared to JETPHOX NLO calculations.
All these measurements and calculations were done for a cone radius $R=0.4$.
To account for the fact that the UE was not subtracted from the isolation cone in the measurements at \sseven, the same procedure described in Ref.~\cite{ALICE:2024kgy} was used: the published \sseven measurement data was scaled down by $\kappa^{\rm iso}$, the proportion of prompt photons which are isolated at the generator level, calculated from PYTHIA $\gamma$-jet events (Eq.~\eqref{eq:efficiency}).
The \sthirteen published measurement is already corrected by this factor.
The uncertainties on the ratios partially cancel, as discussed in Ref.~\cite{ALICE-PUBLIC-2024-003}.
For the ratio of the cross sections at \sseven and \sfive,
the data show a value at the level of 1.5 with a possible slight rise with \ptg. The magnitude of the ratio is in agreement with the NLO pQCD predictions.
The small $\sqrt{s}$ difference and the large experimental uncertainties do not allow to draw a firm conclusion on the rise
with \ptg predicted by the NLO pQCD calculations: from 1.35 at 11--12~\GeVc, up to about 1.5 at 40--60~\GeVc. 
For the \sthirteen over \sfive ratio, the data agree with NLO pQCD calculations
and follow a clear rise with increasing \ptg from about 2.2 at 11 \GeVc to close to 3.5 at 80~\GeVc.
The agreement of the cross-section ratio in data and NLO pQCD calculations shown here and in Ref.~\cite{ALICE:2024kgy} for the ratio \sthirteen over \sseven indicates that the underlying mechanisms in the theoretical approach are valid. 

%~fig~%%%%%%%%%
%%%%%%%%%%

The \ptg-differential cross sections can be modified in \PbPb collisions compared to \pp collisions by initial state modification or cold or hot nuclear matter effects. To quantify these, the nuclear modification factor \raa is calculated as the ratio of the cross sections in \PbPb and \pp collisions normalised by \Ncoll

%%%%%%
\begin{equation}
\label{eq:raa_isogamma}
\raa^{\gamma ~{\rm iso}} =  \frac{1}{\Ncoll} \frac{{\rm d}^{2}\sigma_{\rm Pb-Pb}^{\gamma~{\rm iso}}  /  ({\rm d}p_{\rm T}~{\rm d}\eta) }{ {\rm d}^{2}\sigma_{\rm \pp}^{\gamma~{\rm iso}} /  ({\rm d}p_{\rm T}~{\rm d}\eta)}.
\end{equation}
%%%%%%

This is equivalent to Eq.~\eqref{eq:raa}, but the isolated photon \ptg distribution in \PbPb is already the \ptg-differential cross section calculated using Eq.~\ref{eq:cs}.

\begin{figure}[h]
    \begin{center}
    \includegraphics[width=1.0\textwidth]{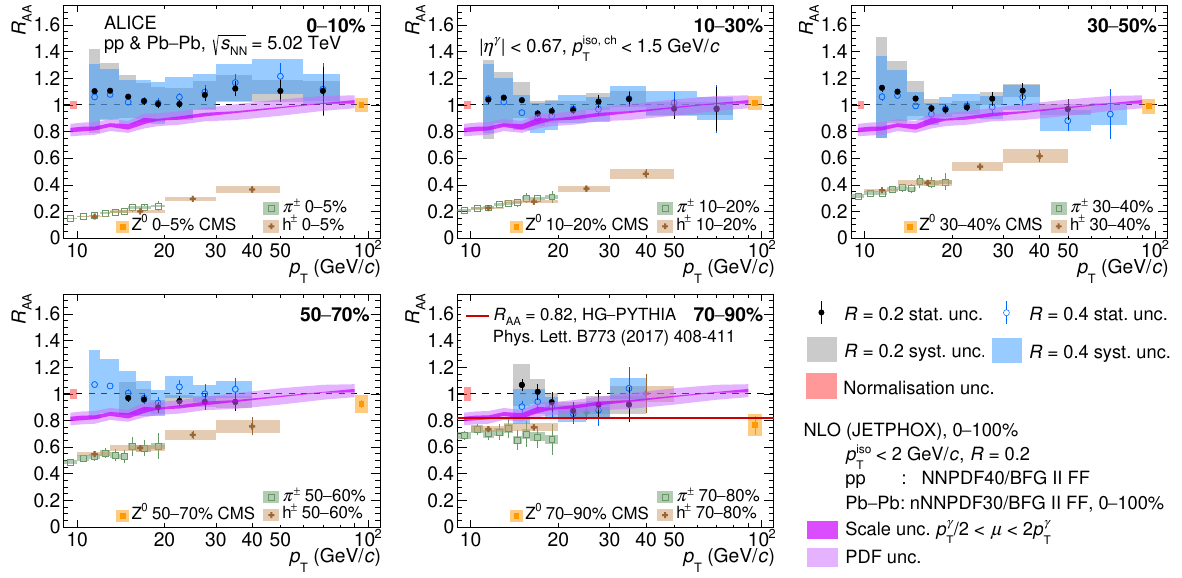}
    \end{center}
    \caption{\label{fig:RaaPhotonPiChargeZNLO}(colour online) 
    Nuclear modification factor \raa for isolated photons at \snnfive for isolation-cone radii $R=0.2$ (black) and $R=0.4$ (blue). 
    Error bars and boxes are the statistical and systematic uncertainties, respectively. 
    The isolated-photon \raa is compared to that of charged particles~\cite{ALICE:2018vuu} and charged pions~\cite{ALICE:2019hno} from ALICE, and to the ratio of the \zz-boson yield in each centrality class to the 0--90\% class measured by CMS~\cite{PhysRevLett.127.102002}. 
    The bands correspond to pQCD calculations with JETPHOX for \PbPb collisions (nPDF) for 0--100\% centrality over pp collisions (PDF). The width of each band corresponds to the scale and PDF uncertainties.
    The normalisation uncertainties are represented as a red box centred at unity.
    The solid line in the most peripheral centrality class 70--90\% at \raa = 0.82 corresponds to the HG-PYTHIA model expectation~\cite{LOIZIDES2017408}.
    }
\end{figure}

Figure~\ref{fig:RaaPhotonPiChargeZNLO} shows the isolated-photon \raa for the five \PbPb centrality classes measured for the cone radii $R=$~0.2 and 0.4. 
In sharp contrast with the charged-particle~\cite{ALICE:2018vuu} and charged-pion~\cite{ALICE:2019hno} \raa also shown in the figure, the isolated-photon \raa are generally compatible with unity. However, for the most peripheral class, a tendency to be below unity will be discussed later.
The strong suppression observed for high-\pt\ hadrons in central \PbPb collisions with respect to \pp collisions, which is due to the jet quenching in the QGP, is therefore not observed for isolated photons, as expected since they do not interact with the QGP. 

In peripheral 70--90\% \PbPb collisions, due to the lower energy density and smaller size of the QGP, the hadron \raa is closer to unity than in more central collisions, but its value of $\raa \approx 0.7$ for $10<\pt<20$~\GeVc is lower than the value expected due to in-medium jet quenching. 
This behaviour was also observed at RHIC energies by the PHENIX Collaboration in Au--Au and d--Au~\cite{PhenixQGP,PhysRevLett.101.232301,PhysRevC.87.034911,abdulameer2023disentangling} collisions at $\snn = 200$~GeV.
Figure~\ref{fig:RaaPhotonPiChargeZNLO} shows that the isolated-photon \raa in the same centrality class tends to be below unity. 
Such a behaviour can be explained by biases in the centrality selection and collision geometry that the Glauber model cannot account for, as discussed in Ref.~\cite{LOIZIDES2017408}. 
In the 70--90\% centrality class, the expected \raa bias calculated with the HG-PYTHIA model~\cite{LOIZIDES2017408} 
is indeed 0.82, i.e. significantly below unity and in agreement within 1$\sigma$ with the measured isolated-photon \raa, as Fig.~\ref{fig:RaaPhotonPiChargeZNLO} shows. 

%~fig~%%%%%%%%%
%%%%%%%%%%

The \zz-boson yield in \PbPb collisions at \snnfive was measured by ATLAS~\cite{Aad:2019lan} and CMS~\cite{PhysRevLett.127.102002} Collaborations, and like the isolated photons, \zz bosons are not affected by the QGP.
Figure~\ref{fig:RaaPhotonPiChargeZNLO} also displays one point for each centrality class with the CMS \zz-boson yield integrated in \pt divided by the one in the 0--90\% centrality class (the systematic uncertainty is taken from the numerator), which is not biased~\cite{LOIZIDES2017408}.
The \zz-boson points are placed along the \pt axis at the boson mass.
The \zz-boson ratios from CMS are in agreement with unity for centrality classes from 0\% to 70\%, but also show a deviation below unity in the most peripheral centrality class 70--90\%. The trends of the isolated photon and CMS \zz-boson ratios are compatible.
This deviation agrees with the HG-PYTHIA model~\cite{PhysRevLett.127.102002}. 
The ATLAS \zz-bosons \raa measurement~\cite{Aad:2019lan} (not shown in the figure because of overlapping centrality class ranges) agrees with unity up to 60\% centrality. 
Above 60\%, the points are significantly above unity, contrary to CMS and ALICE Collaboration observations. 
A possible explanation for this different behaviour is given in Refs.~\cite{Jonas:2021xju, Eskola:2020lee}.
A different visualisation of the \zz-boson results from ATLAS and CMS Collaborations and this isolated-photon \raa can be found in Ref.~\cite{ALICE-PUBLIC-2024-003}.

Finally, the \raa from the JETPHOX NLO calculations for 0--100\% centrality is also shown in Fig.~\ref{fig:RaaPhotonPiChargeZNLO} for $R=0.2$. 
For $R=0.4$, it is almost identical and is not shown for clarity.
The scale uncertainty cancels partially, but the PDF and nPDF uncertainties do not cancel and are fully propagated in the ratio uncertainty,  
which decreases from 6\% at low \ptg to 4\% at high \ptg for both $R$ values.
Like the data, the NLO pQCD \raa is close to unity for $\ptg > 50$~\GeVc. In contrast, a suppression is expected at lower \ptg\ ($\raa \approx 0.8$ at \ptg = 10 GeV/c for 0--100\% centrality), which can be attributed to differences in the proton and nucleus PDFs.
The particularly good agreement of the \PbPb collisions data with NLO pQCD + nPDF in Figs.~\ref{fig:isoPhotonCrossSectionR02} and~\ref{fig:isoPhotonCrossSectionR04} suggests that the data support the \raa calculated with this framework.

\begin{figure}[t]
    \begin{center}
    \includegraphics[width=0.98\textwidth]{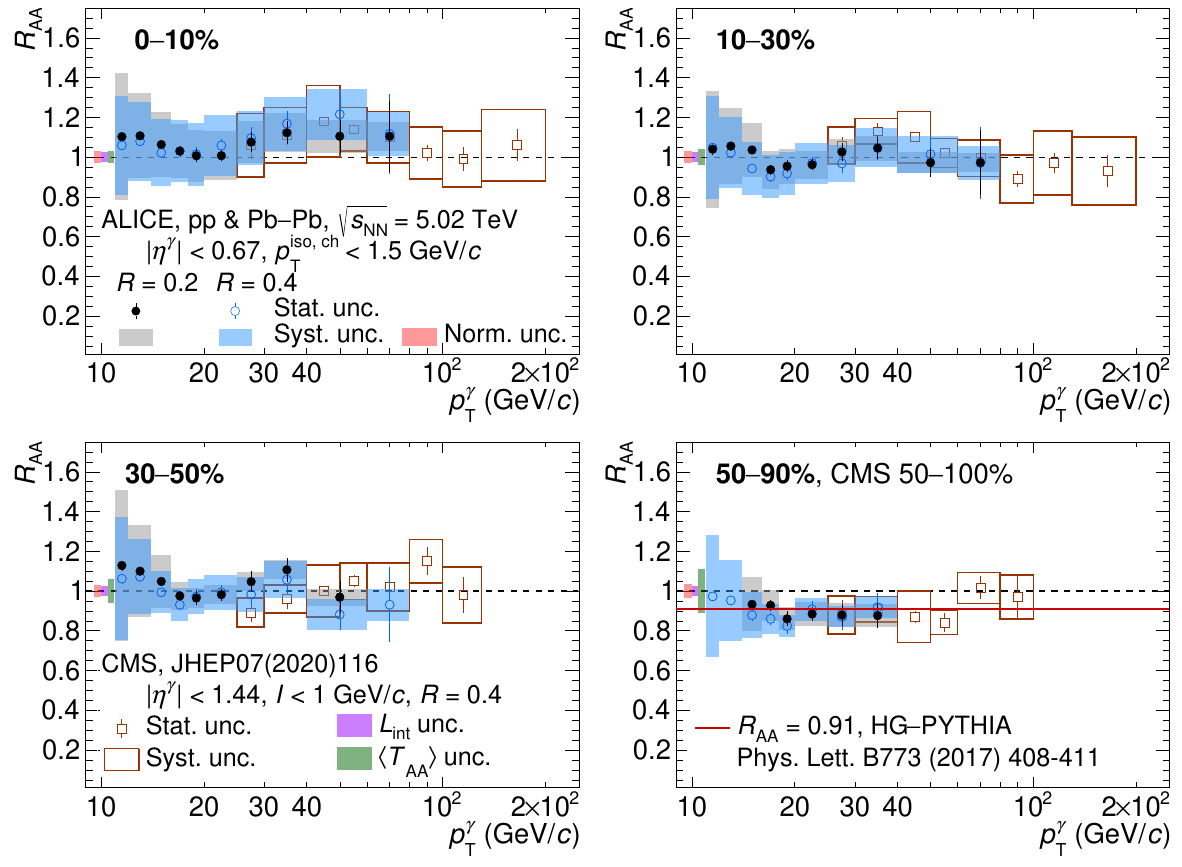}
    \end{center}
    \caption{\label{fig:RaaALICECMS}(colour online) 
    Nuclear modification factor \raa for isolated photons measured by ALICE for isolation-cone radii $R=0.2$ and $R=0.4$  
    and CMS~\cite{Sirunyan:2020} for isolation-cone radius $R=0.4$ at \snnfive and for four centrality classes. 
    Error bars and boxes are the statistical and systematic uncertainties, respectively. 
    The solid line in the peripheral centrality class 50--90\% at \raa = 0.91, is the result of the HG-PYTHIA model~\cite{LOIZIDES2017408}. 
    The ALICE normalisation uncertainties are represented as a red box centred at unity.  
    For CMS, the normalisation uncertainties are displayed as a violet box for the integrated luminosity and a green box for the nuclear overlap function $\langle T_{\rm AA}\rangle=\Ncoll / \sigma_{\rm NN}^{\rm INEL}$.}
\end{figure}

Figure~\ref{fig:RaaALICECMS} shows a comparison of the isolated-photon \raa with the corresponding measurement performed by the CMS Collaboration~\cite{Sirunyan:2020} that starts at $\ptg = 25$~\GeVc. 
Since the CMS most peripheral class covers the 50--90\% range, the same class is reported for ALICE. The cross section, the data-to-theory ratio, and the ratio of spectra with different $R$ can be found in Ref.~\cite{ALICE-PUBLIC-2024-003} for this 50--90\% centrality class, for which the HG-PYTHIA model finds a centrality bias \raa value of 0.91. 
The CMS and ALICE measurements are consistent in all the centrality classes, and they agree with the HG-PYTHIA model in peripheral events. The \raa presented here for the 50--90\% class is unexpectedly close to the one reported for 70--90\%.  This can be attributed to a small overestimation of the purity in the statistically limited 70--90\% sample. For the 50--90\% centrality class, the uncertainties are smaller, and in the range $18<\ptg<30$~\GeVc,  the deviation of the \raa\ from unity is about 2$\sigma$.

%~fig~%%%%%%%%%
%%%%%%%%%%

It is interesting to note that in both experiments the \raa central value is larger than unity between 30 and 60~\GeVc in the 0--10\% centrality class, while still being compatible with unity. More precise measurements would be needed to confirm and understand this trend. 

Reference~\cite{ALICE-PUBLIC-2024-003} also contains ratios of the isolated-photon production cross section in different centrality classes over semi-central (30--50\%) or semi-peripheral (50--70\%) classes. These ratios also agree with unity, as expected, with smaller uncertainties than the \raa.

\section{Conclusions\label{sec:conclusion}}

The isolated-photon \ptg-differential cross section in pp and \PbPb collisions at \snnfive was measured by the ALICE experiment at  $|\etag| <0.67$, 
for different centrality classes in \PbPb collisions and in the transverse momentum range from 10--14~to~40--140~\GeVc. This measurement extends the lower limit of \ptg\ to a smaller value compared to previous measurements by other LHC experiments, which start at $\ptg = 20$ to 25~\GeVc.  

The measured cross sections are compared to NLO pQCD calculations and they agree within uncertainties. 
This agreement with the theory shows that the (n)PDFs used are supported by the data, 
which was also observed in previous ALICE isolated-photon measurements in \pp collisions. 
This is further supported by the agreement with the theory of the production yield ratios in \pp collisions at various \s.
Furthermore, good agreement is observed in all centrality classes for \PbPb collisions, showcasing the validity of \Ncoll scaling for this observable.

The ratio of the \ptg-differential cross sections obtained with different isolation radii is shown for the first time in \PbPb collisions, and is also shown for pp collisions at \sfive and at much lower \ptg than previous measurements in \pp collisions at the LHC: the measurement made by the ATLAS Collaboration at \sthirteen begins at $\ptg=250$~\GeVc. 
In this ratio, an agreement with the NLO pQCD theoretical calculations is found for all the systems.

The isolated-photon nuclear modification factors compare \PbPb and \pp cross-section measurements and are consistent with unity as well as with the theoretical predictions from low to high~\ptg. This indicates first, that isolated photons are not affected by the quark--gluon plasma, and second, data agree with predictions incorporating initial state nuclear effects.
The measured \raa in the most peripheral classes 70--90\% and 50--90\% tends to be below unity, close to 0.9,
in agreement with the HG-PYTHIA model for the centrality selection bias in the experiment, 
and with the observation of \zz-boson apparent suppression in peripheral collisions by CMS.
The isolated-photon \raa is in good agreement with the isolated-photon \raa measured by CMS at high \ptg (larger than 25~\GeVc).

Future ALICE measurements with the LHC Run 3 and 4 campaigns, where a significantly larger accumulated number of collisions is expected, 
in particular for the \PbPb peripheral centrality class, will further improve the measurements and help to further constrain theoretical predictions. 

\newenvironment{acknowledgement}{\relax}{\relax}
\begin{acknowledgement}
\section*{Acknowledgements}

% add specific acknowledgements here 
% ...but please don't remove the line below: funding agencies
% will be acknowledged with a custom tex file handled by EB chairs after Collab Round 2
% Version: 2024-09-09

The ALICE Collaboration would like to thank all its engineers and technicians for their invaluable contributions to the construction of the experiment and the CERN accelerator teams for the outstanding performance of the LHC complex.
The ALICE Collaboration gratefully acknowledges the resources and support provided by all Grid centres and the Worldwide LHC Computing Grid (WLCG) collaboration.
The ALICE Collaboration acknowledges the following funding agencies for their support in building and running the ALICE detector:
A. I. Alikhanyan National Science Laboratory (Yerevan Physics Institute) Foundation (ANSL), State Committee of Science and World Federation of Scientists (WFS), Armenia;
Austrian Academy of Sciences, Austrian Science Fund (FWF): [M 2467-N36] and Nationalstiftung f\"{u}r Forschung, Technologie und Entwicklung, Austria;
Ministry of Communications and High Technologies, National Nuclear Research Center, Azerbaijan;
Conselho Nacional de Desenvolvimento Cient\'{\i}fico e Tecnol\'{o}gico (CNPq), Financiadora de Estudos e Projetos (Finep), Funda\c{c}\~{a}o de Amparo \`{a} Pesquisa do Estado de S\~{a}o Paulo (FAPESP) and Universidade Federal do Rio Grande do Sul (UFRGS), Brazil;
Bulgarian Ministry of Education and Science, within the National Roadmap for Research Infrastructures 2020-2027 (object CERN), Bulgaria;
Ministry of Education of China (MOEC) , Ministry of Science \& Technology of China (MSTC) and National Natural Science Foundation of China (NSFC), China;
Ministry of Science and Education and Croatian Science Foundation, Croatia;
Centro de Aplicaciones Tecnol\'{o}gicas y Desarrollo Nuclear (CEADEN), Cubaenerg\'{\i}a, Cuba;
Ministry of Education, Youth and Sports of the Czech Republic, Czech Republic;
The Danish Council for Independent Research | Natural Sciences, the VILLUM FONDEN and Danish National Research Foundation (DNRF), Denmark;
Helsinki Institute of Physics (HIP), Finland;
Commissariat \`{a} l'Energie Atomique (CEA) and Institut National de Physique Nucl\'{e}aire et de Physique des Particules (IN2P3) and Centre National de la Recherche Scientifique (CNRS), France;
Bundesministerium f\"{u}r Bildung und Forschung (BMBF) and GSI Helmholtzzentrum f\"{u}r Schwerionenforschung GmbH, Germany;
General Secretariat for Research and Technology, Ministry of Education, Research and Religions, Greece;
National Research, Development and Innovation Office, Hungary;
Department of Atomic Energy Government of India (DAE), Department of Science and Technology, Government of India (DST), University Grants Commission, Government of India (UGC) and Council of Scientific and Industrial Research (CSIR), India;
National Research and Innovation Agency - BRIN, Indonesia;
Istituto Nazionale di Fisica Nucleare (INFN), Italy;
Japanese Ministry of Education, Culture, Sports, Science and Technology (MEXT) and Japan Society for the Promotion of Science (JSPS) KAKENHI, Japan;
Consejo Nacional de Ciencia (CONACYT) y Tecnolog\'{i}a, through Fondo de Cooperaci\'{o}n Internacional en Ciencia y Tecnolog\'{i}a (FONCICYT) and Direcci\'{o}n General de Asuntos del Personal Academico (DGAPA), Mexico;
Nederlandse Organisatie voor Wetenschappelijk Onderzoek (NWO), Netherlands;
The Research Council of Norway, Norway;
Pontificia Universidad Cat\'{o}lica del Per\'{u}, Peru;
Ministry of Science and Higher Education, National Science Centre and WUT ID-UB, Poland;
Korea Institute of Science and Technology Information and National Research Foundation of Korea (NRF), Republic of Korea;
Ministry of Education and Scientific Research, Institute of Atomic Physics, Ministry of Research and Innovation and Institute of Atomic Physics and Universitatea Nationala de Stiinta si Tehnologie Politehnica Bucuresti, Romania;
Ministry of Education, Science, Research and Sport of the Slovak Republic, Slovakia;
National Research Foundation of South Africa, South Africa;
Swedish Research Council (VR) and Knut \& Alice Wallenberg Foundation (KAW), Sweden;
European Organization for Nuclear Research, Switzerland;
Suranaree University of Technology (SUT), National Science and Technology Development Agency (NSTDA) and National Science, Research and Innovation Fund (NSRF via PMU-B B05F650021), Thailand;
Turkish Energy, Nuclear and Mineral Research Agency (TENMAK), Turkey;
National Academy of  Sciences of Ukraine, Ukraine;
Science and Technology Facilities Council (STFC), United Kingdom;
National Science Foundation of the United States of America (NSF) and United States Department of Energy, Office of Nuclear Physics (DOE NP), United States of America.
In addition, individual groups or members have received support from:
Czech Science Foundation (grant no. 23-07499S), Czech Republic;
FORTE project, reg.\ no.\ CZ.02.01.01/00/22\_008/0004632, Czech Republic, co-funded by the European Union, Czech Republic;
European Research Council (grant no. 950692), European Union;
ICSC - Centro Nazionale di Ricerca in High Performance Computing, Big Data and Quantum Computing, European Union - NextGenerationEU;
Academy of Finland (Center of Excellence in Quark Matter) (grant nos. 346327, 346328), Finland.

\end{acknowledgement}

%%%%%%%% Bibliography 
\bibliographystyle{utphys}   % Remember we use title in the biblio
\bibliography{bibliography}
%\input {bibliography.tex}  

%%%%%%%%%%%%%%%%%%%%%%%%%%%%%%%%
% Appendices: yours (if any) + authorlist
%%%%%%%%%%%%%%%%%%%%%%%%%%%%%%%%
\newpage
\appendix

%
%\input{} % put your appendices here (if any)
%

%%%%% Authorlist - please do not touch: handled by EB chairs 
\section{The ALICE Collaboration}
\label{app:collab}
% ALICE Collaboration author list for 2024-09-09
\begin{flushleft} 
\small

S.~Acharya\,\orcidlink{0000-0002-9213-5329}\,$^{\rm 126}$, 
A.~Agarwal$^{\rm 134}$, 
G.~Aglieri Rinella\,\orcidlink{0000-0002-9611-3696}\,$^{\rm 32}$, 
L.~Aglietta\,\orcidlink{0009-0003-0763-6802}\,$^{\rm 24}$, 
M.~Agnello\,\orcidlink{0000-0002-0760-5075}\,$^{\rm 29}$, 
N.~Agrawal\,\orcidlink{0000-0003-0348-9836}\,$^{\rm 25}$, 
Z.~Ahammed\,\orcidlink{0000-0001-5241-7412}\,$^{\rm 134}$, 
S.~Ahmad\,\orcidlink{0000-0003-0497-5705}\,$^{\rm 15}$, 
S.U.~Ahn\,\orcidlink{0000-0001-8847-489X}\,$^{\rm 71}$, 
I.~Ahuja\,\orcidlink{0000-0002-4417-1392}\,$^{\rm 37}$
% ...

\section*{Affiliation Notes}

$^{\rm I}$ Deceased\\
$^{\rm II}$ Also at: Max-Planck-Institut fur Physik, Munich, Germany\\
$^{\rm III}$ Also at: Italian National Agency for New Technologies, Energy and Sustainable Economic Development (ENEA), Bologna, Italy\\
$^{\rm IV}$ Also at: Dipartimento DET del Politecnico di Torino, Turin, Italy\\
$^{\rm V}$ Also at: Yildiz Technical University, Istanbul, T\"{u}rkiye\\
$^{\rm VI}$ Also at: Department of Applied Physics, Aligarh Muslim University, Aligarh, India\\
$^{\rm VII}$ Also at: Institute of Theoretical Physics, University of Wroclaw, Poland\\
$^{\rm VIII}$ Also at: An institution covered by a cooperation agreement with CERN\\

\section*{Collaboration Institutes}

$^{1}$ A.I. Alikhanyan National Science Laboratory (Yerevan Physics Institute) Foundation, Yerevan, Armenia\\
$^{2}$ AGH University of Krakow, Cracow, Poland\\
$^{3}$ Bogolyubov Institute for Theoretical Physics, National Academy of Sciences of Ukraine, Kiev, Ukraine\\
$^{4}$ Bose Institute, Department of Physics  and Centre for Astroparticle Physics and Space Science (CAPSS), Kolkata, India\\
$^{5}$ California Polytechnic State University, San Luis Obispo, California, United States\\
$^{6}$ Central China Normal University, Wuhan, China\\
$^{7}$ Centro de Aplicaciones Tecnol\'{o}gicas y Desarrollo Nuclear (CEADEN), Havana, Cuba\\
$^{8}$ Centro de Investigaci\'{o}n y de Estudios Avanzados (CINVESTAV), Mexico City and M\'{e}rida, Mexico\\
$^{9}$ Chicago State University, Chicago, Illinois, United States\\
$^{10}$ China Institute of Atomic Energy, Beijing, China\\
$^{11}$ China University of Geosciences, Wuhan, China\\
$^{12}$ Chungbuk National University, Cheongju, Republic of Korea\\
$^{13}$ Comenius University Bratislava, Faculty of Mathematics, Physics and Informatics, Bratislava, Slovak Republic\\
$^{14}$ Creighton University, Omaha, Nebraska, United States\\
$^{15}$ Department of Physics, Aligarh Muslim University, Aligarh, India\\
$^{16}$ Department of Physics, Pusan National University, Pusan, Republic of Korea\\
$^{17}$ Department of Physics, Sejong University, Seoul, Republic of Korea\\
$^{18}$ Department of Physics, University of California, Berkeley, California, United States\\
$^{19}$ Department of Physics, University of Oslo, Oslo, Norway\\
$^{20}$ Department of Physics and Technology, University of Bergen, Bergen, Norway\\
$^{21}$ Dipartimento di Fisica, Universit\`{a} di Pavia, Pavia, Italy\\
$^{22}$ Dipartimento di Fisica dell'Universit\`{a} and Sezione INFN, Cagliari, Italy\\
$^{23}$ Dipartimento di Fisica dell'Universit\`{a} and Sezione INFN, Trieste, Italy\\
$^{24}$ Dipartimento di Fisica dell'Universit\`{a} and Sezione INFN, Turin, Italy\\
$^{25}$ Dipartimento di Fisica e Astronomia dell'Universit\`{a} and Sezione INFN, Bologna, Italy\\
$^{26}$ Dipartimento di Fisica e Astronomia dell'Universit\`{a} and Sezione INFN, Catania, Italy\\
$^{27}$ Dipartimento di Fisica e Astronomia dell'Universit\`{a} and Sezione INFN, Padova, Italy\\
$^{28}$ Dipartimento di Fisica `E.R.~Caianiello' dell'Universit\`{a} and Gruppo Collegato INFN, Salerno, Italy\\
$^{29}$ Dipartimento DISAT del Politecnico and Sezione INFN, Turin, Italy\\
$^{30}$ Dipartimento di Scienze MIFT, Universit\`{a} di Messina, Messina, Italy\\
$^{31}$ Dipartimento Interateneo di Fisica `M.~Merlin' and Sezione INFN, Bari, Italy\\
$^{32}$ European Organization for Nuclear Research (CERN), Geneva, Switzerland\\
$^{33}$ Faculty of Electrical Engineering, Mechanical Engineering and Naval Architecture, University of Split, Split, Croatia\\
$^{34}$ Faculty of Engineering and Science, Western Norway University of Applied Sciences, Bergen, Norway\\
$^{35}$ Faculty of Nuclear Sciences and Physical Engineering, Czech Technical University in Prague, Prague, Czech Republic\\
$^{36}$ Faculty of Physics, Sofia University, Sofia, Bulgaria\\
$^{37}$ Faculty of Science, P.J.~\v{S}af\'{a}rik University, Ko\v{s}ice, Slovak Republic\\
$^{38}$ Frankfurt Institute for Advanced Studies, Johann Wolfgang Goethe-Universit\"{a}t Frankfurt, Frankfurt, Germany\\
$^{39}$ Fudan University, Shanghai, China\\
$^{40}$ Gangneung-Wonju National University, Gangneung, Republic of Korea\\
$^{41}$ Gauhati University, Department of Physics, Guwahati, India\\
$^{42}$ Helmholtz-Institut f\"{u}r Strahlen- und Kernphysik, Rheinische Friedrich-Wilhelms-Universit\"{a}t Bonn, Bonn, Germany\\
$^{43}$ Helsinki Institute of Physics (HIP), Helsinki, Finland\\
$^{44}$ High Energy Physics Group,  Universidad Aut\'{o}noma de Puebla, Puebla, Mexico\\
$^{45}$ Horia Hulubei National Institute of Physics and Nuclear Engineering, Bucharest, Romania\\
$^{46}$ HUN-REN Wigner Research Centre for Physics, Budapest, Hungary\\
$^{47}$ Indian Institute of Technology Bombay (IIT), Mumbai, India\\
$^{48}$ Indian Institute of Technology Indore, Indore, India\\
$^{49}$ INFN, Laboratori Nazionali di Frascati, Frascati, Italy\\
$^{50}$ INFN, Sezione di Bari, Bari, Italy\\
$^{51}$ INFN, Sezione di Bologna, Bologna, Italy\\
$^{52}$ INFN, Sezione di Cagliari, Cagliari, Italy\\
$^{53}$ INFN, Sezione di Catania, Catania, Italy\\
$^{54}$ INFN, Sezione di Padova, Padova, Italy\\
$^{55}$ INFN, Sezione di Pavia, Pavia, Italy\\
$^{56}$ INFN, Sezione di Torino, Turin, Italy\\
$^{57}$ INFN, Sezione di Trieste, Trieste, Italy\\
$^{58}$ Inha University, Incheon, Republic of Korea\\
$^{59}$ Institute for Gravitational and Subatomic Physics (GRASP), Utrecht University/Nikhef, Utrecht, Netherlands\\
$^{60}$ Institute of Experimental Physics, Slovak Academy of Sciences, Ko\v{s}ice, Slovak Republic\\
$^{61}$ Institute of Physics, Homi Bhabha National Institute, Bhubaneswar, India\\
$^{62}$ Institute of Physics of the Czech Academy of Sciences, Prague, Czech Republic\\
$^{63}$ Institute of Space Science (ISS), Bucharest, Romania\\
$^{64}$ Institut f\"{u}r Kernphysik, Johann Wolfgang Goethe-Universit\"{a}t Frankfurt, Frankfurt, Germany\\
$^{65}$ Instituto de Ciencias Nucleares, Universidad Nacional Aut\'{o}noma de M\'{e}xico, Mexico City, Mexico\\
$^{66}$ Instituto de F\'{i}sica, Universidade Federal do Rio Grande do Sul (UFRGS), Porto Alegre, Brazil\\
$^{67}$ Instituto de F\'{\i}sica, Universidad Nacional Aut\'{o}noma de M\'{e}xico, Mexico City, Mexico\\
$^{68}$ iThemba LABS, National Research Foundation, Somerset West, South Africa\\
$^{69}$ Jeonbuk National University, Jeonju, Republic of Korea\\
$^{70}$ Johann-Wolfgang-Goethe Universit\"{a}t Frankfurt Institut f\"{u}r Informatik, Fachbereich Informatik und Mathematik, Frankfurt, Germany\\
$^{71}$ Korea Institute of Science and Technology Information, Daejeon, Republic of Korea\\
$^{72}$ Laboratoire de Physique Subatomique et de Cosmologie, Universit\'{e} Grenoble-Alpes, CNRS-IN2P3, Grenoble, France\\
$^{73}$ Lawrence Berkeley National Laboratory, Berkeley, California, United States\\
$^{74}$ Lund University Department of Physics, Division of Particle Physics, Lund, Sweden\\
$^{75}$ Nagasaki Institute of Applied Science, Nagasaki, Japan\\
$^{76}$ Nara Women{'}s University (NWU), Nara, Japan\\
$^{77}$ National and Kapodistrian University of Athens, School of Science, Department of Physics , Athens, Greece\\
$^{78}$ National Centre for Nuclear Research, Warsaw, Poland\\
$^{79}$ National Institute of Science Education and Research, Homi Bhabha National Institute, Jatni, India\\
$^{80}$ National Nuclear Research Center, Baku, Azerbaijan\\
$^{81}$ National Research and Innovation Agency - BRIN, Jakarta, Indonesia\\
$^{82}$ Niels Bohr Institute, University of Copenhagen, Copenhagen, Denmark\\
$^{83}$ Nikhef, National institute for subatomic physics, Amsterdam, Netherlands\\
$^{84}$ Nuclear Physics Group, STFC Daresbury Laboratory, Daresbury, United Kingdom\\
$^{85}$ Nuclear Physics Institute of the Czech Academy of Sciences, Husinec-\v{R}e\v{z}, Czech Republic\\
$^{86}$ Oak Ridge National Laboratory, Oak Ridge, Tennessee, United States\\
$^{87}$ Ohio State University, Columbus, Ohio, United States\\
$^{88}$ Physics department, Faculty of science, University of Zagreb, Zagreb, Croatia\\
$^{89}$ Physics Department, Panjab University, Chandigarh, India\\
$^{90}$ Physics Department, University of Jammu, Jammu, India\\
$^{91}$ Physics Program and International Institute for Sustainability with Knotted Chiral Meta Matter (SKCM2), Hiroshima University, Hiroshima, Japan\\
$^{92}$ Physikalisches Institut, Eberhard-Karls-Universit\"{a}t T\"{u}bingen, T\"{u}bingen, Germany\\
$^{93}$ Physikalisches Institut, Ruprecht-Karls-Universit\"{a}t Heidelberg, Heidelberg, Germany\\
$^{94}$ Physik Department, Technische Universit\"{a}t M\"{u}nchen, Munich, Germany\\
$^{95}$ Politecnico di Bari and Sezione INFN, Bari, Italy\\
$^{96}$ Research Division and ExtreMe Matter Institute EMMI, GSI Helmholtzzentrum f\"ur Schwerionenforschung GmbH, Darmstadt, Germany\\
$^{97}$ Saga University, Saga, Japan\\
$^{98}$ Saha Institute of Nuclear Physics, Homi Bhabha National Institute, Kolkata, India\\
$^{99}$ School of Physics and Astronomy, University of Birmingham, Birmingham, United Kingdom\\
$^{100}$ Secci\'{o}n F\'{\i}sica, Departamento de Ciencias, Pontificia Universidad Cat\'{o}lica del Per\'{u}, Lima, Peru\\
$^{101}$ Stefan Meyer Institut f\"{u}r Subatomare Physik (SMI), Vienna, Austria\\
$^{102}$ SUBATECH, IMT Atlantique, Nantes Universit\'{e}, CNRS-IN2P3, Nantes, France\\
$^{103}$ Sungkyunkwan University, Suwon City, Republic of Korea\\
$^{104}$ Suranaree University of Technology, Nakhon Ratchasima, Thailand\\
$^{105}$ Technical University of Ko\v{s}ice, Ko\v{s}ice, Slovak Republic\\
$^{106}$ The Henryk Niewodniczanski Institute of Nuclear Physics, Polish Academy of Sciences, Cracow, Poland\\
$^{107}$ The University of Texas at Austin, Austin, Texas, United States\\
$^{108}$ Universidad Aut\'{o}noma de Sinaloa, Culiac\'{a}n, Mexico\\
$^{109}$ Universidade de S\~{a}o Paulo (USP), S\~{a}o Paulo, Brazil\\
$^{110}$ Universidade Estadual de Campinas (UNICAMP), Campinas, Brazil\\
$^{111}$ Universidade Federal do ABC, Santo Andre, Brazil\\
$^{112}$ Universitatea Nationala de Stiinta si Tehnologie Politehnica Bucuresti, Bucharest, Romania\\
$^{113}$ University of Cape Town, Cape Town, South Africa\\
$^{114}$ University of Derby, Derby, United Kingdom\\
$^{115}$ University of Houston, Houston, Texas, United States\\
$^{116}$ University of Jyv\"{a}skyl\"{a}, Jyv\"{a}skyl\"{a}, Finland\\
$^{117}$ University of Kansas, Lawrence, Kansas, United States\\
$^{118}$ University of Liverpool, Liverpool, United Kingdom\\
$^{119}$ University of Science and Technology of China, Hefei, China\\
$^{120}$ University of South-Eastern Norway, Kongsberg, Norway\\
$^{121}$ University of Tennessee, Knoxville, Tennessee, United States\\
$^{122}$ University of the Witwatersrand, Johannesburg, South Africa\\
$^{123}$ University of Tokyo, Tokyo, Japan\\
$^{124}$ University of Tsukuba, Tsukuba, Japan\\
$^{125}$ Universit\"{a}t M\"{u}nster, Institut f\"{u}r Kernphysik, M\"{u}nster, Germany\\
$^{126}$ Universit\'{e} Clermont Auvergne, CNRS/IN2P3, LPC, Clermont-Ferrand, France\\
$^{127}$ Universit\'{e} de Lyon, CNRS/IN2P3, Institut de Physique des 2 Infinis de Lyon, Lyon, France\\
$^{128}$ Universit\'{e} de Strasbourg, CNRS, IPHC UMR 7178, F-67000 Strasbourg, France, Strasbourg, France\\
$^{129}$ Universit\'{e} Paris-Saclay, Centre d'Etudes de Saclay (CEA), IRFU, D\'{e}partment de Physique Nucl\'{e}aire (DPhN), Saclay, France\\
$^{130}$ Universit\'{e}  Paris-Saclay, CNRS/IN2P3, IJCLab, Orsay, France\\
$^{131}$ Universit\`{a} degli Studi di Foggia, Foggia, Italy\\
$^{132}$ Universit\`{a} del Piemonte Orientale, Vercelli, Italy\\
$^{133}$ Universit\`{a} di Brescia, Brescia, Italy\\
$^{134}$ Variable Energy Cyclotron Centre, Homi Bhabha National Institute, Kolkata, India\\
$^{135}$ Warsaw University of Technology, Warsaw, Poland\\
$^{136}$ Wayne State University, Detroit, Michigan, United States\\
$^{137}$ Yale University, New Haven, Connecticut, United States\\
$^{138}$ Yildiz Technical University, Istanbul, Turkey\\
$^{139}$ Yonsei University, Seoul, Republic of Korea\\
$^{140}$ Affiliated with an institute covered by a cooperation agreement with CERN\\
$^{141}$ Affiliated with an international laboratory covered by a cooperation agreement with CERN.\\

\end{flushleft} 

\end{document}